\immediate\write18{makeindex \jobname.nlo -s nomencl.ist -o \jobname.nls}

\DeclareUnicodeCharacter{202F}{\,}

\documentclass[pdfa,chapterbib]{Thesis}

 \title{Modified Gravity and Regular Black Hole Models}             
 \author{Jose Abraham Pinedo Soto}
 \degree{\PhD}                   
 \department{Physics} 
 \convocationdate{2025}

	\abstracttext{
		The prediction of spacetime singularities—regions of infinite curvature where classical physics breaks down—is one of the most profound challenges in General Relativity (GR). In particular, black hole solutions such as the Schwarzschild and Kerr metrics feature central singularities that signal the incompleteness of the theory in high-curvature regimes. This thesis is devoted to the study of modified theories of gravity that aim to resolve these singularities and construct physically viable models of regular black holes: black holes that retain an event horizon but exhibit no curvature divergence anywhere in the spacetime.

The thesis is structured in three main parts. The first part investigates nonlocal modifications to classical field theories, including generalizations of Maxwell electrodynamics and linearized gravity in arbitrary dimensions. Within this framework, we derive stationary solutions for point-like sources for different choices of form factors specifying different nonlocal models. These solutions are regular due to the smearing effect introduced by nonlocal form factors, which suppress high-momentum contributions to the field and ensure a finite behavior even for point-like sources.

The second and most extensive part focuses on regular black holes in four-dimensional spacetimes, constructed from infinite-derivative gravity theories. Starting from the linearized equations with nonlocal operators, we explore how curvature singularities in Schwarzschild and Kerr black holes can be resolved. We analyze both static and rotating cases. For static configurations, we derive metrics that match the Schwarzschild solution at large distances but exhibit a smooth de Sitter core near the origin. These metrics are obtained by dressing the point-mass source with a nonlocal form factor, effectively spreading the mass-energy over a finite region. In the rotating case, we generalize the procedure to obtain non-singular counterparts to the Kerr metric, where the ring singularity is resolved. These models preserve the essential causal structure of black holes (including event horizons and ergoregions) while eliminating pathological divergences.

The final part of the thesis turns to higher-curvature theories and their spherical reduction. In this setting, we investigate a large class of analytic black hole solutions by specifying their spherically reduced action, depending on basic curvature invariants. Notably, we obtain regular black hole spacetimes that generalize the Hayward model and others, controlling the behavior of curvature invariants via the functional form of the action. We also examine the thermodynamic properties of these solutions, their near-horizon structure, and the role of extremal configurations.

Throughout the thesis, particular care is taken to ensure that the proposed models are physically sensible: they reduce to GR in the appropriate limits and lead to well-behaved curvature scalars. Collectively, these results highlight the viability of regular black hole models in modified gravity theories, providing a promising path toward a consistent semiclassical description of black holes that avoids the need for quantum gravitational resolution of singularities.
	}

	\preface{
I hereby declare that this Ph.D. thesis is the result of original research. All results presented herein were obtained using only the methods and sources explicitly stated, and all references have been properly cited.

The thesis is based on four publications: \cite{Boos_2020,Boos_2021,Frolov:2023jvi,rbh_pfkz}. The research contained in these works was carried out in close collaboration with Professor Valeri Frolov, Dr. Andrei Zelnikov, Dr. Jens Boos, and Alex Koek. The author of this thesis was actively and substantially involved in all stages of the research.

Each chapter—excluding Chapters~\ref{Intro} and~\ref{Concl}, which correspond to the introduction and conclusion, respectively—is closely based on one of these publications. The relevant reference is indicated at the beginning of each chapter. The appendices provide supplementary material that supports various parts of the thesis.

\begin{enumerate}
    \item J. Boos, J. Pinedo Soto and V. P. Frolov “Ultrarelativistic spinning objects in non-local ghost-free gravity,” Phys. Rev. D 101 (2020) no. 12, 124065 

    \item J. Boos, J. Pinedo Soto, and V. P. Frolov  "Ultrarelativistic charged and magnetized objects in nonlocal ghost-free electrodynamics"  Phys.Rev.D 103 (2021) 4, 045013 

    \item V. P. Frolov and J Pinedo Soto "Nonlocal Modification of the Kerr Metric"  Symmetry 15 (2023) 9, 1771 

    \item V. P. Frolov, A. Koek, J. Pinedo Soto and A. Zelnikov "Regular black holes inspired by quasitopological gravity" Phys.Rev.D 111 (2025) 4, 4 
\end{enumerate}

	}

	\thesisquote{``\lipsum*[14]{}''\\-Author}
	\dedication{To Nana}
	\acknowledgementtext{%
		Acknowledgments
		
	}

\begin{document}

 \maketitle                    
 \makeabstract
 \makepreface
 \makededication             


 \onehalfspacing              

 \tableofcontents              
 \listoffigures             

 \bodyoftext                 


 \truedoublespacing            


	\chapter{Introduction}
\label{Intro}

Centuries of scientific progress have led to the understanding that nature is governed by four fundamental forces, with gravity emerging as the dominant interaction on cosmic scales. A complete understanding of gravity is therefore essential to uncovering the universe’s origin and ultimate fate. Remarkably, both black holes and gravitational waves—two phenomena central to modern gravitational physics—were predicted by the theory of General Relativity (GR), formulated by Einstein in 1915. \cite{Einstein:1915ca,Misner:1973prb}. Over the past five decades, our understanding of the astrophysical role of black holes has undergone a profound transformation. This shift has been driven by significant advances in observational precision and the emergence of new observational channels, particularly in the electromagnetic and gravitational wave spectra. A landmark moment occurred on September 14, 2015, when the LIGO observatory made the first direct detection of gravitational waves, originating from the merger of two black holes, each with masses around 30 times that of the Sun \cite{PhysRevLett.116.061102}. Today, multiple gravitational wave detectors, including LIGO, Virgo, and KAGRA, operate as part of a coordinated international network. On March 19, 2025, this network recorded its 200th gravitational wave candidate event during the O4 observing run, with the majority of these detections attributed to binary black hole (BH-BH) coalescences \cite{dicesare2025statuso4runlatest}. These observations provide direct empirical evidence for the existence of hundreds of stellar-mass black holes.

Another milestone was achieved on April 10, 2019, when the Event Horizon Telescope (EHT)—a global network of radio telescopes—released the first horizon-scale image of a black hole, located in the center of the galaxy Messier 87 \cite{bh_image}. This was followed by the release on May 12, 2022, of the first image of Sagittarius A*, the supermassive black hole at the center of our own Milky Way galaxy \cite{SagA_Image}. These groundbreaking EHT observations, in combination with gravitational wave detections and other multi-messenger astrophysical data, reinforce the conclusion that black holes are widely present in the universe. There exists compelling evidence that nearly every massive galaxy hosts a massive or supermassive black hole at its core \cite{SMBH_and_HostGalaxy}.

A major breakthrough in black hole physics occurred in 1974 when Stephen Hawking demonstrated that black holes are not entirely black but emit thermal radiation due to quantum effects near the event horizon \cite{Hawking1974}. Although this radiation is negligible for astrophysical black holes, it could be significant for hypothetical microscopic primordial black holes, whose existence remains speculative but whose role could be important in the early universe \cite{PhysRevD:Carr2010}. Hawking’s discovery established that black holes possess thermodynamic properties—such as temperature and entropy—thereby linking classical gravity, quantum field theory, and thermodynamics. This insight elevated black holes to a central role in theoretical physics, serving as a “Rosetta stone” for uncovering connections between different branches of research. Today, black holes are widely employed as conceptual laboratories in string theory, holography, and other quantum gravity frameworks \cite{STROMINGER199699,MATHUR2005}.

What makes black holes particularly intriguing are the conceptual puzzles they pose. Some arise from internal inconsistencies within classical general relativity. In particular, many solutions to Einstein’s equations—including cosmological and black hole spacetimes—feature singularities: regions where scalar curvature invariants diverge and the classical theory ceases to be predictive. Such singularities signal the breakdown of general relativity and point to the need for a more fundamental theory. The aim of this thesis is to investigate how recently proposed modifications of Einstein’s gravity—especially those involving nonlocality or higher curvature terms—can provide a framework in which these singularities are resolved.

\section{Einstein Gravity and its Black Hole Solutions}

In the Einstein theory of gravity, the gravitational field is described by the four-dimensional metric $g_{\mu\nu}$, which is defined on a smooth manifold representing the spacetime. The line element 
\begin{equation} 
ds^2=g_{\mu\nu} dx^{\mu} dx^{\nu}
\end{equation} 
determines the interval between two close events in spacetime with coordinates $x^{\mu}$ and $x^{\mu}+dx^{\mu}$.
On the other hand, one has the matter content distribution, described by the stress-energy tensor $T_{\mu\nu}$, which affects the geometry of the spacetime. The corresponding metric can be found as a solution of the Einstein field equations
(see e.g. \cite{Misner:1973prb,Frolov:book,Carroll_2019}:
\begin{equation}\label{Einstein_FieldEqs}
R_{\mu\nu} - \frac{1}{2}g_{\mu\nu}R + \Lambda g_{\mu\nu} = \frac{8\pi G}{c^4}T_{\mu\nu} \, .
\end{equation}
Here $R_{\mu\nu}$ is the Ricci tensor, $R=g^{\mu\nu}R_{\mu\nu}$ is the Ricci scalar. These equations encapsulate the theory of General Relativity (GR) and describe how matter impacts the surrounding space. The cosmological constant $\Lambda$ will not be relevant for the problems considered in the thesis; therefore, we will ignore it going forward.
We mainly use units in which Newton’s coupling constant $G$ and the speed of light $c$ are equal to 1: $G=c=1$. We also adopt the metric signature $(-,+,+,+)$.

General Relativity stands as one of the most successful and well-tested physical theories in the history of science. In the weak field limit, physical phenomena predicted by the post-Newtonian approximation, like the precession of the perihelion of Mercury \cite{Einstein:1915bz}, the gravitational redshift of electromagnetic radiation \cite{PhysRevLett:redshift_gammarays,PhysRevLett:redshift_hydrogen},
light deflection \cite{Dyson1920}, Shapiro time delay \cite{PhysRevLett:ShapirotimeDelay}, and frame dragging \cite{Ciufolini2004,PhysRevLett:framedragging} have all been experimentally verified. The existence of black holes and the success of current cosmological models provide confirmation of general relativity in the strong-field regime.

The theory also makes additional strong-field predictions that have been confirmed in recent years, such as decreasing orbital periods due to gravitational damping \cite{Weisberg_2010}, the direct observation of gravitational waves \cite{PhysRevLett.116.061102} and the shape of the black hole shadows \cite{bh_image,SagA_Image}. For a general review on the status of General Relativity and its experimental confirmations, along with future prospects, see \cite{Will2014,Will_2018}.

The first solution to Einstein's field equations describing the vacuum static spherically symmetric spacetime was discovered by Schwarzschild in 1916 \cite{schwarzschild1999gravitationalfieldmasspoint}. It has the form
\begin{equation}\label{SCHW}
\begin{split}
ds^2&=-f dt^2+\dfrac{dr^2}{f}+r^2 d\Omega^2\, ,\\
f&=1-\dfrac{2m}{r}\, ,
\ \ \ 
d\Omega^2=d\theta^2+\sin^2\theta d\phi^2
\, .
\end{split}
\end{equation} 

This metric describes a black hole in an asymptotically flat spacetime. In this solution, $m$ is the black hole mass, and $d\Omega^2$ is the metric of the two-dimensional round sphere.
It is a solution that describes an object which was not initially believed to be physical due to some of its unusual properties. 

However, it was later realized that such an object could exist due to the gravitational collapse of matter, like dust or a massive star, provided all the matter collapses beyond the Schwarzschild radius $r_S = 2m$ \cite{OppVo, OppSny}. Their existence has been confirmed through both indirect and direct observations, along with the assistance of gravitational wave detectors \cite{bh_image,SagA_Image,PhysRevLett.116.061102, PhysRevLett.118.221101, Abbott_2021, M872019}, as we have already discussed. 

In 1968 John Wheeler coined the term "black hole" to describe these objects \cite{Wheeler1968}.

\subsection{Rotating Black Holes}

A stationary axisymmetric solution of the Einstein field equations describing a rotating black hole was later found by Roy Kerr \cite{PhysRevLett.11.237}. The Kerr metric and some of its properties are presented in Appendix~\ref{app:KerrMetric}. It is important to emphasize that the Kerr solution plays a central role in both theoretical and observational aspects of black hole physics. Due to the dynamical nature of astrophysical processes, it is widely expected that most black holes in the universe possess nonzero angular momentum.

Rotation has significant physical consequences. For instance, the spin of a black hole is thought to be one of the primary drivers (along with the magnetic field strength and accretion process) behind the launching of powerful relativistic jets via mechanisms of energy extraction from the black hole’s rotational energy \cite{EMextractionKerr,BHmodelsAGN}. These jets, in turn, influence star formation rates and the thermal history of the interstellar medium in their host galaxies, mainly suppressing star formation by heating and expelling gas \cite{Fabian:GalaxyFormation}. Moreover, the inclusion of spin is essential for accurate modeling of accretion disks, which are central to understanding both active galactic nuclei and X-ray binaries \cite{accretionSBH1,accretionSBH2}.

In addition, one of the outstanding puzzles in modern astrophysics is the origin of supermassive black holes (SMBHs) observed at high redshifts \cite{earlySMBH}. Their spin may encode crucial information about the formation mechanisms that produced them, whether through hierarchical mergers, direct collapse, or sustained accretion processes \cite{Spin:SMBH}. Consequently, measuring black hole spin has become a major observational priority \cite{reynolds2019observingblackholesspin}.

From a theoretical perspective, it is essential to understand how rotating black holes interact with their environment. In realistic astrophysical settings, black holes are not isolated, and their interactions with surrounding matter and fields can give rise to several important physical effects. These include feedback mechanisms, gravitational wave emission, and possible imprints on the distribution of dark matter and galactic structure. Such phenomena are of particular interest for identifying potential observational signatures; see, for example, \cite{BHingalaxies,Frolov:2024xyo,Frolov:2024tiu,BHinDM} and references therein.

In this thesis, we do not address the astrophysical implications or deviations that spinning black holes in modified gravity theories might exhibit compared to General Relativity. Nevertheless, exploring rotating regular solutions from a theoretical perspective is essential for tackling such questions in future work. To that end, Chapter~\ref{Ch4} is devoted to constructing a model for a regular spinning black hole. Our focus is on developing a solution inspired by the framework of infinite derivative gravity, with the goal of producing a model that is best suited for future studies involving realistic astrophysical scenarios, and which hopefully captures some of the features of a complete non-linear infinite derivative gravity theory.

\subsection{The Singularity Problem}

The metric \eqref{SCHW} has two singular points $r=2m$ and $r=0$. The first one is a coordinate singularity that can be excluded by a proper coordinate transformation. For example, one can use the advanced time $v$  instead of the coordinate $t$ 
\begin{equation} 
v=t+F(r)\, ,\ \ \ 
F(r)=\int \dfrac{dr}{f}\, .
\end{equation} 
Then the metric \eqref{SCHW} takes the form
\begin{equation} \label{ADV}
ds^2=-f dv^2+2 dv dr+r^2 d\Omega^2 \, .
\end{equation} 
The coordinates $(v,r,\theta,\phi)$ cover both the black hole exterior (where $f>0$), the black hole interior (where $f<0$), and they are also regular at $r=2m$, where the black hole horizon is located. The metric \eqref{ADV}, however, is still singular at the point $r=0$. It turns out that this singularity cannot be removed by any coordinate transformation and represents a real physical singularity under General Relativity's description of gravity. Similar singularities exist in cosmology at the beginning of the Universe in the Big Bang.

Both types of singularities are, in fact, curvature singularities. In cosmology, where the Weyl tensor vanishes, a divergent scalar curvature invariant is $R_{\mu\nu}R^{\mu\nu}$. For a black hole in an empty spacetime, the Ricci tensor vanishes, so the divergent scalar invariant is now the square of the Weyl tensor $C_{\mu\nu\alpha\beta}C^{\mu\nu\alpha\beta}$. Usually, it is convenient to use the so-called Kretschmann invariant

\begin{equation}
K=R_{\mu\nu\alpha\beta}R^{\mu\nu\alpha\beta}=C_{\mu\nu\alpha\beta}C^{\mu\nu\alpha\beta}+ \frac{4}{D-2} R_{\mu\nu}R^{\mu\nu} - \frac{2}{(D-1)(D-2)}R^2   
\, ,
\end{equation} 
where $R_{\mu\nu\alpha\beta}$ is the Riemann tensor and $D$ is the number of spacetime dimensions, since this curvature scalar is divergent for both types (black hole and cosmology) of solutions.\footnote{
In the famous papers by Penrose and Hawking it was given
a general proof of geodesic incompleteness in cosmological and black hole spacetimes. This incompleteness can be identified with some kind of singularity (See \cite{Penrose:1964wq,Hawking:1976ra,Senovilla:2021pdg,Senovilla:2022vlr}). See also a recent discussion of this subject in the paper by Kerr \cite{kerr2023blackholessingularities}. }

For the Schwarzschild metric \eqref{SCHW}, the Kretschmann  scalar invariant is
\begin{equation}\label{KreschSchw}
K = \frac{48 m^2}{r^6}.
\end{equation}
The expression for the Kretschmann invariant for the Kerr metric can be found in Appendix~\ref{app:KerrMetric}.

The existence of curvature singularities poses a fundamental problem for the theory of General Relativity. At the very foundations of this theory lies the notion of ideal clocks and rulers, which are used to determine an event's location in space and time. Such objects should have some internal structure and be extended, but since the curvature is responsible for tidal forces acting on extended objects, when it infinitely grows, none of the extended objects, such as clocks or rulers, can exist. Under these conditions, our usual concept of spacetime fails, as it is not possible to make any prediction or measurement for any event in it. This issue is sometimes referred to as the theory "breaking down", and it means that General Relativity, predicting the existence of curvature singularities which represent a contradiction to its underlying and fundamental assumptions about spacetime, is incomplete. 

It is generally assumed that General Relativity is an effective theory valid at large scales and low energies. In the small scale and high energy regimes, it should be modified. There are many proposals for how such a modification can be constructed. A general expectation is that the resolution to this problem is to emerge from the fusion of quantum mechanics with General Relativity in a theory known as quantum gravity. Quantum gravity becomes important in scenarios where both theories must coexist, such as it is very small distances from the most potent gravitational force in the universe, the center of a black hole. Formulating a consistent theory of quantum gravity is inherently challenging, as it must adhere to the tested behavior of both quantum mechanics and General Relativity. For this reason, this thesis focuses on theories aiming to rectify the behavior of General Relativity at small distances while maintaining its accuracy at large distances, where it has been extensively validated, our approach is from a classical sense as we do not deal with quantum effects, but it may be regarded as some effective approach that captures some of the classical effects of a more fundamental quantum gravity theory. In this approach, one assumes that the gravitational field is still described by a classical metric $g_{\mu\nu}$ and that the equations that it satisfies are different from the Einstein equations in the spacetime domains where curvature is high. There are many ways in which one can modify gravity and construct an effective action for such a modified theory. One can consider, in addition to the metric, other fundamental fields and accept higher in curvature terms and/or higher derivatives of the curvature in the action. One should naturally expect that these modifications in the low-energy regime reproduce General Relativity and thus pass all of the successful experimental tests of GR. Since there exists a large variety of modified gravity models, it is difficult to summarize all these attempts. Instead of trying to do this, we refer the reader to nice comprehensive reviews \cite{Clifton:2011jh,  Nojiri:2017ncd, PhysRevD:MazumdarHigher}.

\section{Regular Black Hole Models}

In this thesis, we focus mainly on the problem of singularities inside the "standard" black holes, where by standard we mean solutions of General Relativity that possess both an event horizon and a curvature singularity; examples include the Schwarzschild and Kerr metrics. It is important to emphasize that these features are generally regarded as mathematical idealizations, stemming from simplifying assumptions rather than physically realistic descriptions. For instance, the event horizon is not directly observable by any finite-time observer \cite{Horizon:Visser}. In fact, in order to find its position at some given time one needs to know the null geodesics behavior up to an infinite time in the future. For this reason, in the discussions of the dynamical black holes, one often uses apparent or trapped horizons. For stationary vacuum black holes, the outer apparent horizon coincides with the event horizon. Similarly, the singularity is often interpreted as our ignorance of matter behavior under extreme conditions, such as gravitational collapse \cite{crowther2021attitudessingularitiessearchtheory}.

Because of this, it is natural that there exist many publications in which different mechanisms for "smoothing out the singularities" are proposed. To escape the formation of the singularity and to preserve the standard description of gravity as a metric on a smooth spacetime manifold, one often assumes that when the curvature becomes high, the Einstein theory becomes modified by corrections which become important in this regime. Such a modification usually contains a length scale parameter $\ell$ which specifies the "size" of the spacetime domain where the solutions of these new modified gravity equations become significantly different from the solutions of the Einstein equations. In particular, the solutions are modified in the short-distance regime such that they remain regular at the position where there would be a singularity otherwise, or in the case of black holes, at their core. We are interested in exploring such modifications, focusing on constructing black hole solutions with a non-singular core. We refer to these objects as \textit{regular black holes}, and define them by the following criteria:
\begin{itemize}
    \item They possess trapped regions and outer horizons.
    \item They are geodesically complete \footnote{ This is needed in order to cure the singularity from the perspective of the Hawking and Penrose singularity theorems. While we will not explicitly show this property, it is expected that a regular black hole with a regular core will satisfy it. }.
    \item Scalar curvature invariants remain finite at the center, replacing the singularity with a finite core.
\end{itemize} 

This is an approach that has several interesting consequences. In cosmology, for example, instead of the Big Bang, one can expect a bouncing cosmological scenario, in which our expanding Universe has a pre-big-bang past (see for example \cite{Bamba_2014,PhysRevD:loopbounce,Biswas_2010_bounce,Biswas_2012_bounce}). For black holes, it opens the possibility of new universe(s) forming inside of them (see for example \cite{bueno2024, Bambi2014,PhysRevD:Rastgoo,FROLOV1989272,bronnikov_2018,Easson_2001}) as well as traversable wormholes \cite{Simpson_2019, visser:RQBH, Franzin_2021}. In this thesis, we will discuss some aspects of modified gravity theories and regular black hole models.

Before continuing further with our discussion, we will address the following, rather natural, question: For black holes whose horizon size is much larger than the fundamental scale $\ell$, the high curvature region is located deep inside the black hole. Why, if at all, does one need to care about the spacetime properties in this domain, which is invisible for an external observer? Besides general curiosity about the fundamental laws of nature and the desire to have a complete theory of gravity, one might be interested in different physical scenarios, including, for example, bouncing cosmologies. This is particularly interesting because at least some primordial black holes can be formed in the pre-big-bang universe, which could, at least in principle, explain the sizes of the supermassive black holes found at the center of some galaxies \cite{Rees_Volonteri_2006,Volonteri2021,earlySMBH}. We shall not address this problem in the thesis. But even in order to be able to formulate and discuss such a problem, one needs to at least have some model of a regular black hole.  

For small mass black holes, the problem of the singularity is closely connected to the problem of the final state of their quantum evaporation. In particular, models of regular black holes might give an interesting new look at the information loss paradox \cite{Hayward,bronnikov_2018,Easson_2001,Frolov:1979tu,FROLOV1981307,Frolov2014,CarballoRubio2018,PhysRevD:Kunstatter}.
In the discussion of regular black hole models, there exist different approaches, some of which are:
\begin{itemize}
    \item The discussion of properties of regular metrics that possess some required properties, but that in general are not solutions of some effective gravity equations.
    \item  Regular black holes in low-dimensional spacetimes and in dilaton gravity.
    \item Regular black hole metrics obtained as solutions of some dynamical effective gravity equations obtained from some covariant gravity action.
\end{itemize}
A very well-known example of the first type is the so-called Hayward metric \cite{Hayward}
\begin{equation}\label{HAYWARD}
\begin{split}
ds^2&=-f dt^2+\frac{dr^2}{f}+r^2 d\Omega^2\, ,\\
f&=1-\frac{2m r^2}{r^3+2m \ell^2}\, ,
\end{split}
\end{equation} 
The parameter $m$ is the black hole mass as measured by a distant observer, $\ell$ is the length parameter that characterizes the scales where this metric deviates from the Schwarzschild metric. For the case $\ell\to 0$, the metric \eqref{HAYWARD} reduces to the latter, given by \eqref{SCHW}.
Near the center $r=0$, the metric function $f(r)$ has the following asymptotic form
\begin{equation} 
f(r)=1-\frac{r^2}{\ell^2}+\ldots \, ,
\end{equation} 
the spacetime curvature is constant and proportional to $\ell^{-2}$.  It is often said that this metric has a "de Sitter-like core". 

The metric \eqref{HAYWARD} is interesting since it has another important property: Its quadratic in curvature scalar invariants are always bounded by some values $C/\ell^4$, where $C$ is a dimensionless constant which depends on the type of the scalar invariant under consideration, but remarkably it does not depend on the parameter $m$ of the solution. In this sense, this bound of the curvature is universal; this property is known as Markov's limiting curvature conjecture \cite{limcur_markov} (see discussion in section \ref{sec:Hayward}).

Let us emphasize that in a general case, this condition imposes restrictions on the form of the metric of a regular black hole. Let us also note that the metric was proposed ad hoc, without any derivation from any modified gravity equations or modified gravity covariant action. Generalizations of the Hayward metric were discussed in \cite{FrolovMetric}, and its relation to the information loss paradox can be found in \cite{Frolov2014}.

 The Hayward metric is a nice example since it features most of the properties that many regular black models share:
\begin{itemize}
    \item The singularity is replaced by a smooth region with finite curvature invariants, consistent with Markov’s limiting curvature conjecture \cite{limcur_markov}.
    \item A small length scale $\ell$  governs deviations from classical behavior.
    \item The core geometry typically approaches de Sitter space, in line with expectations for high-density matter \cite{Gliner1966}.
    \item The Schwarzschild solution is recovered in the asymptotic limit $r \to \infty$.
    \item The spacetime contains at least two horizons.
\end{itemize}

While these are not strict requirements, they are common to many models in the literature. Notably, recent work has identified classes of theories in which regular black hole solutions can be systematically derived from action principles \cite{QT_BH}.  A discussion of other regular black hole metrics can be found in \cite{nonsingularparadigmblackhole,PRL:Garcia,PRD:Bronnikov,Dymnikova_2004,Berej2006,BALART201414,bronnikov_2018,ZHI-2024}.

Another key challenge lies in understanding the formation and evolution of regular black holes. Several works have addressed dynamical formation, including thin-shell collapse \cite{bueno2024,bueno2024shell} and the Oppenheimer–Snyder model \cite{bueno2025oppenheimersnyder}. Energy-dependent couplings inspired by asymptotic safety have also led to regular black hole formation models \cite{PRL:Bonnano}. Dust collapse models in higher order extensions of General Relativity were discussed in \cite{Buoninfante:2024oyi}.

In 2D dilaton gravity, quantum corrections have been shown to lead to regular collapse outcomes \cite{PRD:Ziprick,PRD:taves}, with potential embeddings in 4D spacetimes. Recent simulations suggest that such models form both trapped and anti-trapped surfaces \cite{Barceló_2015,mdpi:barcelo}, and may lead to unitary evaporation, highlighting their viability as physical black hole alternatives. Different solutions describing regular 2D black holes in dilaton gravity were further discussed in e.g. \cite{PhysRevD:lemos,PhysRevResearch:frolov, PhysRevD:Kunstatter}.

At this point, it is important to point out that regular black hole models like the ones we described are not the only proposed alternative to black holes. For example, there exist objects like the so-called \textit{black hole mimickers}, which are horizonless geometries often proposed in alternative theories to address the singularity issue. These are defined as spacetimes that do not contain trapped regions, instead, they have a finite redshift surface, sometimes referred to as the “surface”, which need not be hard, and they may describe ultracompact stars or traversable wormholes \cite{pnas:gravasars,Visser:gravastars,Mathur:fuzzball,Guo:fuzzball,Visser:DarkStar,PRD:blackshells2,Adler_2025,PRD:bosonstars,PRL:bosonstars,PRD:blackshells,Mazumdar_mimicker}. These objects may dynamically evolve into standard black holes or regular black holes \cite{Lemos_2008,PhysRevD:Casadio}. Specifically, it has been argued that black hole mimickers can form event horizons once a critical mass is reached, regardless of the underlying microscopic theory \cite{PhysRevD:Carballo_HBH,PhysRevD:Visser_PhenoBH}. For more exhaustive taxonomies about the alternatives to standard black holes, see \cite{Carballo-Rubio_2020, carballo:geodesicallycompleteBHs}. 

We will not consider mimickers or other alternatives to standard black holes in this thesis. Instead, we will discuss models of regular black holes based on and inspired by two types of modified gravity: the so-called infinite derivative gravity and quasi-topological gravity. 

\section{Field of Point-Like Sources and Regularizing Mechanisms}

Let us write the metric \eqref{ADV} in the form
\begin{equation} \label{DOUBLE}
ds^2=ds_{\mathrm{flat}}^2+2\Phi dv^2\, ,
\end{equation} 
where $ds_{flat}^2=-dv^2+2dv dr+r^2 d^2\Omega$ is the metric of a flat (Minkowski) spacetime and 
\begin{equation} 
\Phi=-\dfrac{m}{r}\, .
\end{equation} 
One can say that in $(v,r,\theta,\phi)$ coordinates, the Schwarzschild solution of the Einstein equations is a linear superposition of the flat metric and another part associated with the gravitational potential of a point-like particle of mass $m$ in the Newtonian gravity theory. 

Certainly, since the Einstein equations are non-linear, one could not expect that a simple linear superposition of two metrics would give an exact solution of these equations. Interestingly, for the Schwarzschild and Kerr metrics, such a superposition has a well-defined meaning. The approach, which explains and uses this property, known as the Kerr-Schild double copy formalism, is discussed later in Chapter \ref{Ch4} of the thesis; additional information may also be found in Appendix~\ref{app:KerrSchild}. An important feature of this approach is that in both cases it allows one to reduce the problem of solving the complicated nonlinear Einstein equations to solving the linear Poisson equation for a point-like source in a flat space.

At the moment, using the representation \eqref{DOUBLE}, one can conclude that at least formally, the problem of the black hole singularity is somehow connected with the problem of the singularity of the gravitational potential $\Phi$ for a point-like particle in Newtonian gravity. In other words, one can expect that if there exists some mechanism that prevents the curvature singularity formation inside the black hole, there may exist a similar mechanism that regularizes the corresponding Newtonian field of a point-like mass.

A similar problem exists in Maxwell's electrodynamics. Namely, the potential $\Phi$ of a point-like electric charge $e$ 
\begin{equation} 
\Phi=  \frac{e}{r}\, ,
\end{equation} 
is divergent at the point $r=0$ where this charge is located
\footnote{
Here, the electric potential is written in Gaussian units. Later, we shall also use Heaviside units in which the electric potential takes the form $\Phi=e/(4\pi r)$.
}. As a result, the self-energy of a point charge diverges. This is an old problem that has been intensively discussed a long time ago in attempts to develop a classical model of an electron \cite{PhysRev:weisskopf,bovy2006selfenergyelectronquintessentialproblem,Weisskopf_1994,deAssis:2022itz}.

Among other attempts to regularize the field of a point-like source, the following two alternatives have been proposed, both of which require a modification of the theory at small scales:
\begin{itemize}
    \item A non-local modification of the theory, by introducing an infinite number of derivatives acting on the field equations;
    \item A non-linear modification of the theory, considering higher order in powers of the field contributions in the field equations.
\end{itemize}
Let us illustrate how these mechanisms work.

\subsection{Infinite Derivative Model of a Massless Scalar Field}\label{IDscalar}

Later, we will discuss in detail the infinite-derivative modifications
of the  Maxwell theory (Chapter~\ref{Ch2}) and of the linearized gravity theory (Chapter~\ref{Ch3}). 
Here, for illustration purposes, we briefly discuss a simpler case. Namely, we consider a four-dimensional {\it scalar massless field} $\phi$. Consider its action in the presence of an external current $J(x)$, which is:
\begin{equation}\label{ACT}
S[\phi] = \int d^4x\left(\frac{1}{2} \phi \Box  \phi -   J(x)\phi\right) \, .
\end{equation}

For the static case, the corresponding field equation is
\begin{equation}\label{KG}
\Delta \phi = J \, .
\end{equation}
For a static point-like source, $J = g\delta^{(3)}(\vec{r})$, the solution is
\begin{equation}\label{KGSOL}
\phi(r) = - \dfrac{g}{4\pi r} \, .
\end{equation}

Consider the following infinite-derivative (also known as nonlocal) modification of the  action \eqref{ACT}:
\begin{equation}\label{NACT}
S_{NL}[\phi] =\int d^4x\left( \frac{1}{2}\phi f(\Box) \Box \phi - J(x)\phi \right) \, .
\end{equation}
Here, $f(\Box)$ is an object referred to as \emph{form factor}. We assume that $f(0) = 1$ to ensure that in the local limit, when the effects of nonlocality are negligible, the usual local action \eqref{ACT} is recovered. The modified equation of motion is then
\begin{equation}\label{NLscalareq}
f(\Delta)\Delta \phi = J \, .
\end{equation}

We will discuss the possible choices of the form factor in detail later. At the moment, we set $f(\Delta) = e^{-\ell^2 \Delta}$. This particular choice of the form factor has several convenient properties, and it is widely used in the literature. Using the Green function method, we can find the solution for a point-like source
\begin{equation}
    \phi(r) = -gG(r) \, .
\end{equation}
Where $G(r)$ is the Green function that solves the equation
\begin{equation}
    f(\Delta)\Delta G(r) = -\delta^{(3)}(r) \, .
\end{equation}
It can be found by using its Fourier transform, and using the following representation of the $\delta$ function
\begin{equation}\label{delta}
    \delta^{(3)}(\vec{r}) = \frac{1}{(2\pi)^3}\int e^{-i\vec{k}\cdot\vec{r}}d^3\vec{k} \, ,
\end{equation}
one finds that 
\begin{equation}
    G(r) = \frac{1}{(2\pi)^3}\int \frac{e^{-i\vec{k}\cdot \vec{r}} e^{-\ell^2 k^2}}{k^2}d^3\vec{k} 
    \, .
\end{equation}
For our choice of form factor, the integral above can be solved in terms of the error function $\mathrm{erf}(z)$ \cite{NIST2010}, yielding the solution
\begin{equation}
\phi(r) = -\frac{g}{4\pi r}\mathrm{erf}\left(\frac{r}{2\ell}\right) \, .
\end{equation}
This solution is finite at the origin, in contrast with solution \eqref{KGSOL}. In the regime where $r \gg 2\ell$, it reproduces the standard potential \eqref{KGSOL}. Hence, nonlocal effects become relevant only at distances comparable to the scale $\ell$, which has dimensions of length and therefore sets the scale of nonlocality. At large distances, the theory smoothly reduces to the standard local behavior as well. This regularization method via a nonlocal Green function has been extensively used in the literature in the context of nonlocal gravity to, for example, find a mass gap in black hole formation \cite{Frolov_2015}, the study of ultrarelativistic objects \cite{Frolov_2016, Boos_2020, Boos_2021}, collapse of small masses \cite{frolov2015sphericalcollapsesmallmasses} and much more \cite{Giacchini_2019WFL,ScalarKernel,Heredia_2022,Kol__2021,BraneGF}.

There is an alternative way to understand this result. One can write the equation \eqref{NLscalareq} as
\begin{equation}\label{eq_smeared}
    \Delta\phi = g\tilde{\delta}(\vec{r})\, , 
   \quad  \tilde{\delta}(\vec{r})=f^{-1}(\Box)\delta^{(3)}(\vec{r})\, .
\end{equation}
So that the nonlocality is now embedded in the source, having some sort of smearing effect, instead of acting as an operator in the field equations. By using the Fourier transform of the delta function \eqref{delta},
is possible to find an expression for the distribution of the smeared source as
\begin{equation}\label{deltasmeared}
   \tilde{\delta}(r)= \frac{1}{2\pi^2 r}\int_0^{\infty}dk k\sin(kr)e^{-\ell^{2} k^2}=
\dfrac{e^{-r^2/(4\ell^2)}}{8\pi^{3/2} \ell^3}   
     \, .     
\end{equation}

For the form factor we have under consideration, the function $f(-k^2) =\exp(\ell^2 k^2)$, which does not have any zeros, the integral representation \eqref{deltasmeared} converges for large $k^2$. The potential $\phi$ is therefore a solution of the equation \eqref{eq_smeared} with the corresponding smeared distribution of the charge $\tilde{\delta}(\vec{r})$. Since $\phi(r)$ is a function of $r$, the field equation becomes 
\begin{equation}
    \frac{1}{r^2}\frac{\partial}{\partial r} \bigg( r^2 \frac{\partial}{\partial r}\phi(r) \bigg) = g\tilde{\delta}(r)
\end{equation}
The solution can then be written in the form
\begin{equation}
    \phi(r) = g\int_{\infty}^{r} dx\frac{g_{\mathrm{sm}}(x)}{x^2} = - \frac{g}{4\pi r} \mathrm{erf}(r/2\ell)\, , 
\end{equation}
where $g_{\mathrm{sm}}(r)$ is the effective charge function 
\begin{equation}
    g_{\mathrm{sm}}(r) = \int_0^{r} dx x^2  \tilde{\delta}(x) = \frac{\mathrm{erf}(r/2\ell)}{4\pi} - \frac{re^{-r^2/4\ell^2}}{4\ell\pi^{3/2}}\, .
\end{equation}

To summarize these results, one can think that non-locality, through effectively smearing the point-like source, regularizes the field divergence. This alternative approach is also used in the literature of local and non-local gravity models. For applications following this approach, see, for example, \cite{Burzill__2021,Giacchini_2019,Tseytlin_1995,dePaulaNetto:2023vtg,Buoninfante_2018Ring}.

\subsection{Non-Linear Electrodynamics and Regularization}\label{NLelectro}

A similar problem to that of the divergence of the self-energy of a point-like charged particle exists in the standard Maxwell electrodynamics theory as well. Let us remind that Maxwell's equations follow from the action
\begin{equation}\label{MaxwellCh5}
S[A_{\mu}]= - \frac{1}{4}\int d^4x F_{\mu\nu} F^{\mu\nu}\, ,
\end{equation} 
where $F_{\mu\nu}=A_{\mu ,\nu}-A_{\nu ,\mu}$.
For a static spherically symmetric field $A_{\mu}=\delta_{\mu}^t \Phi(r)$, the solutions of the Maxwell equations for a point-like charge $e$ is
\begin{equation} 
\Phi(r)= \frac{e}{4\pi r}\, .
\end{equation} 
The electric field of the charge becomes singular as it approaches $r \to 0$, where the point-like charge is located. This behavior can be improved by considering a properly constructed non-linear generalization of the Maxwell theory. In order to do so, one may consider the two scalar invariants which can be obtained from the field $F_{\mu\nu}$, namely 
\begin{equation} 
I=\dfrac{1}{2}F_{\mu\nu}F^{\mu\nu}\, , \quad  J=\dfrac{1}{4}e^{\mu\nu\alpha\beta}F_{\mu\nu}F_{\alpha\beta}\, ,
\end{equation} 
where $e^{\mu\nu\alpha\beta}$ is the Levi-Civita tensor.
A generic action for a non-linear electrodynamics would be
\begin{equation} 
S[A_{\mu}]=- \frac{1}{4}\int d^4 x L(I,J)\, .
\end{equation} 

Let us consider a specific case of such a non-linear electrodynamics model, which is known as the Born-Infeld theory \cite{BornInfeld1934}. In order to study static solutions, we use the Lagrangian density of this theory
\begin{equation} \label{BIL}
L=b^2\bigg(
1-\sqrt{1+\dfrac{F_{\mu\nu}F^{\mu\nu}}{2b^2}}
\bigg)\, .
\end{equation} 
In the absence of the magnetic field, this Lagrangian density reduces to \cite{alam2022reviewborninfeldelectrodynamics}
\begin{equation} 
L=b^2\bigg(
1-\sqrt{1-\dfrac{\vec{E}^2}{b^2}}  \bigg)\, .
\end{equation} 
where $\vec{E}= -\nabla \Phi$, and $\Phi$ is the electric scalar potential.
Here $b$ is some large constant, which has the meaning of the maximal field strength value. In the limit where $b^2\gg F_{\mu\nu}F^{\mu\nu} $ the Lagrangian \eqref{BIL} reduces to the standard Maxwell expression \eqref{MaxwellCh5}.

Denote the electric displacement field $\vec{D}$ as 
\begin{equation} \label{D_eq}
\vec{D}=  \dfrac{\partial L}{\partial \vec{E}}=\dfrac{\vec{E}}{\sqrt{1-\dfrac{\vec{E}^2}{b^2}}}\, .
\end{equation} 
Then, the field for a point-like charge obeys the equation
\begin{equation} 
\nabla\cdot \vec{D}= e \delta(\vec{r})\, .
\end{equation} 
Since we are considering the static and spherically symmetric case, the solution will only depend on the $r$ coordinate and be in the $\hat{r}$ direction. Such a solution is 
\begin{equation} 
\vec{D}(r)=\dfrac{e}{4\pi r^2} \hat{r}\, .
\end{equation} 
Then, we can use this expression to find the electric field $\vec{E}$ by using \eqref{D_eq}
\begin{equation} \label{EEE}
\vec{E}= \dfrac{e}{\sqrt{(4\pi)^2r^4+(e/b)^2}}\hat{r}  \, .
\end{equation} 
The electric field $\vec{E}$ in this case is bounded to have a maximum value determined by the parameter $b$, it reduces to the standard Coulomb field in the far distance limit, and now it remains finite near $r=0$. The scalar potential $\Phi$ explicit form can be obtained from \eqref{EEE} as 
\begin{equation}\label{xxyy}
    \Phi(r) = \sqrt{\frac{eb}{4\pi}} \int_{y}^{\infty} \frac{dx}{\sqrt{1+x^4}} = b\bigg( C - \frac{1}{2}F(2  \arctan y,\frac{1}{\sqrt{2}}) \bigg) \, ,
\end{equation}
where $y = r\sqrt{4\pi b/e}$, $F(z,k)$ is the elliptic integral of the first kind \cite{NIST2010} and $C$ is some constant which is related to $b$ \cite{alam2022reviewborninfeldelectrodynamics}. This scalar potential $\Phi$ is regular at $r=0$.

This is a concrete example that demonstrates how the inclusion of non-linearity in a given theory can cure the singular behavior of the field of a point-like source by making the field regular at the origin, where the particle is located.

\section{Modified Gravity Models}

These examples illustrate two distinct approaches within the broader landscape of modified gravity theories. As discussed earlier, both aim to regularize field equations, albeit through different mechanisms. In this thesis, we focus specifically on infinite derivative gravity, which achieves regularization through nonlocal operators, as introduced in subsection~\ref{IDscalar}, and on quasitopological gravity, which employs higher-curvature corrections, analogous to what was shown in subsection~\ref{NLelectro}. We now provide a brief review of each framework to lay the groundwork for the constructions developed in the subsequent Chapters.

\subsection{Infinite Derivative Gravity}

Infinite derivative gravity (sometimes referred to as nonlocal gravity) is a class of modified gravity theories that aim to resolve the singularities of general relativity and provide a consistent ultraviolet completion of gravity. These models achieve this by incorporating infinite-derivative operators into the action, leading to nonlocal modifications of the gravitational interaction. The approach has attracted significant attention due to its ability to remain ghost-free at the classical level, regularize spacetime singularities, and enable novel cosmological dynamics such as bouncing universes.

The foundational works found in \cite{NONLOC,Biswas_2014} introduced the most general quadratic curvature action with nonlocal form factors constructed from entire functions of the d'Alembertian operator. These entire functions—typically of exponential type—are chosen to avoid new pole structures in the propagator, thereby eliminating Ostrogradsky ghosts at the linearized level. Ostrogradsky’s theorem \cite{Ostrogradsky1831} shows that if a Lagrangian depends non-trivially on time derivatives higher than second order, the Hamiltonian of the system becomes linear in at least one momentum variable. This linearity means the energy is unbounded from below, leading to inherent instabilities \cite{woodard2015theoremostrogradsky}. In essence, generic higher-derivative dynamical systems cannot describe stable physical theories unless additional constraints or symmetries eliminate the problematic degrees of freedom. The resulting field equations remain second order in the $\ell\to 0$ limit, while incorporating infinite-derivative nonlocality in the full theory. 

At the classical level, infinite derivative ghost-free gravity modifies the Newtonian potential and makes it finite at the origin of a point-like mass.

It was demonstrated that black hole-like solutions in these models possess a non-singular core, typically resembling de Sitter or Minkowski space \cite{Buoninfante_2018}.

In cosmology, these modifications can lead to non-singular bouncing solutions \cite{Biswas_2010_bounce,Biswas_2012_bounce}. The standard Big Bang singularity is replaced by a smooth transition from contraction to expansion. These features are typically achieved without the need for exotic matter or violation of energy conditions. Notably, the presence of nonlocal form factors acts as a regulator that softens the high-curvature regime. Earlier proposals suggested that nonlocal terms could also serve as an effective dark energy \cite{BARVINSKY201212}, offering infrared modifications consistent with late-time cosmic acceleration. More recent work \cite{Koshelev2023} builds on this to develop predictive inflationary models using ghost-free nonlocal actions. These scenarios naturally interpolate between early-time inflation and late-time acceleration, with predictions compatible with CMB observations.

While infinite derivative gravity theories are ghost-free at the tree level, their quantum stability remains a subject of ongoing research. It has been pointed out that nonlocal interactions can reintroduce infinitely many complex poles at the loop level \cite{SHAPIRO201567}. This suggests that ghost freedom may be a feature of the classical or perturbative regime only, and that a more complete UV description might be required to maintain consistency beyond the tree level.

The literature is vast, and a complete review of it is not the point of this section; we refer the reader to the nice references \cite{boosNl,buoninfante_thesis,BambiModestoShapiro2024} for more thorough reviews. 

\subsection{Quasitopological Gravity}

Quasitopological gravity (QTG) traces its origins in \cite{Oliva_2010,Myers:2010ru}, where they extended the Einstein-Hilbert action by including not only the Gauss–Bonnet term  \footnote{In four-dimensional spacetime this quadratic curvature invariant gives a topological surface term in the gravity action. However, its contribution to the action becomes non-trivial for in higher dimensions. Its inclusion into the Einstein–Hilbert action provides a natural, ghost-free correction to Einstein’s gravity and it gives rise to the class of Gauss–Bonnet gravity models \cite{lovelock1971,deruelle2003gaussbonnetgravity,Fernandes_2022GaussBonnet}.} but also a new contribution cubic in curvature, proposing an extension of GR named cubic gravity. Their approach involved constructing all possible cubic contractions of the Riemann tensor and then tuning the corresponding coupling constants so that the resulting equations of motion remained second order when evaluated on an anti–de Sitter (AdS) background. The motivation was to broaden the landscape of AdS/CFT dualities beyond Einstein gravity, while retaining a manageable set of field equations. The term \emph{quasi-topological} arises because, although the cubic curvature term contributes nontrivially to the dynamics, it yields a trivial contribution in six dimensions—despite not being a topological invariant.

It was later demonstrated \cite{Bueno:2016cu} that there exists a unique theory of cubic curvature gravity—referred to as \textit{Einsteinian Cubic Gravity}—that propagates a transverse, massless, ghost-free graviton on maximally symmetric spacetimes. This construction was then extended to include quartic curvature terms \cite{Bueno:2017ho}, preserving the same desirable features.

These developments is what led to the introduction of \emph{Generalized Quasitopological gravity} (GQTG) \cite{Hennigar:2017ego}, a broader class of theories that preserve several appealing properties. Initially formulated at cubic order, GQTG is defined to satisfy the following criteria for spherically symmetric vacuum spacetimes:
\begin{enumerate}
    \item It admits a smooth Einstein gravity limit with a cosmological constant.
    \item The spacetime geometry is described by a single metric function.
    \item The field equations reduce to a second-order differential equation in the vacuum, or under specific matter couplings.
    \item The linearized theory around a maximally symmetric background propagates only the massless graviton, just as in General Relativity.
\end{enumerate}

This class of theories was later extended to include arbitrary powers of curvature \cite{Bueno:2020gq}, and there is mounting evidence that all higher-order curvature gravities can be recast as GQTG theories via appropriate field redefinitions \cite{Bueno:2019ltp}. A complete classification of these theories was developed in \cite{Bueno:2022res,Moreno:2023rfl}, where it was shown that at each order in curvature, there exists a unique QTG model whose spherically symmetric vacuum metric equations reduce to an algebraic equation for the metric function, and that these equations of motion are algebraic is the defining property of QTG. While the framework of GQTG is more robust since it deals with equations of motion, which are up to second order in field derivatives, we will focus on this later class of theories with algebraic solutions.

More recently, the field has been focused on exploring the dynamical formation processes of the regular black hole models found under the Quasitopological gravity theories \cite{bueno2024,bueno2024shell,bueno2025oppenheimersnyder}, suggesting that a complete description for the formation of regular black holes under the collapse of different matter configurations is possible. Thermodynamic aspects have been recently discussed in \cite{cisterna2025,Hennigar:2025ftm}. Such models where upgraded to regular black holes with inner event horizons in \cite{DiFilippo:2024mwm}, however it was demonstrated as well that such construction is only feasible for specific values of the mass of the black hole. For this reason, the stability of these models remains not well understood.

While QT gravity is defined for static, spherically symmetric metrics, recent progress has been made in extending it to time-dependent cases. In \cite{ling2025bigbounceblackbounce}, a bouncing cosmological model within the framework of QT gravity is analyzed using the Friedmann–Lemaître–Robertson–Walker ansatz. Furthermore, in \cite{GeometricInflation} it is shown that QT gravity can describe regular cosmological scenarios. These results indicate that the model is also capable of producing regular metrics for time-dependent ansätze.

\section{Thesis Outline}

This thesis is organized into five Chapters, each addressing a different aspect of the problem of singularities in classical gravity and the construction of regular black hole solutions in modified theories of gravity. 

\begin{itemize}

    \item \textbf{Chapter 2} explores infinite-derivative modifications of Maxwell's electrodynamics. We introduce nonlocal operators via form factors and study the field generated by point-like electric and magnetized sources in arbitrary spacetime dimensions. The resulting solutions exhibit regular behavior near the origin, demonstrating the effectiveness of nonlocality as a regularization mechanism.

    \item \textbf{Chapter 3} generalizes the previous approach to the gravitational case by studying linearized Einstein gravity and its infinite-derivative extensions. We analyze the field produced by point-like sources in both four and higher dimensions, using Green function techniques. The Chapter demonstrates how nonlocal gravity suppresses divergences and yields finite curvature solutions that generalize the linearized Schwarzschild solution.

    \item \textbf{Chapter 4} is devoted to the construction of regular black hole metrics by drawing from the mechanisms of nonlocal gravity models and the Kerr-Schild double copy formalism. Starting from solutions of linear equations, we obtain a regular solution for a static spherically symmetric black hole, which may be interpreted as a correction to the Schwarzschild metric, and then generalize the procedure to obtain a non-singular rotating black hole that reduces to the Kerr metric when the scale of non-locality goes to zero. The Chapter also analyzes the structure, event horizon deformation, and curvature regularity of these regular black holes.

    \item \textbf{Chapter 5} presents an alternative approach, this time taking quasi-topological gravity and its regular black holes from higher-curvature corrections as inspiration. We use non-linear Lagrangians to construct some sort of generalization in the context of two-dimensional dilaton gravity. We construct analytic solutions with bounded curvature invariants and well-defined thermodynamic properties. Special attention is given to the Hayward metric and its generalizations, as well as to the constraints required to achieve regularity.

    \item \textbf{Chapter 6} summarizes the main results and discusses the implications of both approaches. We compare infinite-derivative and higher-curvature models, discuss their physical viability, and outline potential future directions. 
\end{itemize}

	\chapter{Infinite Derivative Electrodynamics}
\label{Ch2}

\section{Introduction}

In this Chapter, following the paper \cite{Boos_2021}, we discuss higher and infinite derivative modifications of the Maxwell theory. For generality, we consider the case when the number of spacetime dimensions $D\ge 4$.

We use Lorentz and gauge invariance to demonstrate that such a modified theory contains a function $f(\Box)$ called a form factor. We define the appropriate choices of the form factor so that the theory does not contain any new unphysical degrees of freedom.

Since in the presence of the form factor the field equations are linear, the principle of superposition, similarly to the Maxwell case, remains valid. We use the method of Green functions to obtain solutions to these models. After applying the Fourier transform, the modified field equations for the Green function reduce to an algebraic equation. Using a solution of this equation, one obtains an integral representation for the Green function of the modified theory. Finally, we study the obtained solutions of the modified Maxwell theory and demonstrate that in the presence of the form factor, they are regular for point-like sources.

\section{Maxwell Theory}

As a starting point, it will be instructive to review the standard four-dimensional Maxwell theory and derive solutions for the field equations of a static point-like source.

\subsection{Field Equations}

Assuming a flat spacetime background, one can use Cartesian coordinates $X^{\mu} = (t, \mathbf{x})$, in which the Minkowski metric takes the form:
\begin{equation}
ds^2 = \eta_{\mu\nu} dX^{\mu} dX^{\nu} = -dt^2 + \sum_{i=1}^{3} (dx^i)^2\, .
\end{equation}
However, in what follows, we will also employ curved coordinates. For this purpose, it is convenient to express Maxwell's equations in a generally covariant form. Denoting the metric by $g_{\mu\nu}$, we write:
\begin{equation}
ds^2 = g_{\mu\nu} dx^{\mu} dx^{\nu}\, .
\end{equation}
We use covariant derivatives, denoted by $\nabla_{\mu}$ or equivalently by $(\ldots)_{;\mu}$. In Cartesian coordinates, the covariant derivative simplifies to:
\begin{equation}
\nabla_{\mu} = \frac{\partial}{\partial X^{\mu}}\, .
\end{equation}

The Maxwell equations follow from the variation of the action:
\begin{equation}\label{MwAction}
S_M[A_{\mu}] = -\int d^4x \sqrt{-g}\left(\frac{1}{4} F_{\mu\nu} F^{\mu\nu} - A_{\mu} j^{\mu} \right)  \, ,
\end{equation}
with respect to $A_{\mu}$. Here, $F_{\mu\nu}$ is the electromagnetic field strength tensor:
\begin{equation}\label{sTensor}
F_{\mu\nu} = \nabla_{\mu} A_{\nu} - \nabla_{\nu} A_{\mu}\, .
\end{equation}
We adopt Heaviside units and set the speed of light equal to unity, $c \equiv 1$. Varying the action \eqref{MwAction} yields:
\begin{equation}\label{maxeq}
\nabla_{\nu} F^{\nu\mu} = \frac{1}{\sqrt{-g}} \partial_{\nu} (\sqrt{-g} F^{\nu\mu}) = j^{\mu}\, .
\end{equation}
In Cartesian coordinates, this reduces to:
\begin{equation}\label{FFj}
\frac{\partial}{\partial X^{\nu}} F^{\nu\mu} = j^{\mu}\, .
\end{equation}
Substituting the field strength $F$ in terms of the vector potential $A$ into \eqref{FFj} gives:
\begin{align}\label{NLM1}
\Box A^{\nu} - \nabla^{\nu} \nabla_{\mu} A^{\mu} = j^{\nu}\, .
\end{align}
The Maxwell equations are invariant under the gauge transformation $A_{\mu} \to A_{\mu} + \partial_\mu \lambda$, where $\lambda$ is an arbitrary function. Using this gauge freedom, one may impose the Lorenz gauge condition $\nabla_{\mu} A^{\mu} = 0$, leading to:
\begin{align}
\Box A_{\mu} = j_{\mu}\, .
\end{align}

\subsection{Stationary Field of Point-Like Objects}\label{field-eq:greensol}

Since Maxwell’s equations are linear, it suffices to solve them for a stationary point-like source. Although we will not consider this case here, it is worth noting that the field of a more general object could then be obtained by integrating this solution over its spatial volume, weighted by the corresponding charge and magnetic moment distributions. 

Let $X^{\mu} = (t, \mathbf{x})$ denote Cartesian coordinates. Consider the stationary conserved current:
\begin{align}\label{jjj}
j^{\mu} = q \delta^{\mu}_t \delta^{(3)}(\mathbf{x}) + \delta^{\mu}_i M^{ik} \partial_k \delta^{(3)}(\mathbf{x})\, .
\end{align}
Here $q$ is the charge of the point particle, and $M_{ik} = -M_{ki}$ is a constant antisymmetric matrix characterizing its intrinsic magnetic moment. This current satisfies the conservation law $\partial_{\mu} j^{\mu} = 0$. By applying rigid rotations of the Cartesian coordinates, this three-dimensional matrix can be written as $M^{ik} = e^{ika} \hat{m}_{a}$ where $e^{ika}$ is the Levi-Civita symbol and $\hat{m}_{a} = (m,0,0)$. The vector $\hat{m}_a$, directed along $X$-axis, is the vector of the magnetic moment. For this choice the matrix $M^{ij}$ takes 
the following canonical form \cite{Cayley_2009,Zumino}:
\begin{equation}
\mathbf{M} =
\begin{pmatrix}
0 & 0 & 0 \\
0 & 0 & m \\
0 & -m & 0
\end{pmatrix}\, .
\end{equation}
We denote the corresponding orthogonal coordinates (Darboux basis) as $(\xi, \vec{x}_{\perp})$, where $\vec{x}_{\perp}$ spans the two-dimensional plane $\Pi$ orthogonal to the $\xi$-axis. This basis is particularly useful when generalizing the results of this Chapter to arbitrary dimensions.

Consider an electric current loop lying in the plane $\Pi$, centered at $\xi = \vec{x}_{\perp} = 0$. If $I$ is the current and $S$ the loop's area, then the magnetic dipole moment is given by $m = I S$. The point-like source current in \eqref{jjj} can be regarded as the limiting case where the loop radius tends to zero while maintaining a finite dipole moment.

We express the electromagnetic potential as:
\begin{align}
A_{\mu}(\mathbf{x}) = \delta_{\mu}^t \varphi(\mathbf{x}) + \delta_{\mu}^i A_i(\mathbf{x})\, ,
\end{align}
and impose the Lorenz gauge condition $\partial_{\mu} A^{\mu} = 0$. The Maxwell equations then reduce to:
\begin{align}
\Delta \varphi &= -q \delta^{(3)}(\mathbf{x}) \, , \label{aa1} \\
\Delta A_i &= M_i{}^k \partial_k \delta^{(3)}(\mathbf{x}) \, . \label{aa2}
\end{align}
Here, $\Delta$ denotes the flat-space Laplacian:
\begin{equation}
\Delta = \sum_{i=1}^{3} \partial_{x_i}^2\, .
\end{equation}
Let $G(\mathbf{x})$ be the Green's function for the Laplacian:
\begin{equation}
\Delta G(\mathbf{x}) = -\delta^{(3)}(\mathbf{x})\, ,
\end{equation}
with the explicit form:
\begin{equation}\label{localGreen4d}
G(\mathbf{x}) = \frac{1}{4\pi |\mathbf{x}|}\, .
\end{equation}
The solutions to equations \eqref{aa1} and \eqref{aa2} are:
\begin{align}
\varphi(\mathbf{x}) = q \, G(\mathbf{x})\, , \quad
A_i(\mathbf{x}) = -M_i{}^k \partial_k G(\mathbf{x})\, .
\end{align}
The potential 1-form $\mathbf{A} = A_{\mu} dX^{\mu}$ then becomes:
\begin{align}
\begin{split}\label{APOTLocal}
A_{\mu} dx^{\mu} &= \varphi \, dt + A_i \, dx^i \\
&= q \, G(r) \, dt - M_i{}^k \partial_k G(\mathbf{x}) \, dx^i\, .
\end{split}
\end{align}

\section{Infinite Derivative Modification of Maxwell Equations} \label{NLMaxwell}

Having reviewed the standard solutions for charged and magnetized point-like sources in Maxwell’s theory, we now turn to constructing higher- and infinite-derivative extensions of the theory that modify its behavior at short distances.

\subsection{Action and Field Equations}

For generality purposes, we assume that the number $D=d+1$ of spacetime dimensions is arbitrary and that $D\ge 4$. Let us consider a $D$-dimensional Minkowski spacetime with Cartesian coordinates $X^\mu = (t, \mathbf{x})$, where $\mathbf{x} = (x^i)$ and $i = 1, \dots, d$ with $d = D - 1$. The metric takes the form
\begin{align}
ds^2 = -dt^2 + \sum_{i=1}^d (dx^i)^2 = -dt^2 + d\mathbf{x}^2 \, .
\end{align}

We introduce a vector field $A_\mu$ that satisfies linear field equations derived from an action. We impose the following conditions on the form of the action:
\begin{enumerate}
    \item The action $S[A_\mu]$ is a scalar and can be written as
    \begin{equation} \label{SA}
    S[A_\mu] = \int d^D x \sqrt{g} \, L \, ,
    \end{equation}
    where $L$ is the Lagrangian density and $g=-\det(g_{\mu\nu})$.
    \item The Lagrangian density $L$ is bilinear in $A_\mu$, i.e.,
    \begin{equation} \label{LA}
    L = A_\mu \, \mathcal{O}^{\mu\nu} A_\nu \, ,
    \end{equation}
    where $\mathcal{O}^{\mu\nu}$ is a symmetric rank-two differential operator constructed from the metric $g_{\mu\nu}$ and covariant derivatives $\nabla_\mu$.
\item The action is gauge invariant under transformations of the form
    \begin{equation} \label{GA}
    A_\mu(x) \to \hat{A_\mu}(x) = A_\mu(x) + \partial_\mu \lambda(x) \, ,
    \end{equation}
    where $\lambda(x)$ is an arbitrary smooth function.
\end{enumerate}

In flat spacetime, the only available symmetric rank-two tensors constructed from the metric $g_{\mu\nu}$ and covariant derivatives  $\nabla_\mu \nabla_\nu$. Thus, the most general form of the operator $\mathcal{O}^{\mu\nu}$ consistent with the above conditions is
\begin{align} \label{OO}
\mathcal{O}^{\mu\nu} = h(\Box) g^{\mu\nu} + f(\Box) \nabla^\mu \nabla^\nu \, ,
\end{align}
where $f(\Box)$ and $h(\Box)$ are functions of the d'Alembertian operator $\Box = \nabla^\mu \nabla_\mu$.\footnote{Although we are working in flat spacetime, we retain the covariant formulation for generality and convenience. In Cartesian coordinates, $\nabla_\mu = \partial/\partial X^\mu$ and the covariant derivatives commute. We also assume the standard property $\nabla_\lambda g_{\mu\nu} = 0$.} Since we are interested in deriving field equations from the action, we are free to integrate by parts, neglecting boundary terms.

Let $S[\hat{A_\mu}]$ denote the action \eqref{SA} evaluated on the gauge-transformed field \eqref{GA}. Then the gauge variation of the action is \cite{Boos_2021}
\begin{align}
\delta_\lambda S &\equiv S[\hat{A_\mu}] - S[A_\mu] = \int d^D x \sqrt{g} \, J \, , \\
J &= -\lambda ( h(\Box) + f(\Box)\Box ) \nabla^\mu A_\mu + \frac{1}{2} \lambda \Box ( h(\Box) + f(\Box)\Box ) \lambda \, . \label{var}
\end{align}

Gauge invariance requires that $\delta_\lambda S = 0$ for arbitrary $\lambda$. The second term in \eqref{var}, which is quadratic in $\lambda$, vanishes for arbitrary $\lambda$ only if
\begin{align} \label{ca}
h(\Box) = -f(\Box)\Box \, .
\end{align}
Under this condition, the first term in \eqref{var} also vanishes identically. Hence, the condition \eqref{ca} ensures that the action is gauge invariant.

The resulting gauge-invariant action for the nonlocal, higher-dimensional Maxwell theory becomes
\begin{align} \label{SSS}
S[A_\mu] = -\frac{1}{2} \int d^D x \sqrt{g} \, A_\mu f(\Box) \left( g^{\mu\nu} \Box - \nabla^\mu \nabla^\nu \right) A_\nu \, .
\end{align}

After integrating by parts, the action can be rewritten in a manifestly gauge-invariant form involving the field strength tensor:
\begin{align}
\begin{split} \label{DA}
S[A_\mu] &= -\int d^D x \sqrt{g} \left[ \frac{1}{4} F_{\mu\nu} f(\Box) F^{\mu\nu} - j^\mu A_\mu \right] \, , \\
F_{\mu\nu} &= \partial_\mu A_\nu - \partial_\nu A_\mu \, .
\end{split}
\end{align}
Here, we have included an interaction term with an external conserved current $j^\mu$ satisfying $\nabla_\mu j^\mu = 0$. The function $f(\Box)$ is referred to as a \emph{form factor}.

The standard $D$-dimensional Maxwell theory arises as a special case of the action \eqref{DA} when, in addition to equations \eqref{SA}-\eqref{GA}, we require that the field equations derived from the action involve at most second-order derivatives. This requirement implies that $f(\Box)$ is a constant, which can be set to unity by a suitable rescaling of $A_\mu$. The action for the local Maxwell field then reads:
\begin{align} \label{MDA}
S[A_\mu] = -\int d^D x \sqrt{g} \left[ \frac{1}{4} F_{\mu\nu} F^{\mu\nu} - j^\mu A_\mu \right] \, .
\end{align}

The field equations derived from the modified action \eqref{DA} are
\begin{align}
f(\Box) \nabla_\mu F^{\mu\nu} &= j^\nu \, , \\
\partial_{[\rho} F_{\mu\nu]} &= 0 \, .
\end{align}
The second equation reflects the Bianchi identity and ensures the existence of a potential $A_\mu$ such that $F_{\mu\nu} = \partial_\mu A_\nu - \partial_\nu A_\mu$ locally. Substituting this expression into the first equation yields
\begin{align} \label{NLM}
f(\Box) \left( \Box A^\nu - \nabla^\nu \nabla_\mu A^\mu \right) = j^\nu \, .
\end{align}

We may now fix the gauge freedom by choosing the Lorenz gauge condition,
\begin{align}
\nabla_\mu A^\mu = 0 \, ,
\end{align}
which simplifies the field equations to
\begin{align} \label{MaxNLfield}
f(\Box) \Box A_\mu = j_\mu \, .
\end{align}
Throughout this Chapter, we work exclusively in the Lorenz gauge. Having derived our field equations, a few comments about the form factors $f(\Box)$ are in order in the next section

\subsection{Form Factors}\label{sec:formfactors}

All of the form factors, in the infrared limit, should satisfy the condition $f(0)=1$ so that they reproduce the Maxwell theory. Generally, there are two types of form factors that we can consider: polynomial or non-polynomial.

\subsubsection{Polynomial Form Factors}

One may consider the form factors which are a polynomial function of the $\Box$ operator. These would correspond to a higher derivative modification of the field equations, which regularizes the solutions; see for example \cite{Pauli-Villars}. However, this choice can be problematic; in order to further justify this, it is convenient to use the following procedure. Let us first consider an analytic function $f(z)$ of a complex variable $z$. Such a function can be written in the form
\begin{equation} \label{serf}
f(z)=\sum_{n=0}^{\infty} f_n z^n\, .
\end{equation}
The operator $f(\Box)$ is obtained from $f(z)$ by substituting $z\to \Box$.
In the case when only a finite number $q$ of terms is present in the series \eqref{serf}, the form-factor function is a polynomial of order $q$ 
\begin{equation}
P_q(z)=z^q+f_{q-1}z^{q-1}+\ldots f_1 z+f_0\, .
\end{equation}
The corresponding field equations are indeed a higher derivative modification of the original theory. In a general case, such models have the following common problem. A polynomial of order $q$ has $q$ zeros $z_j$, $j=1,\ldots q$ in the complex plane of a variable $z$ \footnote{Some of these zeroes can coincide. For simplicity, we do not consider such cases here}. As a result, the inverse function of $z P_q (z)$ that enters the equation \eqref{MaxNLfield} has $q+1$ poles so that one has
\begin{equation}\label{CCC}
\frac{1}{z P_q(z)}=\frac{C_0}{z}+\sum_{j=1}^q \frac{C_j} {z-z_j}\, .
\end{equation}
Each of these poles corresponds to a propagating degree of freedom of the field $A_{\mu}$. For the local theory, $P_q(z)=1$ so that $C_0=1$ and $C_{j\ge 1}=0$.

In a general case $q\ge 1$, if one combines all the terms on the right-hand side, one gets the following expression
\begin{equation}
\frac{C_0}{z}+\sum_{j=1}^q \frac{C_j}{z-z_j}=\frac{Q_q}{z P_q(z)}\, .
\end{equation}
Here, $Q_q(z)$ is a polynomial of order $q$. The relation \eqref{CCC} implies that, in fact, $Q_q(z)=1$. This imposes $q+1$ conditions on the coefficients $C_0$ and $C_j$, which uniquely determine them. In particular, the condition that $Q_q(z)$ does not contain $z^q$ implies that
\begin{equation} \label{C0j}
C_0+\sum_{j=1}^q C_j=0\, .
\end{equation}
Let us keep the condition $C_0\ge 0$ for a nontrivial polynomial $P_q(z)$. Then, relation \eqref{C0j} shows that at least some of the residues $C_{j\ge 1}$ should be negative. The corresponding unphysical modes are known as ghosts.

\subsubsection{Non-Polynomial Form Factors}

It has been proposed that this fundamental problem of the higher derivative models can be solved by considering special infinite derivative modifications of the corresponding local theory \cite{NONLOC,Biswas_2014}. Namely, one can consider form factors $f(z)$ which are regular and do not have zeroes in the whole complex plane. Consider the function $f(z)=\exp[P(z)]$, where $P(z)$ is an entire function. Notice that this form factor has no zeroes since $P(z)$ is everywhere holomorphic and therefore the exponent is non-zero and well defined in the whole complex plane. This means that the inverse function $zf(z)$ does not have new poles.

The most common choice for these types of form factors used in the literature is
\begin{equation}
    f(z) = e^{P(z)} \, ,
\end{equation}
where $P(z)$ is a polynomial. There are two useful forms of these polynomials\footnote{ Other form factors, which are tailored for constructing certain infinite derivative quantum gravity theories, have also been studied in \cite{Kuzmin:1989sp,Modesto_2012,tomboulis1997}}

\begin{enumerate}
    \item The Gaussian form factor $P(z) = -z\ell^2$, where $\ell$ is some characteristic length scale where the effects of non locality become relevant.
    \item The generic form factor $P(z) = -(z\ell^2)^N$ which is a generalization of the Gaussian form factor 
\end{enumerate}
 
To make our consideration more concrete, we restrict ourselves to considering the form factors of the following form
\begin{align}\label{FormFactorGauss}
f(\Box) = \exp\left[ (-\ell^2\Box)^N \right] \, , \quad \ell > 0 \, .
\end{align}
Here $N$ is a positive integer number. These models are sometimes referred to as ``ghost-free'' as they do not introduce new unphysical degrees of freedom, and one uses the notation $GF_N$ to identify them. All these models contain a parameter
$\ell > 0$, which has dimensions of length. It characterizes the scale of nonlocality where its effects become relevant. These exponential form factors are a popular choice in the infinite derivative gravity literature \cite{boosNl,buoninfante2022blackholesnonlocalgravity,buoninfante_thesis,ScalarKernel,IDG_convolutional}, and they capture a wide set of different infinite derivative gravity models, each model being defined by the choice of $N =1,2,3,... \, $.

These form factors are chosen so that in the limiting case of $\ell \to 0$ one recovers the local theory since $f(0) = 1$, as mentioned earlier. This condition also guarantees that the residue of the pole of $zf(z)$ at $z=0$ is 1, and that such a theory correctly reproduces the properties of the corresponding local theory in the infrared regime. We will look for solutions to the electromagnetic field equations with this type of non-polynomial form factor in the next section.

Experimental constraints on the scale of nonlocality have been explored through a variety of approaches. From quantum electrodynamics, it is estimated that the nonlocality scale satisfies $\ell < 10^{-20}\mathrm{m}$ \cite{minlength}. In the context of infinite derivative modifications to the Standard Model \cite{PhysRevD:MazumdarSM}, it has been argued that upcoming experiments at the LHC Phase 2 run—scheduled to begin operations in 2029 \cite{CMS-TDR-016}—may be sensitive to nonlocal effects at this same scale \cite{Su2021}. In particular, in \cite{PhysRevD:MazumdarSM}, the authors show that nonlocal modifications to the heavy-boson propagator can enhance the cross section for producing boosted Higgs bosons.

Efforts to detect nonlocality at even smaller length scales, in the range of $10^{-22}\mathrm{m}$ to $10^{-26}\mathrm{m}$, have been proposed using optomechanical oscillator experiments \cite{PhysRevD:Belenchia}. Nonlocal effects have also been considered as a possible explanation for the discrepancy between Standard Model predictions and experimental measurements of the muon's anomalous magnetic moment. For such a contribution to be viable, the lower bound on the nonlocality scale must again be around $10^{-20}\mathrm{m}$. Weaker constraints—on the order of $10^{-15}\mathrm{m}$—can be derived from modifications to the Lamb shift and electrostatic interactions, particularly when comparing theoretical predictions with data from the muonic hydrogen anomaly \cite{capolupo2023phenononlocalQED}.

\section{Static Infinite Derivative Green Functions}\label{SID_Green}

In order to find solutions that describe the electromagnetic field produced by stationary sources, we assume that the current $ j^{\nu} $, which appears on the right-hand side of equation~\eqref{MaxNLfield}, is time-independent. Additionally, we assume that no free, propagating electromagnetic waves are present.

Since the field configuration does not depend on time, the d'Alembert operator $ \Box $ may be replaced by the Laplace operator:
\begin{align}
\label{dLaplace}
\Delta = \nabla^2 = \sum_{i=1}^d \partial_i^2\, .
\end{align}

Here, as earlier, $d=D-1$ denotes the number of spatial dimensions. In both local and infinite derivative electrodynamics, solutions to the field equations for stationary sources can be constructed using the appropriate Green functions. In the infinite derivative case, the Green function $ \mathcal{G}_d $ satisfies the  equation
\begin{align}
\label{GF}
f(\Delta) \Delta \mathcal{G}_d(\mathbf{x}' - \mathbf{x}) = -\delta^{(d)}(\mathbf{x}' - \mathbf{x}) \, .
\end{align}
Throughout this section, we denote by $ \mathcal{G}_d(\mathbf{x}) $ the Green function in the infinite derivative theory. In the local limit, this reduces to the standard Green function $ G_d(\mathbf{x}) $ (see for example, \eqref{localGreen4d}) for the four-dimensional case. A convenient representation for the static Green function $ \mathcal{G}_d $ can be obtained via Fourier transformation. Following \cite{Frolov_2016}, we denote the momentum vector in $ d $ spatial dimensions by $ \mathbf{k} = (k_1, \ldots, k_d) $, and write
\begin{equation}
\mathcal{G}_d(\mathbf{x}' - \mathbf{x}) = \int \frac{d^d \mathbf{k}}{(2\pi)^d} e^{i \mathbf{k} \cdot (\mathbf{x} - \mathbf{x}')} \bar{\mathcal{G}}_d(\mathbf{k}) \, .
\end{equation}
Here, $ \mathbf{k} \cdot \mathbf{x} = \sum_{j=1}^d k_j x^j $. Substituting this expression into equation~\eqref{GF}, and using the Fourier representation of the delta function,
\begin{equation}
\delta^{(d)}(\mathbf{x}' - \mathbf{x}) = \int \frac{d^d \mathbf{k}}{(2\pi)^d} e^{i \mathbf{k} \cdot (\mathbf{x} - \mathbf{x}')} \, ,
\end{equation}
we find
\begin{equation}
\bar{\mathcal{G}}_d(\mathbf{k}) = \frac{1}{k^2 f(-k^2)} \, .
\end{equation}

Let $ \theta $ be the angle between the vectors $ \mathbf{k} $ and $ \mathbf{x} - \mathbf{x}' $, and define
\begin{equation}
\mathbf{k} \cdot (\mathbf{x} - \mathbf{x}') = kr \cos\theta \,, \quad k = |\mathbf{k}| \,, \quad r = |\mathbf{x} - \mathbf{x}'| \, .
\end{equation}
The Green function then becomes
\begin{equation}
\mathcal{G}_d(x - x') = \Omega_{d-2} \int_0^\infty \frac{dk}{(2\pi)^d} \frac{k^{d-3}}{f(-k^2)} \int_0^\pi d\theta\, \sin^{d-2} \theta\, e^{ikr\cos\theta} \, ,
\end{equation}
where
\begin{equation}
\Omega_{d-2} = 2\frac{\pi^{(d-1)/2}}{\Gamma\left(\frac{d-1}{2}\right)} \, ,
\end{equation}
is the surface area of the unit $ (d-2) $-sphere $ S^{d-2} $.

Evaluating the angular integral yields a Bessel function, resulting in the expression
\begin{equation}\label{FNLG}
\mathcal{G}_d(x, x') = \frac{1}{2\pi} \int_0^\infty \frac{dk}{k f(-k^2)} \left( \frac{k}{2\pi r} \right)^{\frac{d}{2}-1} J_{\frac{d}{2}-1}(kr) \, .
\end{equation}
Since $ \mathcal{G}_d(x,x') $ depends only on the radial distance $ r $, it can be written as
\begin{equation}
\label{NLGreen}
\mathcal{G}_d(r) = \frac{1}{(2\pi)^{d/2} r^{d-2}} \int_0^\infty dz\, \frac{z^{\frac{d-4}{2}}}{f(-z^2/r^2)} J_{\frac{d}{2}-1}(z) \, ,
\end{equation}
valid for $ d \geq 3 $. 
Let us emphasize that the obtained expressions \eqref{FNLG} and \eqref{NLGreen} give an integral representation for the static Green functions for an arbitrary form factor, provided that $f(-k^2)$ decreases rapidly enough at large $k$, so that the integrals are convergent at the upper limit.

In the local Maxwell theory, where $ f = 1 $, the integral in equation~\eqref{NLGreen} evaluates to a constant, reproducing the standard result:
\begin{equation}
\label{eq:LocalGF}
G_d(x, x') = \frac{\Gamma\left(\frac{d}{2} - 1\right)}{4\pi^{d/2} r^{d-2}} \, .
\end{equation}

The Green functions in different spatial dimensions are related by the recursive formula
\begin{align}
\label{eq:ch2:recursion}
\mathcal{G}_{d+2}(r) = -\frac{1}{2\pi r} \frac{\partial \mathcal{G}_d(r)}{\partial r} \, ,
\end{align}
as derived in \cite{Frolov_2016}. This allows one to generate higher-dimensional Green functions from known lower-dimensional ones, such as those in $ d = 3 $ and $ d = 4 $.

In what follows, we mainly focus on $GF_1$ models with Gaussian form factor and use the notation $\mathcal{G}_d(r)$ for the corresponding static Green functions in $d$-dimensional flat space. In the uncommon cases when we consider the $GF_N$ models, the corresponding Green functions will be denoted by $\mathcal{G}_d^N(r)$. For instance, in the simplest ghost-free model $ \mathrm{GF}_1 $, one finds
\begin{align}\label{GF1_EQ}
\mathcal{G}_3(r) &= \frac{1}{4\pi r} \, \text{erf}\left( \frac{r}{2\ell} \right) \, , \\
\mathcal{G}_4(r) &= \frac{1}{4\pi^2 r^2} \left[ 1 - e^{-r^2/(4\ell^2)} \right] \, ,
\end{align}
where $ \text{erf}(z) $ is the error function \cite{NIST2010}. In the limit $ \ell \to 0 $, these expressions reduce to their local counterparts. Other examples are
\begin{align}\label{GFdN_EQ}
\mathcal{G}_3^2(r) &= \frac{1}{6\pi^2\ell} \bigg[3\Gamma\bigg(\frac{5}{4}\bigg)  {}_1F_3 \bigg(\frac{1}{4};\frac{1}{2},\frac{3}{4},\frac{5}{4};y^2 \bigg) - 2y\Gamma\bigg(\frac{3}{4}\bigg) {}_1F_3 \bigg(\frac{3}{4};\frac{5}{4},\frac{3}{2},\frac{7}{4};y^2 \bigg)\bigg] \, , \\
\mathcal{G}^2_4(r) &= \frac{1}{64\pi^2 y\ell^2} \left[ 1 - {}_0F_2\bigg(\frac{1}{2}, \frac{1}{2};y^2 \bigg) + 2\sqrt{\pi}y \ {}_0F2\bigg(1,\frac{3}{2};y^2 \bigg) \right] \, .
\end{align}
Where $y=(r/4\ell)^2$ and ${}_aF_b$ is the hypergeometric function \cite{NIST2010}. A collection of Green functions $ \mathcal{G}_d^N(r) $ for various $ \mathrm{GF}_N $ models can be found in \cite{Boos_2020}.

A key feature of Green functions in ghost-free infinite derivative theories is their regularity at $ r = 0 $. This can be verified by analyzing the small-$ r $ behavior of the Bessel function in equation~\eqref{FNLG}. Further details and discussion can be found in \cite{boosNl}.

\subsection{Stationary Solutions for Point-Like Sources}

Let us consider the field of a point-like charged and magnetized source in four-dimensional spacetime. We consider a stationary, conserved external current given by
\begin{align}\label{CUR}
j{}^\mu = \delta{}^\mu_t q \delta{}^{(3)}(\mathbf{x}) + \delta{}^\mu_i M{}^{ik} \partial_k \delta{}^{(3)}(\mathbf{x}) \, .
\end{align}
As earlier,  $ q $ denotes the electric charge of a point particle, and $ M_{ik} = -M_{ki} $ is a constant antisymmetric matrix that characterizes the particle’s intrinsic magnetic moment. The electromagnetic potential can be decomposed as
\begin{align}\label{emPotential_decomp}
A_\mu(\mathbf{x}) = \delta{}_\mu^t \varphi(\mathbf{x}) + \delta{}_\mu^i A_i (\mathbf{x})\, ,
\end{align}
and the corresponding field equations in the Lorentz gauge become
\begin{align}
f(\Delta)\Delta \varphi &= -q \delta{}^{(3)}(\mathbf{x}) \, , \\
f(\Delta)\Delta A_i &= M{}_i{}^k \partial_k \delta{}^{(3)}(\mathbf{x}) \, .
\end{align}
The solutions to these equations, expressed in terms of the Green function, are
\begin{align}
\varphi(\mathbf{x}) = q \, \mathcal{G}_3(\mathbf{x}) \, , \quad
A_i(\mathbf{x}) = -M{}_i{}^k \partial_k \mathcal{G}_3(\mathbf{x}) \, .
\end{align}
Thus, the electromagnetic potential 1-form $ \mathbf{A} = A{}_\mu \, d X{}^\mu $ takes the form
\begin{align}
\begin{split}\label{APOT}
A_\mu \, d X{}^\mu &= \varphi \, d t + A_i \, d x^i \\
&= q \, \mathcal{G}_3(r) \, d t - M{}_i{}^k \partial_k \mathcal{G}_3(\mathbf{x}) \, d x^i \, .
\end{split}
\end{align}
Using the identity
\begin{align}
\partial_i \mathcal{G}_d(\mathbf{x}) = \frac{x_i}{r} \, \partial_r \mathcal{G}_d(\mathbf{x}) = -2\pi x_i \, \mathcal{G}_{d+2}(\mathbf{x}) \, ,
\end{align}
we may also write the potential in the alternative form
\begin{align}
\begin{split}\label{APOT2}
A_\mu \, d X{}^\mu = q \, \mathcal{G}_3(r) \, d t + 2\pi M{}_i{}^k x_k \, \mathcal{G}_{5}(\mathbf{x}) \, d x^i \, .
\end{split}
\end{align}

\subsubsection{Higher Dimensional Solutions}

For $D>4$, solutions for the field of a point-like source can also be obtained by using the corresponding Green functions.
\eqref{NLGreen}. In order to obtain the solutions for equations \eqref{MaxNLfield}, one may consider the following external current:
\begin{align}\label{appCUR}
j{}^\mu = \delta{}^\mu_t \, q \, \delta{}^{(d)}(\mathbf{x}) + \delta{}^\mu_i \, M{}^{ik} \, \partial_k \delta{}^{(d)}(\mathbf{x}) \, ,
\end{align}
where $ q $ denotes the electric charge of a point particle, and $ M_{ik} = -M_{ki} $ is a constant antisymmetric matrix characterizing the particle’s intrinsic magnetic moment.

Let us write $ d = 2k + \sigma $, where $ \sigma = 0 $ for even $ d $, and $ \sigma = 1 $ for odd $ d $. Any antisymmetric $ d \times d $ matrix can be brought to a block-diagonal form through an orthogonal transformation in $ d $-dimensional space~\cite{Gantmacher:1959:TMa,Gantmacher:1959:TMb,prasolov1996}. In the corresponding coordinate system, the matrix $ \mathbf{M} $ takes the canonical form (see~\cite{pinedo2022gravitational}):
\begin{equation}\label{MMMM}
\mathbf{M} = \begin{bmatrix}
    \mathbf{m}_1 & 0 & \dots  & 0 & 0 \\
    0 & \mathbf{m}_2 & \dots  & 0 & 0 \\
    \vdots  & \vdots  & \ddots  & \vdots  & \vdots \\
    0 & \dots  & \dots & \mathbf{m}_k & 0 \\
    0 & 0 & \dots  & 0 & 0 \\
\end{bmatrix} \, ,
\end{equation}
where each $ \mathbf{m}_a $ is a $ 2 \times 2 $ block of the form
\begin{equation}
\mathbf{m}_a = \begin{bmatrix}
    0 & m_a \\
   -m_a & 0 \\
\end{bmatrix} \, .
\end{equation}
The final row and column of zeros appear only in the case $ \sigma = 1 $ and are absent when $ \sigma = 0 $. The parameters $ m_a $ represent the independent components of the magnetic moment. In four spacetime dimensions ($ d = 3 $), only one such block appears, corresponding to a single magnetic moment component.

For instance, in the case of the field equations for a point-like charged and magnetized source, they take the form
\begin{align}\label{field-eq-highdimg-maxwell}
f(\Delta)\Delta \varphi &= -q \, \delta{}^{(d)}(\mathbf{x}) \, , \\
f(\Delta)\Delta A_i &= M{}_i{}^k \, \partial_k \delta{}^{(d)}(\mathbf{x}) \, .
\end{align}
Whose solutions may be written as 
\begin{align}
\varphi(\mathbf{x}) = q \, \mathcal{G}_d(\mathbf{x}) \, , \quad
A_i(\mathbf{x}) = -M{}_i{}^k \, \partial_k \mathcal{G}_d(\mathbf{x}) \, .
\end{align}
Repeating the same steps from above, we can find that the electromagnetic potential 1-form
\begin{align}
\begin{split}
A_\mu \, d X{}^\mu &= \varphi \, d t + A_i \, d x^i \\
&= q \, \mathcal{G}_d(r) \, d t - M{}_i{}^k \, \partial_k \mathcal{G}_d(\mathbf{x}) \, d x^i \, .
\end{split}
\end{align}
An the alternative equivalent expression is
\begin{align}
\begin{split}\label{appAPOT2}
A_\mu \, d X{}^\mu = q \, \mathcal{G}_d(r) \, d t + 2\pi M{}_i{}^k \, x_k \, \mathcal{G}_{d+2}(\mathbf{x}) \, d x^i \, .
\end{split}
\end{align}

\section{Conclusions}

In this Chapter, we have extended the classical theory of electrodynamics by incorporating higher and infinite-derivative modifications into Maxwell's equations. After reviewing the standard covariant formulation of Maxwell theory, we introduced a class of higher and infinite derivative generalizations characterized by analytic form factors. These modifications preserve the Lorentz invariance while providing a proper UV modification of electromagnetic interactions.

In general, introducing higher derivatives in the field equations can be dangerous, as they may introduce unphysical field solutions due to the appearance of poles in the Green function, which correspond to degrees of freedom that break unitarity, referred to as ghosts. We considered a special class of functions that can avoid this problem while maintaining the regularization features. 

In particular, we derived the field equations corresponding to these infinite derivative models and constructed the appropriate Green functions in both four and higher-dimensional spacetimes. Using these tools, we investigated the electromagnetic fields generated by static, point-like charged and magnetized sources and how the new structure, due to the infinite derivative modification, regularizes the field behavior near the source. 

This analysis provides a clear example of how nonlocal operators regularize singularities in field theory, a mechanism we now extend to gravity in the following chapter

	\chapter{Gravity and Its Higher and Infinite Derivative Modifications}
\label{Ch3}

\section{Introduction}

In this Chapter, following \cite{Boos_2020}, we focus on gravitational fields in the weak-field regime, where spacetime deviations from flatness are small. Within this approximation, we employ the framework of linearized gravity. Following the spirit of Chapter \ref{Ch2}, we then discuss a higher and infinite derivative modification of the Einstein-Hilbert action and find the field equations in the linearized curvature regime for a weak gravitational field. We use these equations to find solutions for point-like spinning masses, exhibiting solutions which are regular at the positions of the source, in contrast with the local linearized Einstein gravity.

\section{Linearized Einstein Gravity}

As a starting point, we will explore the four dimensional linearized Einstein gravity equations and find their solutions.

\subsection{Four-Dimensional Action and Field Equations}

To derive the equations of linearized gravity, we begin by expanding the spacetime metric around flat Minkowski space:
\begin{equation}
g_{\mu\nu} = \eta_{\mu\nu} + h_{\mu\nu} \, ,
\end{equation}
where $\eta_{\mu\nu}$ is the flat background metric and $h_{\mu\nu}$ denotes a small perturbation. The linearized field equations for $h_{\mu\nu}$ can be obtained from the Einstein equations (see equation~\eqref{Einst:eq_appA}). There are two equivalent methods to arrive at the same result.

Let $X^{\mu}$ be Cartesian coordinates such that $\eta_{\mu\nu} = \mathrm{diag}(-1,1,1,1)$. To obtain the linear equations, we expand the Einstein-Hilbert action (equation~\eqref{Einstein_FieldEqs}) on a flat spacetime background to second order in $h_{\mu\nu}$. It can be shown (see, e.g., \cite{Misner:1973prb,Carroll_2019}) that the quadratic part of the gravitational action is given by
\begin{equation}\label{LEE}
S_g = -\frac{1}{2\kappa} \int d^4X \left( -\frac{1}{2} h^{\mu\nu} \Box h_{\mu\nu}
+ h^{\mu\nu} \partial_\mu \partial_\alpha h^\alpha{}_\nu
- h^{\mu\nu} \partial_\mu \partial_\nu h
+ \frac{1}{2} h \Box h \right) \, ,
\end{equation}
where $h = \eta^{\mu\nu} h_{\mu\nu}$ is the trace of the perturbation and $\kappa$ is the gravitational coupling.

Varying this action with respect to $h_{\mu\nu}$ and including the matter term yields the linearized Einstein equations \cite{Misner:1973prb,Carroll_2019}
\begin{align}
\begin{split}
\label{GEQNa}
G_{\mu\nu} \equiv &\ \Box h_{\mu\nu}
- \partial_\sigma \left( \partial_\nu h_\mu{}^\sigma + \partial_\mu h_\nu{}^\sigma \right)
+ \eta_{\mu\nu} \left( \partial_\rho \partial_\sigma h^{\rho\sigma} - \Box h \right)
+ \partial_\mu \partial_\nu h \\
&= -2\kappa T_{\mu\nu} \, .
\end{split}
\end{align}
The same result can be obtained directly by expanding the Einstein tensor $G_{\mu\nu}$ to linear order in $h_{\mu\nu}$.

\subsection{Gauge Invariance}

Let us consider the effect of an infinitesimal coordinate transformation on the perturbed metric. Starting with
\begin{equation}
g_{\mu\nu} = \eta_{\mu\nu} + h_{\mu\nu} \quad \Rightarrow \quad g^{\mu\nu} = \eta^{\mu\nu} - h^{\mu\nu} \, ,
\end{equation}
we perform a coordinate shift
\begin{equation}\label{gaugeshift}
X^{\mu} \rightarrow \bar{X}^{\mu} = X^{\mu} + \xi^{\mu}(X) \, ,
\end{equation}
where $\xi^{\mu}$ is a small vector field. Under this transformation, the contravariant metric becomes
\begin{align}
\bar{g}^{\mu\nu} &= \frac{\partial \bar{X}^{\mu}}{\partial X^{\alpha}} \frac{\partial \bar{X}^{\nu}}{\partial X^{\beta}} g^{\alpha\beta} \nonumber \\
&= \left( \delta^\mu_\alpha + \xi^\mu{}_{,\alpha} \right) \left( \delta^\nu_\beta + \xi^\nu{}_{,\beta} \right) \left( \eta^{\alpha\beta} - h^{\alpha\beta} \right) \nonumber \\
&= \eta^{\mu\nu} - \bar{h}^{\mu\nu} + \cdots \, ,
\end{align}
where we have omitted terms of second or higher order in $\xi$ or $h$. The transformed perturbation is given by
\begin{equation}\label{gauh}
\bar{h}_{\mu\nu} = h_{\mu\nu} - \partial_\mu \xi_\nu - \partial_\nu \xi_\mu \, . 
\end{equation}
 This transformation reflects the gauge freedom of linearized gravity. Since the linearized action and equations are derived from a covariant theory, they are invariant under the infinitesimal coordinate transformation \eqref{gaugeshift}. Under this transformation the gravitational field perturbation $h_{\mu\nu}$ changes as $h_{\mu\nu} \to h_{\mu\nu}+2\xi_{(\mu,\nu)}$. This field transformation is in fact a gauge transformation, preserving the dynamics of the perturbation.  This invariance can also be verified directly at the level of the linearized theory (see, e.g., \cite{Misner:1973prb})
\subsection{Field Equations in de Donder Gauge}

The linearized equations \eqref{GEQNa} can be simplified by introducing the trace-reversed perturbation:
\begin{align}\label{TRh_eq}
\hat{h}_{\mu\nu} = h_{\mu\nu} - \frac{1}{2} \eta_{\mu\nu} h \, .
\end{align}
Its inverse relation is
\begin{align}\label{inv}
h_{\mu\nu} = \hat{h}_{\mu\nu} - \frac{1}{2} \eta_{\mu\nu} \hat{h} \, ,
\end{align}
where $\hat{h} = \eta^{\mu\nu} \hat{h}_{\mu\nu}$.

Imposing the de Donder (harmonic) gauge condition,
\begin{align}\label{DDgauge_eq}
\partial^\mu \hat{h}_{\mu\nu} = 0 \, ,
\end{align}
the field equations reduce to a remarkably simple form:
\begin{align}
\label{eq:eom}
\Box \hat{h}_{\mu\nu} = -2\kappa T_{\mu\nu} \, .
\end{align}
The conservation of the energy-momentum tensor, $\partial_\mu T^{\mu\nu} = 0$, ensures that the gauge condition is preserved under time evolution.

\section{Linearized Higher-Dimensional Einstein Gravity}

Higher-dimensional generalizations of Einstein gravity are of considerable interest, especially in the context of string theory, which is naturally formulated in spacetimes with $D > 4$ dimensions \cite{tong2012lecturesstringtheory}. The field equations in $D$ spacetime dimensions retain the same form as in four dimensions, and can be derived from the action \eqref{EE_appA} using the same procedures.

As before, we define the number of spatial dimensions as $d = D - 1$. The main difference in higher dimensions appears in the inverse relation between $h_{\mu\nu}$ and $\hat{h}_{\mu\nu}$, which now reads:
\begin{align}\label{invHD}
h_{\mu\nu} = \hat{h}_{\mu\nu} - \frac{1}{d - 1} \eta_{\mu\nu} \hat{h} \, .
\end{align}
After imposing the de Donder gauge condition,
\begin{align}
\partial^\mu \hat{h}_{\mu\nu} = 0 \, ,
\end{align}
the linearized Einstein equations become
\begin{align}
\Box \hat{h}_{\mu\nu} = -2\varkappa T_{\mu\nu} \, .
\end{align}
Where we now use the $D$ dimensional gravitational coupling $\varkappa$, which is defined in Appendix \ref{app:SphReduction}. As in the four-dimensional case, the gauge condition is consistent with the conservation law $\partial_\mu T^{\mu\nu} = 0$.

\subsection{Stationary Field of Point-Like Sources in Four Spacetime Dimensions}

Let us now consider the four-dimensional case. We denote the Cartesian coordinates in flat spacetime  by $X^{\mu}$
\begin{equation}
X^{\mu} = (t, \mathbf{x}) \quad \text{with} \quad \mathbf{x} = (x^1, x^2, x^3)\, .
\end{equation}
We consider sources with a stationary stress-energy tensor, meaning that $T_{\mu\nu}$ does not depend on time $t$. Additionally, we assume that no freely propagating gravitational waves are present. Under these conditions, the linearized metric perturbation $h_{\mu\nu}$ is also stationary, and the d'Alembert operator in the field equations reduces to the three-dimensional Laplacian:
\begin{align}
\label{gstat}
\Delta \hat{h}_{\mu\nu} = -2\kappa T_{\mu\nu} \, .
\end{align}

These equations are linear, therefore it suffices to compute the field of a point-like source. Solutions for sources with a more general shape can be obtained by integrating this elementary solution against a suitable stress-energy distribution. The stress-energy tensor of a point-like spinning particle is given by
\begin{align}\label{set_stat}
T_{\mu\nu} = \delta^t_\mu \delta^t_\nu \, m\,\delta^{(3)}(\mathbf{x}) + \delta^t_{(\mu} \delta^i_{\nu)} \, j_{i}{}^{j} \partial_j \delta^{(3)}(\mathbf{x}) \, ,
\end{align}
where $m$ is the particle’s mass and $j_{ij} = -j_{ji}$ is a constant antisymmetric matrix representing angular momentum.

As with the electromagnetic case, we can bring $j_{ij}$ into a canonical form using Darboux coordinates. In this basis, the antisymmetric $3 \times 3$ matrix $\mathbf{j}$ takes the form
\begin{equation}\label{JJJ}
\mathbf{j} =
\begin{pmatrix}
0 & 0 & 0 \\
0 & 0 & j \\
0 & -j & 0
\end{pmatrix}\, .
\end{equation}

From the structure of \eqref{set_stat}, only the components $\hat{h}_{00}$ and $\hat{h}_{0i}$ of the metric perturbation are nonzero. We define
\begin{equation}
\phi = \hat{h}_{00} \quad \text{and} \quad A_i = \hat{h}_{0i}\, .
\end{equation}
The corresponding solutions, which can be found via the Green function method as outlined in section \eqref{field-eq:greensol} are
\begin{align}
\phi(r) &= \frac{\kappa m}{4\pi r} \, , \\
A_i(\mathbf{x}) &= -\frac{\kappa j_{ij} x^j}{4\pi r^3} \, ,
\end{align}
where $r = |\mathbf{x}|$ and $\kappa = 8\pi G$.

Note that $\phi = -2\varphi$, where $\varphi$ is the Newtonian potential of a point mass $m$. Using Darboux coordinates $(t, z, y, \hat{y})$ in which $\mathbf{j}$ has the form \eqref{JJJ}, the non-zero components of $A_i$ become
\begin{align}
\mathbf{A} &= (0, A_y, A_{\hat{y}}) \, , \\
A_y &= \frac{\kappa j}{4\pi r^3} \hat{y} \, , \quad A_{\hat{y}} = -\frac{\kappa j}{4\pi r^3} y \, .
\end{align}

We can express the metric perturbation in spherical coordinates $(t, r, \phi, \theta)$ using the standard coordinate transformation:
\begin{equation}
\begin{split}
y &= r \cos\phi \sin\theta \, , \\
\hat{y} &= r \sin\phi \sin\theta \, , \\
z &= r \cos\theta \, .
\end{split}
\end{equation}
In these coordinates, the metric perturbation becomes
\begin{align}
\hat{h}_{\mu\nu} dx^\mu dx^\nu = \frac{\kappa m}{4\pi r} dt^2 - \frac{\kappa j \sin^2 \theta}{2\pi r} dt\, d\phi \, .
\end{align}

The full perturbed metric is then
\begin{align}
\begin{split}\label{PertMetric}
ds^2 &= -\left(1 - \frac{2m}{r}\right) dt^2 - \frac{4j \sin^2 \theta}{r} dt\, d\phi \\
&\quad + \left(1 + \frac{2m}{r}\right) \left[dr^2 + r^2(d\theta^2 + \sin^2\theta\, d\phi^2)\right] \, .
\end{split}
\end{align}
In the Newtonian limit, which is used to study the non-relativistic particle motion, we may neglect the $2m/r$ term in the spatial part of the metric, leading to the simplified form
\begin{align}
\label{PertMetric_1}
ds^2 = -\left(1 - \frac{2m}{r}\right) dt^2 - \frac{4j \sin^2 \theta}{r} dt\, d\phi + dr^2 + r^2(d\theta^2 + \sin^2\theta\, d\phi^2) \, .
\end{align}

This metric closely resembles the asymptotic form of the Kerr solution, which describes a rotating black hole in vacuum, see Appendix~\ref{app:KerrMetric}. At large distances $r \to \infty$, the Kerr metric reduces to \eqref{PertMetric_1}. These results can also be extended to an arbitrary number of spacetime dimensions as it will be shown in the next subsection 

\subsection{Stationary Field of Point-Like Sources in Higher Dimensions}

We can proceed in a way which is completely analogous to that of the four-dimensional linearized Einstein gravity in order to derive stationary solutions in spacetimes of arbitrary dimensionality. Like in the previous Chapter, we introduce the following notation for Cartesian coordinates
\begin{equation}
X^{\mu} = (t, \mathbf{x}) \quad \text{with} \quad \mathbf{x} = (x^1, x^2, \dots, x^d) \, .
\end{equation}
 
We again focus on the stationary case where the stress-energy tensor is time-independent. Consequently, the field equations take the same form as in equation~\eqref{gstat}, with the only difference being that the Laplacian now acts on all $d$ spatial dimensions.

The stress-energy tensor for a point-like spinning particle in $d$ spatial dimensions is given by
\begin{align}
\label{eq:SEtensor}
T_{\mu\nu} = \delta^t_\mu \delta^t_\nu \, m\,\delta^{(d)}(\mathbf{x}) + \delta^t_{(\mu} \delta^i_{\nu)} \, j_{i}{}^j \, \partial_j \delta^{(d)}(\mathbf{x}) \, ,
\end{align}
where $\delta^{(d)}(\mathbf{x})$ denotes the $d$-dimensional Dirac delta function and ${j}_{ij}$ is a constant antisymmetric matrix parameterizing its angular momentum. The structure of the metric perturbation remains unchanged, so the only nonvanishing components are
\begin{equation}
\phi = \hat{h}_{00} \quad \text{and} \quad A_i = \hat{h}_{0i} \, , \quad i = 1,2,\dots,d \, .
\end{equation}

The solutions in an arbitrary number of dimensions $d\ge 3$ can be found using the corresponding Green functions $G_d(r)$
\begin{align}
\begin{split}
\label{eq:solution}
\phi(r) &= 2\varkappa \frac{d-2}{d-1} \, m \, G_d(r) \, , \\
A_i(\mathbf{x}) &= -2\pi\varkappa \, j_i{}^j x^j \, G_{d+2}(r) \, .
\end{split}
\end{align}

Using the explicit form of $G_d(r)$ from equation~\eqref{eq:LocalGF}, we obtain
\begin{align}
\phi(r) &= \frac{\Gamma\left(\tfrac{d}{2}\right)}{(d-1)\pi^{d/2}} \frac{\varkappa m}{r^{d-2}} \, , \\
A_i(\mathbf{x}) &= -\frac{\Gamma\left(\tfrac{d}{2}\right)}{2\pi^{d/2}} \frac{\varkappa j_i{}^j x^j}{r^d} \, .
\end{align}

\section{The Higher and Infinite Derivative Gravity Models}\label{IDGravity}

Infinite derivative gravity was originally proposed to avoid the appearance of unphysical degrees of freedom, known as ghosts, which typically arise in the perturbative quantization of quadratic gravity. These ghost instabilities can be eliminated if the action includes non-polynomial functions of derivatives (as explained in the previous Chapter), which also improve the behavior of the theory in the ultraviolet (UV) regime. The inclusion of such non-polynomial operators makes the gravitational theory nonlocal \cite{tomboulis1997}. This model with quadratic curvature terms included in the gravity action can be written in the form
\cite{Biswas_2014}
\begin{equation}\label{NL_FullAction}
    S_{\mathrm{NL}} = \frac{1}{2\kappa}\int d^4x \sqrt{-g} \left( R + R\mathcal{F}_1(\Box)R + R_{\mu\nu}\mathcal{F}_2(\Box)R^{\mu\nu} + R_{\mu \nu \rho \sigma}\mathcal{F}_3(\Box)R^{\mu\nu\rho\sigma} \right) \, ,
\end{equation}
where $\mathcal{F}_1(\Box)$, $\mathcal{F}_2(\Box)$, and $\mathcal{F}_3(\Box)$ are form factors. Of primary interest are those which are non-polynomial functions of the d'Alembertian operator $\Box = \nabla_{\mu}\nabla^{\mu}$ but in principle they can also be polynomial functions at the cost of introducing ghosts. These functions are assumed to be analytic around $z = 0$ to ensure a smooth and well-defined infrared limit. They can be expanded as Taylor series:
\begin{equation}
    \mathcal{F}_i(\Box) = \sum_{n=0}^{\infty}f_{i,n}\Box^n \, , \quad i = 1,2,3 \, .
\end{equation}

Using the Bianchi identities, it is possible to show that the following relation is valid
\cite{Barvinsky:1990up}
\begin{equation}\label{RiemannID}    R_{\mu\nu\rho\sigma}\Box^nR^{\mu\nu\rho\sigma} = 4 R_{\mu\nu}\Box^nR^{\mu\nu} - R\Box^nR + O(\mathcal{R}^3)  + \mathrm{div.} \, .
\end{equation}
Since the form factors $F_i$ can be expanded in a power series of the operator $\Box$, this identity implies that the Riemann tensor squared contribution in \eqref{NL_FullAction}
is irrelevant as we are neglecting cubic in curvature terms  $O(\mathcal{R}^3)$. As a result, the action \eqref{NL_FullAction} can be written in the form
\cite{BambiModestoShapiro2024}
\begin{equation}\label{NL_Action}
    S_{\mathrm{NL}} = \frac{1}{2\kappa}\int d^4 x \sqrt{-g} \left(R + \frac{1}{2}\left[ R \mathcal{F}_1(\Box) R + R_{\mu\nu} \mathcal{F}_2(\Box) R^{\mu\nu} + O(\mathcal{R}^3) \right] \right) \, .
\end{equation}
Further analysis reveals that the equations can be simplified even more by imposing the following condition \cite{Modesto:2014eta,Barvinsky:1990up}
\begin{equation}
    F(\Box) := \mathcal{F}_2(\Box) = -2\mathcal{F}_1(\Box) \, .
\end{equation}
Then the action \eqref{NL_Action} takes the following simple form
\begin{equation}
    S_{\mathrm{NL}} = \frac{1}{2\kappa}\int d^4 x \sqrt{-g}(R + G_{\mu\nu}F(\Box)R^{\mu\nu}) \, ,
\end{equation}
with $G_{\mu\nu}$ being the Einstein tensor. Varying this action with respect to the metric while keeping now only the linear in curvature terms yields the following gravity equations in the presence of matter \cite{NL_InitialCondsandEOM}
\begin{equation}
    (1 + F(\Box)\Box)G_{\mu\nu} + g_{\mu\nu}\nabla^{\sigma}\nabla^{\rho}F(\Box)G_{\sigma\rho} 
    - 2\nabla^{\sigma}\nabla_{(\mu}F(\Box)G_{\nu)\sigma}  = \kappa T_{\mu\nu} \, ,
\end{equation}
where the form factor $F(\Box)$ is expressed as a power series:
\begin{equation}
    F(\Box) = \sum_{n=0}^{\infty}c_n\Box^n \, ,
\end{equation}
and as usual,
\begin{equation}
    T_{\mu\nu} := \frac{2}{\sqrt{-g}}\frac{\delta S_{\mathrm{matter}}}{\delta g^{\mu\nu}} \, .
\end{equation}
A detailed discussion of these equations can be found in \cite{NL_InitialCondsandEOM,Biswas_2014,Koshelev_2013}.

\section{Linearized Higher and Infinite Derivative Gravity Field Equations}

We consider linear metric perturbations of the form
\begin{equation}
    g_{\mu\nu} = \eta_{\mu\nu} + h_{\mu\nu} \, ,
\end{equation}
and expand the action \eqref{NL_Action} to second order in the metric perturbation $h_{\mu\nu}$. The linearized action becomes \cite{NONLOC}
\begin{align}
    S_{\mathrm{LNL}} &= \frac{1}{4}\int d^4x \left( \frac{1}{2}h_{\mu\nu}f(\Box)\Box h^{\mu\nu} - h^{\ \sigma}_{\mu}f(\Box)\partial_{\sigma}\partial_{\nu}h^{\mu\nu} + hf(\Box)\partial_{\mu}\partial_{\nu}h^{\mu\nu} \right. \\
    & \left. -\frac{1}{2}hf(\Box)\Box h \right) \, ,
\end{align}
where $h := \eta_{\mu\nu}h^{\mu\nu}$ and
\begin{equation}
    f(\Box) := 1 + \frac{1}{2}F(\Box)\Box \, .
\end{equation}

Varying this action with respect to $h_{\mu\nu}$ leads to the following linearized field equations in Cartesian coordinates:
\begin{align}\label{Linearized:NL:EOM}
    f(\Box)\bigg(\Box h_{\mu\nu} - \partial_{\sigma}(\partial_{\nu}h_{\mu}^{\ \sigma}  - \partial_{\mu}h_{\nu}^{\ \sigma}) +\eta_{\mu\nu}(\partial_{\rho}\partial_{\sigma}h^{\rho\sigma} -\Box h) + \partial_{\mu}\partial_{\nu}h \bigg) 
     = -2\kappa T_{\mu\nu} \, .
\end{align}
Once again, we may impose the de Donder gauge, defined by relations \eqref{inv} and \eqref{DDgauge_eq}, in which the field equations simplify to 
\begin{equation}
    f(\Box)\Box\hat{h}_{\mu\nu} = -2\kappa T_{\mu\nu} \, .
\end{equation}
Here, the stress-energy tensor for the perturbation is defined as
\begin{equation}
    T_{\mu\nu} \simeq \frac{2}{\kappa}\frac{\delta S_{\mathrm{matter}}}{\delta h^{\mu\nu}} \, ,
\end{equation}
and satisfies the conservation law $\partial_{\mu}T^{\mu\nu} = 0$.

These linearized equations can then be used to determine the solutions for a higher or infinite derivative gravity model for specific matter configurations. Some examples in the literature include thin shell configurations \cite{Frolov_2015} or spherically symmetric collapse scenarios \cite{frolov2015sphericalcollapsesmallmasses, BraneGF}.

\section{Linearized Solutions in Infinite Derivative Gravity}
An important result in infinite derivative gravity is the finding of solutions in the linear regime, since one can obtain the appropriate Newton limit. Consider the following four dimensional metric ansatz
\begin{equation}
    ds^2 = -(1 + 2\phi)dt^2 + 2\kappa h_{0i}dx^idt + (1-2\phi)d\mathbf{x}^2
\end{equation}
where $i=1,2,3$ is a spatial index such that $d\mathbf{x}^2 = dx^2+dy^2+dz^2$, $\phi$ is some scalar potential. Therefore, one may consider the nonrelativistic limit, in which a pressureless source is such that $T = \eta^{\mu\nu}T_{\mu\nu}\simeq-T_{00}$ whose non zero components are $T_{00}$ and $T_{0i}$, unless it is a static point-like source in which case the second term is zero.
The field equations for a stationary source take the form
\begin{align}\label{Lin_NL}
    f(\Delta)\Delta\phi &= \frac{\kappa}{2}T_{00} \, , \\
    f(\Delta)\Delta h_{0i} &= - 2 \kappa T_{0i} \,  . 
\end{align}
It is customary to assume a de Donder gauge $\partial_\mu h^{\mu\nu} = 0$ whenever $T_{0i} \neq 0 $. One may check that the equations above are exactly the linearized GR equations if one sets $f=1$, which is also equivalent to $\mathcal{F}_1$ and $\mathcal{F}_2$ vanishing from the action \eqref{NL_Action}. One may find a similar form for the field equations for the different choices of the form factors $\mathcal{F}_1$ and $\mathcal{F}_2$ in which one allows more freedom; however this would lead to extra field equations \cite{buoninfante2022blackholesnonlocalgravity}.

At this point it is useful to remind ourselves that the equations above may be solved by either polynomial or non-polynomial form factors $f(\Delta)$, in the next section and throughout this Chapter we focus on the later since we are looking for solutions which do not introduce any new poles to the Green function.

\subsection{Stationary Solutions of Linearized Infinite Derivative Gravity}

Having established the $4D$ case, we now generalize to an arbitrary number of dimensions $D$ to explore the broader structure of solutions. The field equations of linearized gravity with an infinite number of derivatives, written in a compact way by using \eqref{invHD}, for a stationary source, take the form
\begin{equation}\label{fieldeq:lingravNL}
    f(\Delta)\Delta \hat{h}_{\mu\nu} = -2\varkappa T_{\mu\nu} \, .
\end{equation}

In Cartesian coordinates, these equations are similar to those for the electromagnetic potential in the infinite derivative Maxwell theory, which we presented in the previous Chapter \ref{Ch2}. We employ the same strategy to find the solutions as we described earlier.

Assuming a stationary source, the d'Alembert operator reduces to the spatial Laplacian:
\begin{align}
\Delta = \delta^{ij} \partial_i \partial_j \, , \qquad \mathcal{D} = f(\Delta) \Delta \, .
\end{align}
We will work with a form factor $f$ of the $GF_N$ type, as specified in equation~\eqref{FormFactorGauss}.

For the $GF_N$ models, the solution for a point-like source can be written in terms of the corresponding Green function $\mathcal{G}^N_d$, which satisfies the equation
\begin{align}
\label{eq:gf-def}
\mathcal{D} \mathcal{G}^N_d(\mathbf{x}, \mathbf{x}') = -\delta^{(d)}(\mathbf{x} - \mathbf{x}') \, .
\end{align}
These Green functions are the same as those that were introduced for the infinite derivative Maxwell theory (see section \ref{SID_Green}). Considering a source of the form \eqref{eq:SEtensor}, the solutions are 
\begin{align}
\begin{split}\label{eq:NLsolution_anyD}
\phi(r) &= 2\varkappa \frac{d-2}{d-1} {m} \, \mathcal{G}^N_d(r) \, , \\
{A}_{i}(\mathbf{x}) &= -2\pi\varkappa {j}{}_{ij} x^{j} \, \mathcal{G}^N_{d+2}(r) \, .
\end{split}
\end{align}

These solutions remain regular at $r=0$ since this is a general property of the infinite derivative Green function in all dimensions, and they reproduce the linearized Einstein solutions in the case when $f \to 1$. 

\subsection{Point-Like Particle in Four Spacetime Dimensions}

We will once again focus on the $D=4$ case fist. The stress-energy tensor of a point-like spinning particle is given by
\begin{align}\label{GRTT}
T_{\mu\nu} = \delta^t_\mu \delta^t_\nu \, m \, \delta^{(3)}(\mathbf{x}) + \delta^t_{(\mu} \delta^i_{\nu)} \, j_i{}^j \, \partial_j \delta^{(3)}(\mathbf{x}) \, ,
\end{align}
where $m$ is the particle's mass, and $j_{ij}$ is a constant antisymmetric matrix characterizing its angular momentum. A solution to the metric perturbation equations \eqref{solh}--\eqref{solA} sourced by this tensor takes the form\footnote{In this case of four spacetime dimensions, this solution describes the gravitational field of a spinning ring, as discussed in \cite{Buoninfante_2018Ring}.}
\begin{align}
\begin{split}
\label{eq:NLsolution}
\phi(r) &= \kappa m \, \mathcal{G}_3(r) \, , \\
A_i(\mathbf{x}) &= -2\pi\kappa \, j_{ij} x^j \, \mathcal{G}_{5}(r) \, .
\end{split}
\end{align}

Because all static nonlocal Green functions in $GF_N$ models are regular at $r=0$, the same holds true for the metric functions in \eqref{eq:NLsolution}.

At large distances, the fields asymptotically match the standard results from linearized General Relativity \cite{Myers:1986un}:
\begin{align}
\phi(r) &\sim  \frac{\kappa m}{4 \pi r} \, , \\
A_i(\mathbf{x}) &\sim -\frac{1}{4\pi} \frac{\kappa j_{ij} x^j}{r^3} \, .
\end{align}
We adopt the sign convention for $A_i$ such that, one recovers the familiar Lense--Thirring expression ($j_{xy} = j$ and $\kappa = 8\pi G$):
\begin{align}
A_i(\mathbf{x}) \, dx^i \sim \frac{2Gj}{r^3} (x \, dy - y \, dx) = \frac{2Gj}{r} \sin^2\theta \, d\varphi \, .
\end{align}

As an explicit example, in the simplest $GF_1$ model, read:
\begin{align}
\begin{split}
\phi(r) &= \frac{\kappa m}{4\pi r} \mathrm{erf}\left(\frac{r}{2\ell}\right) \, , \\
A_i(\mathbf{x}) &= -\frac{\kappa j_{ij} x^j}{4\pi r^3} \left[ \mathrm{erf}\left(\frac{r}{2\ell}\right) - \frac{r}{\sqrt{\pi} \ell} e^{-(r/2\ell)^2} \right] \, .
\end{split}
\end{align}

\subsection{Field of a Distribution of Matter}

Since this thesis focuses on the problem of singularities and their smearing, we will mainly discuss solutions for point-like sources. However, since the models we are considering in this Chapter are linear, one can use these solutions to obtain the field of extended objects, such as neutron stars. 

Consider a generic stationary distribution of spinning matter has the following stress-energy tensor (see, e.g.,~\cite{BHFrolZel}) 
\begin{align}
\label{TTT}
T_{\mu\nu} = \rho(\mathbf{x}) \delta^t_\mu \delta^t_\nu + \delta^t_{(\mu} \delta^i_{\nu)} \, \partial_j j_i{}^j(\mathbf{x}) \, ,
\end{align}
where $\rho(\mathbf{x})$ is the mass density and $j_i{}^j(\mathbf{x})$ is the angular momentum density.
The solution to equation~\eqref{Lin_NL} for this source takes the form
\begin{align}
\mathbf{h} &= h_{\mu\nu} dX^\mu dX^\nu \, , \label{solh} \\
\mathbf{h} &= \phi \left(dt^2 + \frac{1}{d-2} \delta_{ij} dx^i dx^j \right) + 2 A_i dx^i dt \, , \nonumber \\
\phi(\mathbf{x}) &= 2\varkappa \frac{d-2}{d-1} \int d^d y \, \rho(\mathbf{y}) \, \mathcal{G}_d(\mathbf{x} - \mathbf{y}) \, , \label{solPhi} \\
A_i(\mathbf{x}) &= \varkappa \int d^d y \, j_i{}^j(\mathbf{y}) \frac{\partial \mathcal{G}_d(\mathbf{x} - \mathbf{y})}{\partial x^j} \, . \label{solA}
\end{align}

Due to translational invariance, the Green function depends only on the difference of its arguments:
\begin{align}
\mathcal{G}_d(\mathbf{x}, \mathbf{x}') = \mathcal{G}_d(\mathbf{x} - \mathbf{x}') = \mathcal{G}_d(r) \, ,
\end{align}
where $r = |\mathbf{x} - \mathbf{x}'|$. Using the recursion property of the Green function in equation~\eqref{eq:ch2:recursion}, the vector potential simplifies to
\begin{align}
A_i(\mathbf{x}) = -2\pi\varkappa \int d^d y \, j_{ij}(\mathbf{y}) (x^j - y^j) \, \mathcal{G}_{d+2}(\mathbf{x} - \mathbf{y}) \, .
\end{align}
This illustrates how solutions can be found for some distribution of spinning matter by using the infinite derivative Green function. One interesting example is the $4D$ gravitational field of a rotating ring, which was discussed in \cite{Buoninfante_2018Ring}.

\section{Conclusions}

In this Chapter, we have analyzed a class of higher and infinite-derivative, ghost-free gravitational theories denoted by $\mathrm{GF}_N$. These theories attend the singularity problem in that they regularize the solutions otherwise singular in Einstein gravity. 

We keep our focus on the special class of non-polynomial form factors as we look for solutions that do not introduce any unphysical degrees of freedom via poles in the Green function that solves the field equations.

By first deriving the field equations in the higher and infinite derivative linearized gravity regime, we showed how the results from the previous Chapter, namely the modified Green function, can be used to find gravitational solutions. We considered stationary, rotating point-like sources and showed explicitly how their solutions remain regular at the position of the source. Confirming that these types of models are good candidates to address the singularity problem from General Relativity, which we will investigate in the next Chapter by constructing regular black hole models.

	\chapter{Regular Black Hole Models From Infinite Derivative Gravity}
\label{Ch4}

\section{Introduction}

In the previous Chapter, we demonstrated how infinite-derivative modifications to linearized gravity can effectively eliminate field singularities. A natural question is: can similar techniques be applied within a fully covariant, non-linear framework of gravity?

Unfortunately, addressing this question proves to be highly nontrivial. Even when one restricts the analysis to actions that are quadratic in curvature, the resulting field equations become exceptionally complex (see, e.g., \cite{Biswas_2014}). A central challenge comes from the fact that the d'Alembert operator $\Box$, when defined on a general curved background, depends nontrivially on the metric itself, complicating the analysis considerably.

In this Chapter, we explore an approach aimed at constructing regular black hole metrics inspired by the nonlocal regularization mechanisms of infinite-derivative gravity \cite{Frolov:2023jvi}. Our strategy is based on the double copy method, a framework in which solutions to the Einstein equations are sought in the Kerr–Schild form\footnote{The term double copy refers to the fact that solutions of the Einstein field equations can be constructed from two copies of a corresponding gauge-theory solution: the electromagnetic potential $A^{\mu}$ and a scalar potential $\Phi$. For further details, see Appendix \ref{app:KerrSchild} }:
\begin{equation}
g_{\mu\nu} = \eta_{\mu\nu} + \Phi l_{\mu}l_{\nu} , .
\end{equation}
Here, $\eta_{\mu\nu}$ denotes the flat Minkowski metric, $\Phi$ is a scalar potential, and $\mathbf{l}$ is a null vector tangent to a shear-free geodesic congruence. A remarkable property of this ansatz is that certain exact solutions of the full non-linear Einstein equations—most notably the Kerr metric—can be obtained by solving a linear equation for the potential $\Phi$. Specifically, one only needs to solve the Laplace equation in flat space coordinates $\vec{r} = (X, Y, Z)$:
\begin{equation}\label{FFF}
\Delta \Phi = 4\pi j , ,
\end{equation}
where $j$ is a source term. For the Schwarzschild solution, this corresponds to a point-like source $j = m \delta^{(3)}(\vec{r})$. In the case of the Kerr solution, the source is analytically continued to complexified spatial coordinates.

The double copy approach has garnered significant interest in recent years, with numerous applications across classical and quantum gravity (see, e.g., \cite{BHdoublecopy,ball2023,LUNA2015272,Chawla_2023,dcgluon,Bahjat-Abbas:2017htu,PhysRevD:Goldberg,PhysRevLett:Bern,Borsten2020}). A brief summary of this method and its relevance to our work is provided in Appendix~\ref{app:KerrSchild}. 

Since the basic equation \eqref{FFF} is linear and defined in flat space, it can be naturally extended to include higher—and even infinite—derivatives. This generalization is achieved by replacing the Laplacian $\Delta$ in equation \eqref{FFF} with an operator of the form $f(\Delta)\Delta$, where $f(\Delta)$ is an appropriately chosen form factor. Solutions to this modified equation with a source term $j = m \delta^3(\vec{r})$ were obtained and analyzed in the previous Chapter. By employing these solutions and transforming back to standard static, spherically symmetric coordinates, one obtains metrics that reduce to the Schwarzschild solution in the limit where the nonlocality parameter $\ell \to 0$. In this sense, the resulting metrics can be interpreted as nonlocal perturbations of the classical Schwarzschild black hole.

It is important to emphasize that the modified metric derived in this manner is not an exact solution of a fully nonlocal (infinite-derivative) gravitational theory because of the fact that the full field equations are not being solved and it is not guaranteed that the double copy formalism extends to modified theories of gravity. Nevertheless, one can explicitly verify that the proposed modification eliminates the curvature singularity at $r = 0$, and therefore constitutes a concrete model of a regular black hole.

In this Chapter (based on \cite{Frolov:2023jvi}), we examine this approach in detail. We begin with the case of static, spherically symmetric black holes and subsequently extend the analysis to rotating black holes. Because the position of the point-like source is analytically continued into the complexified spatial coordinates, it becomes necessary to develop a formalism for solving the modified Laplace equation with a nonlocal form factor. This framework is presented in Section 4.3. Finally, in Section 4.4, we apply these tools to construct a regular, modified Kerr black hole metric and analyze its geometric and physical properties.

\section{Infinite Derivative Modification of the Schwarzchild Metric}
Before considering the case of a rotating black hole, let us discuss the simpler case of a static spherically symmetric black hole. In order to define our ansatz, we consider the flat metric
\begin{equation}
    ds_0^2 = -dt^2+ dr^2 + r^2d\Omega^2 \, , \quad   d\Omega^2 = (d\theta^2 + \sin^2\theta d\phi^2) \, ,
\end{equation}
and denote by $\mathbf{l}$ the following 1-form
\begin{equation}
    l_{\mu}dx^{\mu} = -dt + \epsilon dr \, , \quad l^{\mu} = (1,\epsilon,0,0) \, .
\end{equation}
The corresponding vector $l^{\mu}$ is null. For $\epsilon = -1$ it is tangent to incoming radial null rays, and for $\epsilon = 1$ it is tangent to outgoing ones. We use the ansatz in Kerr-Schild form
\begin{equation}
    ds^2 = ds_0^2 + \Phi(l_{\mu}dx^{\mu})^2 \, .
\end{equation}
And consider $\Phi = \Phi(r)$ to be some arbitrary function. Let us first make a few remarks about this ansatz, which explicitly takes the form
\begin{equation}\label{KerrSchildSchw}
    ds^2 = -(1-\Phi)dt^2 - 2\epsilon\Phi dtdr + (1+\Phi)dr^2 + r^2d\Omega^2 \, .
\end{equation}
Notice that the non-diagonal term $g_{tr}$ can be excluded via the coordinate transformation
\begin{equation}
    dt = dt_s - \epsilon \frac{\Phi}{1-\Phi}dr \, ,
\end{equation}
so that it takes the following form in $(t_s,r,\theta,\phi)$ coordinates
\begin{equation}\label{Schw_metric}
    ds^2 = -(1-\Phi)dt_s^2 + \frac{dr^2}{1-\Phi} + r^2d\Omega^2 \, .
\end{equation}
It is important to note that the coordinate $t_s$ is different from $t$. In fact, for $\epsilon = -1$ one has that
\begin{equation}
    dt= dv-dr \, , \quad dv = dt_s + \frac{dr}{1-\Phi} \, .
\end{equation}
Where $v$ is the advanced time coordinate. Therefore, the coordinates $(t,r,\theta,\phi)$ describe the black hole in Kerr-Schild coordinates not only in the exterior of the event horizon but in the interior as well, remaining regular at the horizon. In order to recover the Schwarzschild metric, one should use the function $\Phi=2\varphi$, where
\begin{equation} 
\varphi=-\frac{m}{r} \, ,
\end{equation} 
is the gravitational potential of a point mass $m$ in the Newtonian gravity
\begin{equation} 
\Delta\varphi(r)=4\pi m\delta^3(\vec{r})\, .
\end{equation} 
This means that $\Phi(r)$ satisfies the equation
\begin{equation}\label{flat_lapCH4}
    \Delta\Phi(r) = -8\pi m\delta^{(3)}(\vec{r}) \, . 
\end{equation}
In this equation, both the Laplacian $\Delta$ and the delta function $\delta^{(3)}(\vec{r})$ are defined in the flat spacetime $ds_0^2$. One can write the solution to this equation in terms of the Green function 
\begin{equation}
    \Phi = - 8\pi m G_0(\vec{r}) \, .
\end{equation}
Which satisfies the equation
\begin{equation}\label{eqreal_Green}
    \Delta G_0(\vec{r}) = \delta^{(3)}(\vec{r}) \, .
\end{equation}

In this Chapter, we will only consider four-dimensional spacetime. For simplicity, we omit the subscript $d$ in the notations for the Green functions adopted in the previous Chapters. The index $0$ in the notation $G_0(r)$ now indicates that the corresponding object is calculated in the flat space in Newtonian gravity.

In order to obtain a modification of this metric inspired by infinite derivative gravity, we will consider the potential $\Phi$ to be a solution of the following equation now 
\begin{equation}
    f(\Delta)\Delta\Phi = - 8 \pi m \delta^{(3)}(\vec{X}) \, ,
\end{equation}
where the function $f(\Delta)$ is a form factor akin to those in infinite derivative gravity, which we have already discussed in depth in Chapter \eqref{Ch2}. The integral representation for this Green function is given in \eqref{FNLG}. Particular expressions for some cases of $GF_N$ models are given in equations \eqref{GF1_EQ} - \eqref{GFdN_EQ}. The analytic expression of the potential $\Phi^{(N)}$ takes the form
\begin{equation}
    \Phi^{(N)} = 4\pi m \mathcal{G}^{N}_3 \, .
\end{equation}

It is interesting to note that for all $N$ the potentials $\Phi^{(N)}$ are finite at $r=0$, with the following asymptotic form \cite{boosNl}
\begin{align}
    \Phi^{(N)} &= \Phi_0^{(N)} + \Phi_1^{(N)}r^2 + O(r^4) \, , \\
    \Phi_0^{(N)} &= \frac{2m}{\pi N \ell}\Gamma \bigg(\frac{1}{2N} \bigg) \, , \quad \Phi_2^{(N)} = - \frac{4m}{3N\ell^3}\Gamma \bigg( \frac{3}{2N}\bigg) \, .
\end{align}
It is important to note that the coefficients $\Phi_0^{(N)}$ are finite and positive for all models in the $\mathrm{GF}_N$ class. When this function $\Phi^{(N)}$ is used in the metric \eqref{KerrSchildSchw}, the resulting geometry corresponds to a regular black hole solution which, at the linearized level, coincides with the results of \cite{NONLOC}. This agreement serves as a strong indication of the consistency and validity of the procedure, at least in the case of static and spherically symmetric black holes. We will now proceed to generalize this construction in order to obtain solutions describing rotating regular black holes.

\section{The Kerr-Schild Metric for a Rotating Black Hole}\label{Sec_KerrSchild}

Since we will make use of the double copy mechanism to construct our solution, we will briefly examine the Kerr-Schild form of the Kerr metric, which is (see Appendix~\ref{app:KerrMetric})
\begin{equation}\label{KerrSchildMetric} 
ds^2=ds_0^2+\Phi_0 (l_{\mu}dx^{\mu})(l_{\nu}dx^{\nu}).
\end{equation} 
Here $ds_0^2$ is a flat metric which in the oblate spheroidal coordinates and it has the form
\begin{equation}\label{flatsphero}
\begin{split}
ds_0^2&= - dt^2 + \Sigma\left( \dfrac{dr^2}{\Delta^0_r}+\dfrac{dy^2}{\Delta_y}\right)+\dfrac{\Delta^0_r\Delta_y}{a^2}d\phi^2\, ,\\
\Sigma&=r^2+y^2 \, , \quad  \Delta^0_r=r^2+a^2 \, , \quad  \Delta_y=a^2-y^2\, .
\end{split}
\end{equation}
In these coordinates, the scalar function $\Phi_0$ and the coordinate $y$ are of the form
\begin{equation}\label{LocalPhi}
    \Phi_0 = \frac{2mr}{\Sigma} \, , \quad 
    y=a\cos\theta\, .
\end{equation}
And the $l^{\mu}$ are null tangent vectors to a shear-free congruence of null geodesics, explicitly
\begin{equation}\label{Nullvector} 
\ell_{\mu} dx^{\mu} = -dt + \epsilon\frac{\Sigma}{\Delta^0_r} dr - \frac{\Delta_y}{a}d\phi \,.
\end{equation} 
where $\epsilon = \pm1$. The following statements are valid for both metrics $ds^2$ and $ds_0^2$
\begin{itemize}
    \item The contravariant components of the vector $\mathbf{l}$ in $(t,r,y,\phi)$ coordinates are $l^{\mu} = \bigg(1,\epsilon,0,\frac{a}{\Delta^0_r} \bigg)$ ;
    \item $\mathbf{l}$ is a null vector $\mathbf{l} = l_{\mu}l^{\mu} = 0$ ; 
    \item $\mathbf{l}$ satisfies the relation $l^{\mu}_{\ ; \mu} = \epsilon\frac{2r}{\Sigma} $ ;
    \item $\mathbf{l}$ also satisfies $l_{(\mu ;\nu)}l^{(\mu ; \nu)} - \frac{1}{2}(l^{\mu}_{\ ; \mu})^2 = 0 $.
\end{itemize}
The last property implies that the congruence of null vectors $\mathbf{l}$ is shear-free. This property is discussed e.g. in 
\cite{sommers1976,Frolov1979}. Such null shear free geodesic congruence is related to the light cones with apex on the world-line in the complex plane. Their twist is a measure of how far the complex world-line is from the real slice \cite{Newman:2004ba}.

This form of the metric is related to the standard Boyer–Lindquist coordinates through the following coordinate transformation:
\begin{equation}
\begin{aligned}\label{KerrSch_to_Boyer}
dt &= dt_{BL} - \epsilon\frac{2mr}{\Delta_r} dr, \\
d\phi &= -d\phi_{BL} + \epsilon \frac{2mar}{(r^2 + a^2)\Delta_r} dr.
\end{aligned}
\end{equation}
Where $\Delta_r = r^2-2mr + a^2 $.

\section{Infinite Derivative Green Function for a Source in Complex Spacetime}

In order to obtain a solution for the potential $\Phi$ which can be used to describe a rotating black hole we find ourselves in the need of extending our approach and apply it to a source that now sits in complex Minkowski spacetime, mirroring the Kerr-Schild double copy mechanism for constructing the Kerr metric. This problem was discussed in earlier publications in the context of Maxwell's theory \cite{Newman:2004ba,newman1973,PhysRevD:Newman2002,Complex_Coulomb}. The delta function in a complex flat spacetime was defined in \cite{Complex_delta,Smagin2014,Generalized_delta,Generalized_delta2}, where its properties were also studied. We shall use these results in this section.

\subsection{Complex Delta Function}

Let us now discuss the scalar potential $\Phi_0$ which enters the metric \eqref{KerrSchildMetric}. It is easy to check that it satisfies the Laplace equation 
\begin{equation}
    \Delta \Phi_0 = 0 \, .
\end{equation}
It turns out that $\Phi_0$ is actually a solution of this equation for a point-like source in complex space, which may be written as 
\begin{equation}
    \Phi_0 = -8\pi m \mathrm{Re}[G_0(X,Y,Z+ia)] \, ,
\end{equation}
where $G_0(X,Y,Z+ia)$ is an analytical extension in the complex domain of the Cartesian coordinates of the fundamental solution of the Laplace equation 
\cite{Generalized_delta}. We follow \cite{Complex_delta,Generalized_delta} in order to define the delta function describing the position of our source to obtain the solution for $G_0(X,Y,Z+ia)$. We denote 
\begin{equation}
    \mathcal{Z} = Z +ia \, .
\end{equation}
Then, a generalized delta function $\tilde{\delta}(\mathcal{Z})$ for some complex argument $\mathcal{Z}$ is defined as
\begin{equation}\label{complex_delta}
    \tilde{\delta}(\mathcal{Z}) = \lim_{\sigma \to \infty} \int_{-\infty}^{\infty} e^{-i\mathcal{Z}p}e^{-p^2/2\sigma^2}dp \, .
\end{equation}
Where $\sigma$ is constant. Notice that a Gaussian exponent containing $\sigma$ was introduced to obtain convergence on the integral over $p$. Therefore, the limit $\sigma \to \infty$ should be taken at the end of the calculations. The integral above may be performed, yielding
\begin{equation}
    \tilde{\delta}(\mathcal{Z}) = \lim_{\sigma \to \infty}\sqrt{\frac{\sigma}{\pi}}e^{-\sigma \mathcal{Z}^2} \, .
\end{equation}
It is important to mention that this expression diverges as $\sigma\to \infty$ in the quadrants $|\mathrm{Re(\mathcal{Z})}|\leq |\mathrm{Im}(\mathcal{Z})|$ and converges to zero everywhere else. As long as both endpoints in the integration contour are in the convergent sector, then the definition in \eqref{complex_delta} can be used.
Consider some function $f(z)$ of the complex variable $z$, which is analytic in all of the complex plane and that it decreases sufficiently rapidly at large distances along the real axis. Then, as shown in \cite{Complex_delta,Generalized_delta}, the following equation is valid
\begin{equation}
    \int_{-\infty}^{\infty} f(x)\tilde{\delta}(x-z)dx = f(z)\, .
\end{equation}
One may also check that $\tilde{\delta}(-\mathcal{Z}) = \tilde{\delta}(\mathcal{Z})$. At this point, it is also useful to define the real part of this complex delta function, as we will be using it when constructing our model
\begin{equation}
    \delta_R(\mathcal{Z}) = \frac{1}{2}(\tilde{\delta}(\mathcal{Z}) + \tilde{\delta}(\bar{\mathcal{Z}})) = \lim_{\sigma \to \infty}\frac{1}{2\pi}\int_{-\infty}^{\infty}\cos(pz)e^{ap}e^{-p^2/2\sigma^2}dp \, .
\end{equation}
The relation $\delta_R(z+ia) = \delta_R(-z+ia)$ is also satisfied, meaning that this object is an even function of $z$. Further properties of this generalized delta function and its applications may be found in \cite{Complex_delta,Generalized_delta,Generalized_delta2,Smagin2014,Fourier_Beny}.

\subsection{Local Green Function for a Point-Like Source in Complex Spacetime}
With these tools, we are now equipped to obtain an expression for the Green function $G_0(X,Y,\mathcal{Z})$ which is a solution of 
\begin{equation}\label{eqGreen_complx}
    \Delta G_0(X,Y,\mathcal{Z}) = \delta(X)\delta(Y)\tilde{\delta}(\mathcal{Z}) \, .
\end{equation}
Denote $\vec{\rho} = (X,Y)$ and $\vec{\eta} = (\eta_X,\eta_Y)$. Then
\begin{align}
    \delta(X)\delta(Y) &= \frac{1}{(2\pi)^2}\int e^{-i\vec{\eta}\cdot \vec{\rho}}d^2\vec{\eta} \, , \\
    G_0(X,Y,\mathcal{Z}) &= \frac{1}{(2\pi)^2}\int e^{-i\vec{\eta}\cdot\vec{\rho}}\tilde{G_0}(\vec{\eta},\mathcal{Z})d^2\vec{\eta} \, .
\end{align}
Where the function $\tilde{G}_0(\vec{\eta},\mathcal{Z})$
is defined as 
\begin{equation}
 \tilde{G}_0(\vec{\eta,\mathcal{Z}}) = \lim_{\sigma \to \infty}\frac{1}{2\pi}\int_{-\infty}^{\infty}e^{-i\mathcal{Z}p}e^{p^2/2\sigma^2}\tilde{G}_0(\eta,p)dp \, .
\end{equation}
Using \eqref{eqGreen_complx} one is able to find that 
\begin{equation}
    \tilde{G_0}(\eta,\rho) = - \frac{1}{\eta^2+p^2} \, .
\end{equation}
Where $\eta^2 = \eta_X^2 + \eta_Y^2$. These results may be combined to give
\begin{align}
    G_0(\rho,\mathcal{Z}) &= -\frac{1}{(2\pi)^3}\int d^2\eta e^{-i\vec{\eta}\cdot \vec{\rho}}Y_0(\eta,\mathcal{Z}) \, , \\
    Y_0(\eta,\mathcal{Z}) &= \lim_{\sigma \to \infty}\int_{-\infty}^{\infty}dp \frac{e^{-p^2/2\sigma^2e^{-ip\mathcal{Z}}}}{\eta^2 + p^2} \, .
\end{align}
Let $\vec{\eta}\cdot\vec{\rho} = \eta \rho \cos\phi$ and $d^2\eta = \eta d\eta d\phi$, then since the range of $\phi$ is $[0,2\pi]$
\begin{equation}\label{Bessel_rel}
    \int_0^{2\pi}d\phi e^{-i\rho\eta\cos \phi} = 2\pi J_0(\eta\rho) \, .
\end{equation}
Therefore
\begin{equation}
    G_0(\rho,\mathcal{Z}) = -\frac{1}{4\pi^2}\int_0^{\infty}d\eta \eta Y_0(\eta,\mathcal{Z})J_0(\eta\rho) \, .
\end{equation}
Taking the integral over $p$ for $Y_0$ is possible by using 
\begin{equation}
    Y_0 = \frac{\pi}{\eta}\lim_{\sigma \to \infty} \mathrm{exp} \bigg(\frac{\eta^2}{\sqrt{2}\sigma^2} - \eta \mathcal{Z} \bigg)\bigg[1 - \mathrm{erf}(\eta/\sqrt{2}\sigma)  \bigg] \, .
\end{equation}
At this point, the limit $\sigma \to \infty$ can be taken, yielding the result
\begin{equation}
    Y_0 = \frac{\pi e^{-\eta \mathcal{Z}}}{\eta} \, .
\end{equation}
Using this equation, it is possible to write 
\begin{equation}
    G_0(\eta,\mathcal{Z}) = - \frac{1}{4\pi} \int_0^{\infty}d\eta e^{-\eta \mathcal{Z}} J_0(\eta \rho) \, .
\end{equation}
Which can finally be integrated to give 
\begin{equation}\label{Green_Complex_Cyl}
    G_0(\eta,\mathcal{Z}) = - \frac{1}{4\pi \sqrt{\rho^2 + \mathcal{Z}^2}} \, .
\end{equation}
It is easy to check that 
\begin{equation}
    \rho^2 + \mathcal{Z}^2 = (r+ia\cos\theta)^2 \, .
\end{equation}
We will choose the following prescription
\begin{equation}
    \sqrt{\rho^2 + \mathcal{Z}^2} = r+ia\cos\theta \, ,\quad r \in[0,\infty] \, , \quad \theta \in [0,\pi] \, .
\end{equation}
Hence, we can write equation \eqref{Green_Complex_Cyl} in spheroidal coordinates as 
\begin{equation}\label{localGreen}
    G_0(r,\theta) = - \frac{1}{4\pi}\frac{1}{r+ia\cos\theta} = -\frac{1}{4\pi}\frac{r-ia\cos\theta}{r^2 + a^2\cos^2\theta} \, .
\end{equation}
This equation confirms our statement about $\Phi_0$, in which we said that
\begin{equation}
    \Phi_0 = -8\pi m \mathrm{Re}[G_0(r,\theta)] = \frac{2mr}{r^2+a^2\cos^2\theta} \, .
\end{equation}
Similar solutions for a point source in complex space have been found for the Maxwell equations, the associated electromagnetic field and its properties are studied in \cite{Complex_Coulomb}. The potential has also been used to construct a Newtonian analogue of the Kerr metric \cite{Newtonian_Analogue}. 

\subsection{Complex Potential in an Infinite Derivative Model}

Now, we aim to repeat the procedure above to write a solution for the potential $\Phi$, which we will use instead of $\Phi_0$ in order to modify the Kerr metric. To that end, we will consider the following modified version of equation \eqref{eqGreen_complx}
\begin{equation}\label{eq_NLGreenCompx}
    f(\Delta)\Delta G(X,Y,\mathcal{Z}) = \delta(X)\delta(Y)\tilde{\delta}(\mathcal{Z}) \, .
\end{equation}
Where $f$ is a form factor like the ones discussed in Chapters \ref{Ch2} and \ref{Ch3}. Once we can find a solution for the equation \eqref{eq_NLGreenCompx}, we define the nonlocal potential $\Phi$ as
\begin{equation}
    \Phi = -8\pi m \mathrm{Re}[G(X,Y,\mathcal{Z})] \, .
\end{equation}
So that now, the problem is to find the nonlocal Green function $G(X,Y,\mathcal{Z})$ and in order to solve it, we proceed as in the previous section. Namely
\begin{equation}
    G(X,Y,\mathcal{Z}) = \frac{1}{(2\pi)^2} \int e^{-i\vec{\eta}\cdot\vec{\rho}}\tilde{G}(\vec{\eta},\mathcal{Z})d^2\vec{\eta} \, ,
\end{equation}
with the following representation of the function $\tilde{G}(\vec{\eta},\mathcal{Z})$
\begin{equation}
    \tilde{G}(\vec{\eta},\mathcal{Z}) = \lim_{\sigma \to \infty} \frac{1}{2\pi} \int_{-\infty}^{\infty} e^{-i\mathcal{Z}p}e^{-p^2/2\sigma^2}\tilde{G}(\eta,p)dp \, .
\end{equation}
Using the equation \eqref{eq_NLGreenCompx} one is able to find
\begin{equation}
    \tilde{G}(\eta,\rho) = - \frac{1}{f(-(\eta^2 + p^2))(\eta^2 + p^2)} \, .
\end{equation}
These results may be combined to obtain the following relation
\begin{align}
    G(\rho,\mathcal{Z}) &= - \frac{1}{(2\pi)^3} \int d^2\eta e^{-i\vec{\eta} \cdot \vec{\rho}} Y(\eta,\mathcal{Z}) \, , \\
    Y(\eta,\mathcal{Z}) &= \lim_{\sigma \to \infty} \int_{-\infty}^{\infty} dp \frac{e^{-p^2/2\sigma^2} e^{-ip\mathcal{Z}}}{f(-(\eta^2 + p^2))(\eta^2 + p^2)} \, .
\end{align}
Once again, equation \eqref{Bessel_rel} is useful since it allows us to find
\begin{equation}\label{int_complexGreen}
    G(\rho,\mathcal{Z}) = -\frac{1}{4\pi^2}\int_0^{\infty}d\eta \eta Y(\eta,\mathcal{Z})J_0(\eta\rho) \, .
\end{equation}
For any $\mathrm{GF}_N$ model the integral in $Y(\eta,\mathcal{Z})$ will contain an exponentially decreasing factor which is proportional to $\mathrm{exp}[-(\ell^2(\eta^2+p^2))^N]$ that guarantees convergence of the integral. This also ensures that we can take the limit $\sigma \to \infty$ in the integrand. For the rest of this Chapter, we will focus on the $\mathrm{GF}_1$ case, for which our form factor takes the form
\begin{equation}
    f(\eta^2 + p^2) = e^{-\alpha(\eta^2 + p^2)} \, , \quad  \alpha = \ell^2 \, .
\end{equation}
we may explicitly write 
\begin{equation}
    Y(\eta,\mathcal{Z}) = 2e^{-\alpha\eta^2}\int_0^{\infty}dpe^{-\alpha p^2}\frac{\cos(p\mathcal{Z})}{\eta^2 + p^2} \, .
\end{equation}
This choice is made since the Green function can be found exactly in an explicit form, which we find via the following relation
\begin{equation}
    \frac{d}{d\alpha}Y = -A \, ,
\end{equation}
where
\begin{equation}
    A = 2e^{-\alpha\eta^2} \int_0^{\infty}dp e^{-\alpha p^2}\cos(p\mathcal{Z}) = \sqrt{\frac{\pi}{\alpha}} e^{-\alpha\eta^2}e^{-\mathcal{Z}/4\alpha} \, .
\end{equation}
We may differentiate equation \eqref{int_complexGreen} with respect to $\alpha$ and find 
\begin{equation}
    \frac{dG}{d\alpha} = \frac{1}{4\pi^2}\int_0^{\infty}d\eta \eta AJ_0(\eta\rho) \, .
\end{equation}
The integral on the LHS can be taken to find
\begin{equation}\label{Green_Heat}
    \frac{dG}{d\alpha} = K(\vec{X};\alpha) = \frac{\mathrm{exp} \bigg(-\frac{\rho^2 + \mathcal{Z^2}}{4\alpha} \bigg)}{8 \pi^{3/2}\alpha^{3/2}} \, .
\end{equation}
This may be integrated over $\alpha$, and keeping in mind that $\alpha = \ell^2$ gives
\begin{equation}
    G(\rho,\mathcal{Z}) = - \frac{1}{4\pi}\frac{\mathrm{erf \bigg(\sqrt{\rho^2 + \mathcal{Z}^2} /2\ell\bigg)}}{\sqrt{\rho^2 + \mathcal{Z}^2}} \, .
\end{equation}
We also note that 
\begin{equation}
    \rho^2 + \mathcal{Z}^2 = (r+iy)^2 \, , \quad y = a\cos\theta \, .
\end{equation}
One has 
\begin{equation}
    G(r,y) = -\frac{1}{4\pi}\frac{\mathrm{erf}(\frac{r+iy}{2\ell})}{r+iy} \, .
\end{equation}
Where $\mathrm{erf}(z)$ is the error function of a complex variable $z$. For further information on its definition and properties one can check \cite{ATLAS_functions}. We are making use of the following properties 
\begin{equation}
    \mathrm{erf}(-z) = -\mathrm{erf}(z) \, , \quad \overline{\mathrm{erf}(\zeta)} = \mathrm{erf}(\bar{\zeta}) \, .
\end{equation}
We write our solution in terms of the variable 
\begin{equation}
    \zeta= \frac{r+iy}{2\ell} \, ,
\end{equation}
hence
\begin{equation}
    G(r,y) \equiv G(\zeta) = - \frac{1}{8\pi\ell}\frac{\mathrm{erf}(\zeta)}{\zeta} \, .
\end{equation}
Therefore the solution also obeys the following relations 
\begin{equation}\label{complxGreen_IDS}
    G(-\zeta) = G(\zeta) \, , \quad \overline{G(\zeta)} = G(\bar{\zeta}) \, .
\end{equation}
Since we need to take the real part of the Green function to be able to write an expression for the potential $\Phi$
\begin{equation}
    \Phi = -4\pi m G_R \, .
\end{equation}
where
\begin{equation}
    G_R(\zeta) = 2\mathrm{Re}(G(\zeta)) = G(\zeta) + \overline{G(\zeta)} \, .
\end{equation}
We observe that in the $\mathcal{R}_+$ domain, that is where $r>|y|$, the error function remains finite at infinity. For a fixed value of $r$ and $y$ one has that 
\begin{equation}
    \lim_{\ell \to 0} \mathrm{erf}\bigg( \frac{r+iy}{2\ell}\bigg) = 1 \, .
\end{equation}
Therefore
\begin{equation}
    \lim_{\ell \to 0}G(r,y) = -\frac{1}{4\pi}\frac{1}{r+iy} \, .
\end{equation}
This means that in the local limit, when the characteristic scale of non locality $\ell$ goes to zero, one recovers the local Green function \eqref{localGreen} as a limit of this newly constructed complex nonlocal Green function. 
This property is not valid in the $\mathcal{R}_-$ domain where $r<|y|$, said explicitly, one does not recover the local Green function from its nonlocal version in the limit $\ell \to 0$.

Let us now discuss the case $r<|y|$ in more detail. At the boundary surface $\partial\mathcal{R}$ that separates the  $\mathcal{R_+}$ and $\mathcal{R}_-$ domains, that is when $r = |y|$, the function $G_R(\zeta)$ takes the value 
\begin{equation}
    G_R(\zeta)|_{\partial\mathcal{R}} = G[r(1-i\lambda)] + G[r(1+i\lambda)] \, ,
\end{equation}
where $\lambda = \mathrm{sgn}(y)$. Let us denote 
\begin{equation}
    \tilde{G}(\zeta) = G(i\zeta) \, , \quad \tilde{G}_R(\zeta) = \tilde{G}(\zeta) + \overline{\tilde{G}(\zeta)} \, .
\end{equation}
Using the identities \eqref{complxGreen_IDS}, one may find that on the sphere $\partial\mathcal{R}$
\begin{equation}
    G_R(\zeta)|_{\partial\mathcal{R}} = \tilde{G}_R(\zeta)|_{\partial\mathcal{R}} \, .
\end{equation}
We use $\tilde{G}(\zeta)$ to define the potential $\Phi$ in the domain $\mathcal{R_-}$. We obtain the following expression for the potential $\Phi$ that is valid in both domains $\mathcal{R}_{\pm}$
\begin{equation}\label{nonlocalPhi}
    \Phi = \mu \mathrm{Re}\bigg( \frac{\mathrm{erf}(\zeta)}{\zeta}\bigg) \, ,
\end{equation}
where $\mu = m/\ell$ and
\begin{equation}
    \zeta  = \begin{cases}
 \frac{r+iy}{2\ell} \, ,\quad r>|y| \\
 \frac{y+ir}{2\ell} \, , \quad r<|y|
             \end{cases}
\end{equation}

\begin{figure}[!hbt]
    \centering
      \includegraphics[width=0.7\textwidth]{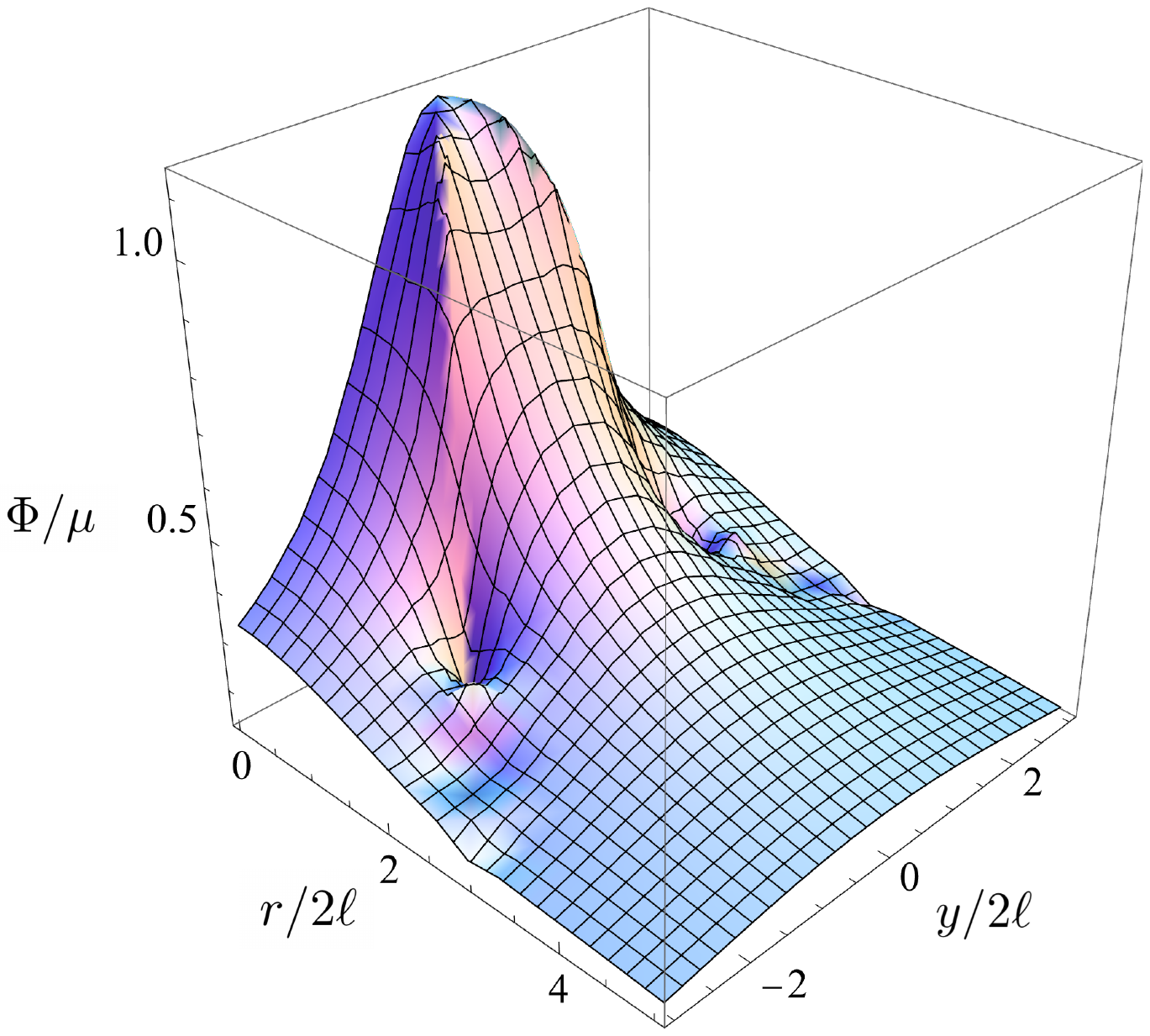}
    \caption{\label{F3} Plot of a potential $\Phi/\mu$ as a function of $(r/2\ell,y/2\ell)$.}
\end{figure}

A plot of the potential is shown in Figure \ref{F3}, it is continuous at $\partial\mathcal{R}$ and has a correct local limit when $\ell \to 0$. At this point, it is useful to introduce the complementary error function
\begin{equation}
    \mathrm{erfc}(z) = 1 - \mathrm{erf}(z) \, ,
\end{equation}
so that we can write the potential $\Phi$ as
\begin{equation}\label{potential_NL}
    \Phi = (r,y) = \Phi_0(r,y) + \Psi(r,y) \, .
\end{equation}
Where $\Phi_0$ is the potential given by the local theory and is given by \eqref{localGreen} and
\begin{equation}\label{Psi_eq}
    \Psi = - \mu \mathrm{Re}\bigg(\frac{\mathrm{erfc}(\zeta)}{\zeta} \bigg) \, .
\end{equation}
This representation is useful because the function $\Psi$ contains all of the contribution from nonlocality to the potential $\Phi$.

We are now ready to construct our model, but before we delve into the discussion of its properties we make one final remark about our nonlocal Green function. The function $K$ which we used to find a solution for $G(X,Y,\mathcal{Z})$ in \eqref{Green_Heat} is of the form
\begin{equation}
    K(\vec{X};\alpha) = \frac{\mathrm{exp}(-\vec{X}/4\alpha)}{8\pi^{3/2}\alpha^{3/2}} \, , \quad \vec{X}^2 = X^2 + Y^2 + (Z+ia)^2 \, .
\end{equation}
This function obeys the equation
\begin{equation}
    \frac{\partial K}{\partial \alpha} - \Delta K = 0 \, .
\end{equation}
Where $\Delta = \partial^2_X + \partial_Y^2 + \partial_Z^2 $ is the Laplacian in flat space. This means that $K$ can be considered as a heat kernel in a space with interval $\vec{X}^2$. This method of obtaining solutions in higher and infinite derivative linearized gravity via heat kernels has been extensively studied by several authors \cite{boosNl,ScalarKernel,Frolov_2015,Boos_2020}.
The real part of the interval $\vec{X}^2$ is positive in the $\mathcal{R}_+$ domain and negative in the $\mathcal{R}_-$ domain. For this reason, the problem with the definition of the Green function in the $\mathcal{R}_-$ domain is similar to the problem of defining the heat kernel in Minkowski space with a Lorentzian signature. This problem can be solved by using a complex parameter $\alpha$ and choosing the proper branch of the corresponding complex function. See \cite{DeWitt:1964mxt,DeWitt:1975ys} for more details.

\subsection{Properties of the Potential}

Having found a solution, we can now discuss the properties of the potential $\Phi$ in detail. We start by noting that Figure \ref{F3} shows oscillations along the lines $r = |y|$, corresponding to the values of the potential $\Phi$ on the ring sphere $\partial \mathcal{R}$, which are displayed more clearly in Figure \ref{VP_RS}. These oscillations grow stronger as they approach the ring $\partial \mathcal{D}$ at $r = y = 0$, but beyond $r \sim 3\ell$ they disappear and the potential steadily approaches a finite value. 

\subsubsection{Potential $\Phi$ at the Ring $\partial\mathcal{D}$}
In order to obtain the value of the potential $\Phi_{\mathrm{ring}}$ at the ring $\partial\mathcal{D}$, defined by $r=y=0$, we make use of the following expansion of the error function \cite{ATLAS_functions,Abramowitz1965}
\begin{equation}
    \mathrm{efc}(\zeta) = \frac{2\zeta}{\sqrt{\pi}} + 0(\zeta^2) \, .
\end{equation}
From which we can obtain 
\begin{equation}\label{Kerrpotential_ring}
    \Phi_{\mathrm{ring}} = \frac{2\mu}{\sqrt{\pi}} \, .
\end{equation}
The potential at the ring is finite and  independent of the rotation parameter $a$.

\subsubsection{Potential $\Phi$ at the Symmetry Axis $\theta = 0$}
\begin{figure}[!hbt]
    \centering
      \includegraphics[width=0.7\textwidth]{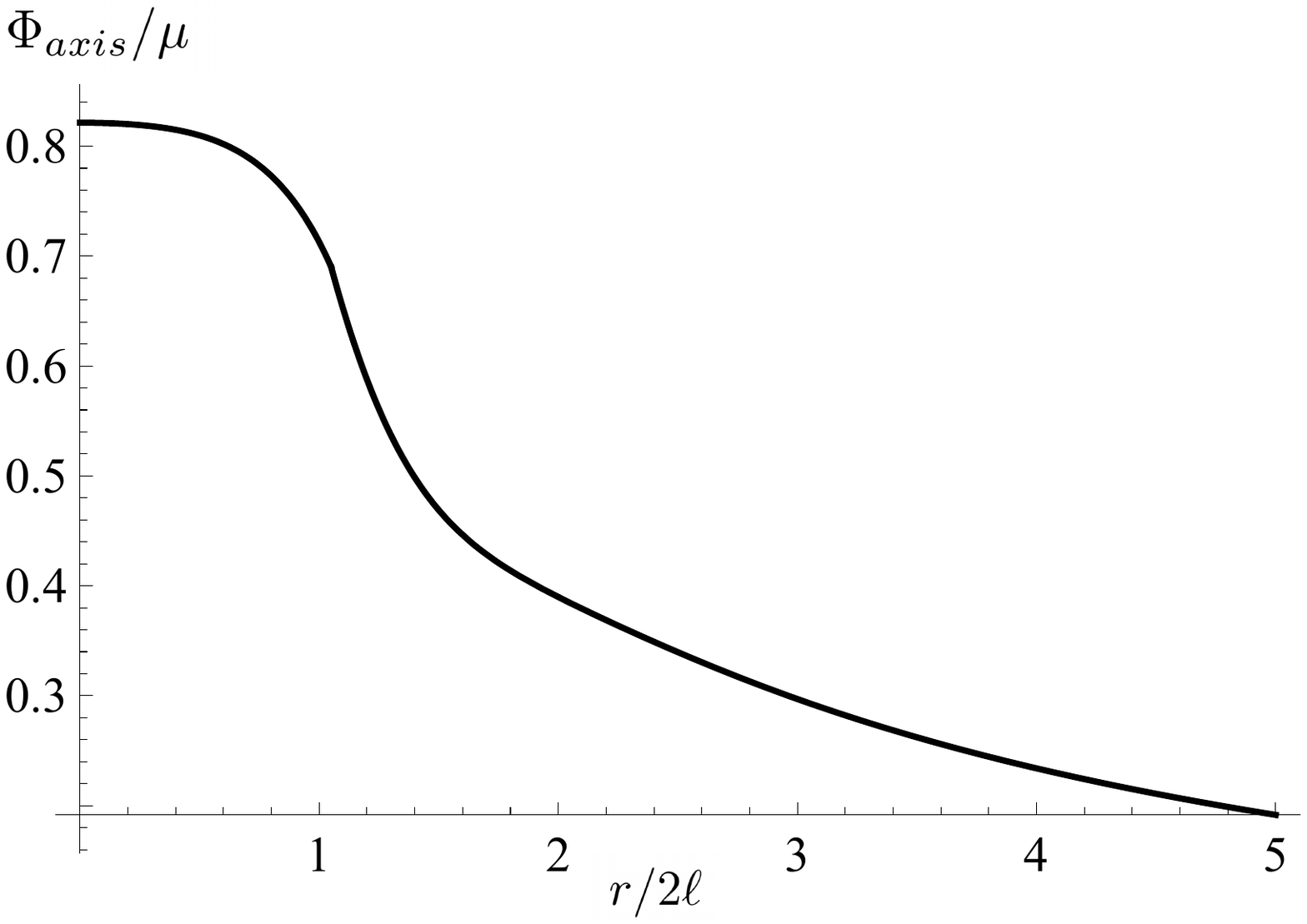}
      \vspace{-0.8cm}
    \caption{\label{PAXIS} Plot of the potential $\Phi_{axis}$ along the $\theta = 0$ axis
    as a function of $r/2\ell$ for $a/2\ell = 1.05$. }
\end{figure}

We consider now the value of the potential at the symmetry axis defined by $\theta = 0$, or equivalently $\theta = \pi$. One has that
\begin{equation}\label{Kerrpotential_axis}
    \Phi_{\mathrm{axis}} = \mu \mathrm{Re}\bigg( \frac{\mathrm{erf}(\zeta)}{\zeta}\bigg) \, ,
\end{equation}
where 
\begin{equation}
    \zeta = \begin{cases}
        \frac{r+ia}{2\ell}\, , \quad r> |y| \\
        \frac{a+ir}{2\ell} \, , \quad r< |y|
    \end{cases}
\end{equation}
The plot of $\Phi_{\mathrm{axis}}$ is shown in figure \ref{PAXIS}

\subsubsection{Potential $\Phi$ on the Disc $\mathcal{D}$}
The disc $\mathcal{D}$ is defined by the equation $r=0$, with the coordinates $0<|y|<a$ and $\phi \in (0,2\pi)$ on the disc. Since $y=a\cos\theta$, the angle $\theta$ on the disc takes values in the interval $(-\pi<\theta<\pi)$.

\begin{figure}[!hbt]
    \centering
\includegraphics[width=0.7\textwidth]{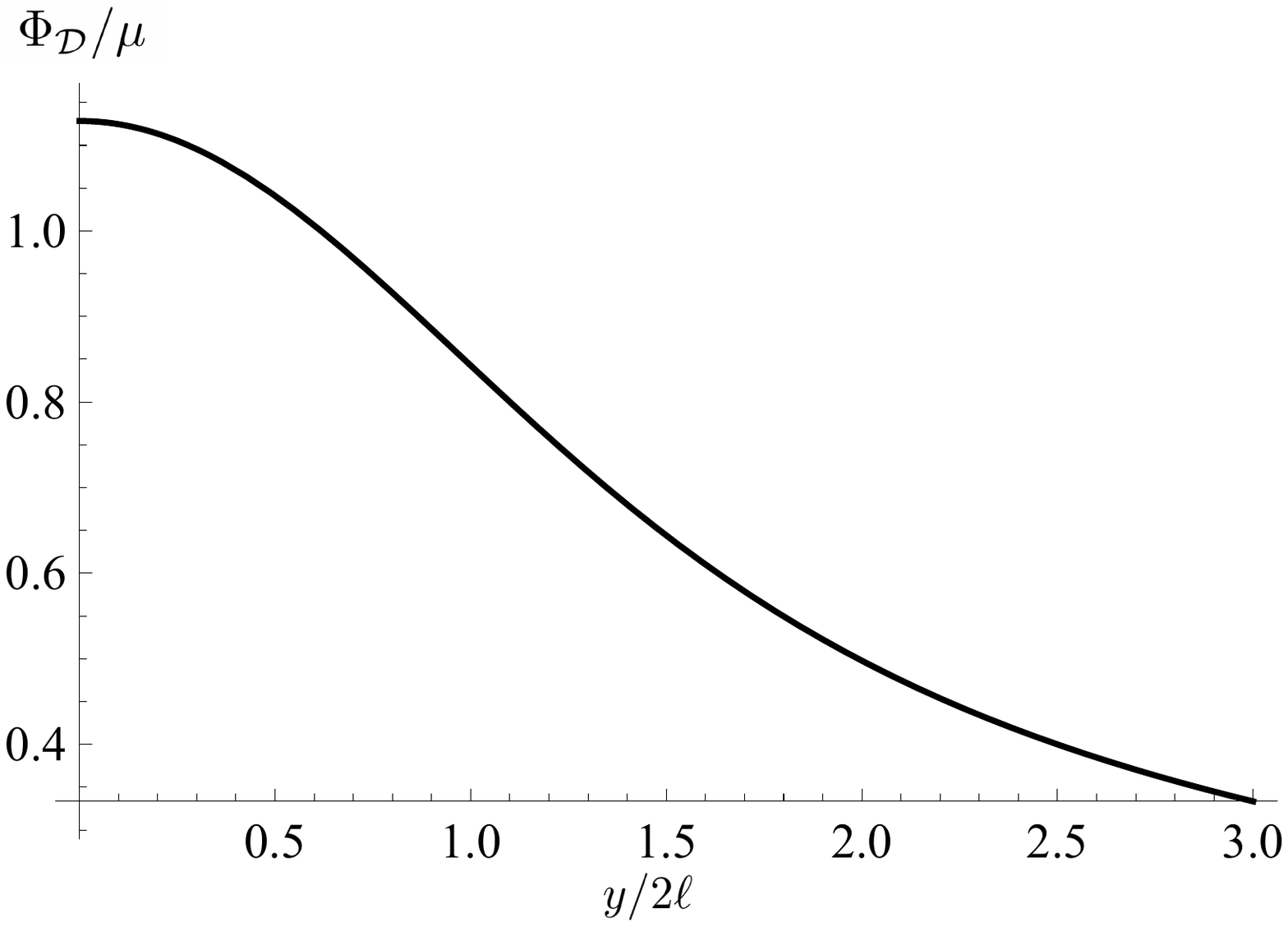}
      \vspace{-0.8cm}
    \caption{\label{VP_DISC} Plot of  the potential  $\Phi_{\mathcal{D}}$  on the disc $\mathcal{D}$ as a function of $\hat{y}=y/2\ell$. }
\end{figure}

We may evaluate the potential to obtain 
\begin{equation}
    \Phi_{\mathcal{D}} = \mu \mathrm{Re}\bigg(\frac{\mathrm{erf}(\zeta_0)}{\zeta_0} \bigg) \, ,
\end{equation}
where $\zeta_0 = y/(2\ell)$. The plot of $\Phi_{\mathcal{D}}$ is shown in figure \ref{VP_DISC}.

Notice that $r=y=0$ is the equation that defines the ring, and the potential is described by equation \eqref{Kerrpotential_ring}. For some disc of radius $a$, the region where $|y|>a$ should be omitted in plot \eqref{VP_DISC}. At the center of this disc, the value of $\Phi_{\mathcal{D}}$ coincides with the limit $r=0$ of the potential $\Phi_{\mathrm{axis}}$ given by \eqref{Kerrpotential_axis}

\subsubsection{Potential $\Phi$ on the Sphere $\partial\mathcal{R}$}

\begin{figure}[!hbt]
    \centering
      \includegraphics[width=0.7\textwidth]{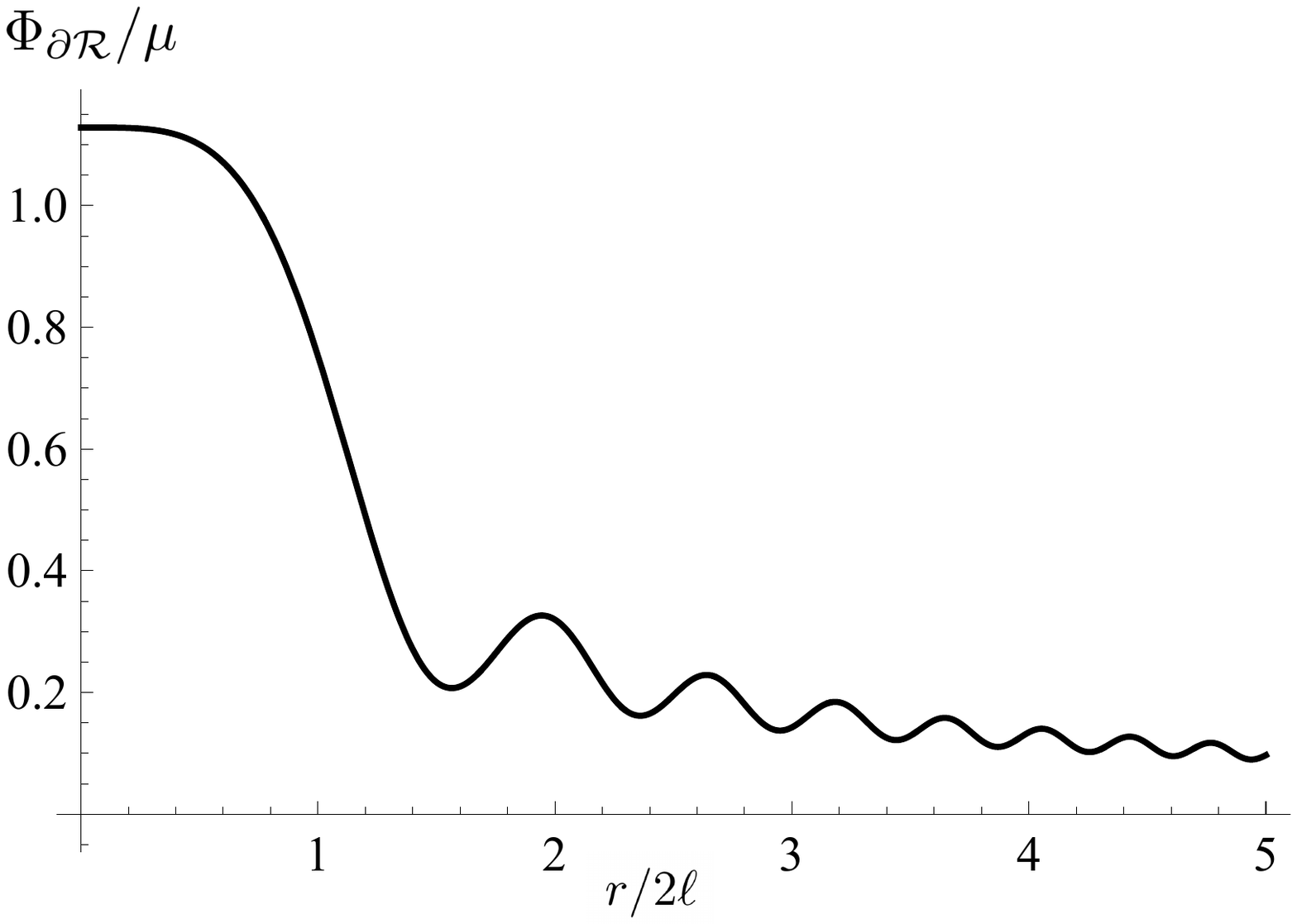}
      \vspace{-0.8cm}
    \caption{\label{VP_RS} Plot of the potential $\Phi$ on the surface $\partial\mathcal{R}$
    as a function of $\hat{r}=r/2\ell$.}
\end{figure}

The sphere $\partial\mathcal{R}$ is defined by the equation $r=|y|$, where the potential $\Phi_{\partial\mathcal{R}}$ is given by
\begin{equation}
    \Phi_{\partial\mathcal{R}} = \mu \mathrm{Re}\bigg(\frac{\mathrm{erf}(\zeta_0 )}{\zeta_0} \bigg) \, , \quad \zeta_0 = (1+i)\frac{r}{2\ell} \, .
\end{equation}
The plot of the potential $\Phi_{\partial\mathcal{R}}$ is shown in figure \ref{VP_RS}. For $r=0$, on the ring $\partial\mathcal{D}$, the potential $\Phi_{\partial\mathcal{R}}$ coincides with equation \eqref{Kerrpotential_ring}.

\subsubsection{Small $\ell$ Limit}

In the limit when $\ell$ is small then the correction term $\Psi$ is small as well, as one would expect. Since the argument of the function $\Psi$ becomes large, for both cases $r>|y|$ and $r<|y|$ we can use the following asymptotic form of the complementary error function \cite{ATLAS_functions}
\begin{equation}
    \mathrm{erfc}(\zeta) = \frac{1}{\sqrt{\pi}\zeta}e^{-\zeta^2} + \cdots \, .
\end{equation}
The nonlocal contribution to the potential $\Phi$ for small $\ell$ is given by the term
\begin{equation}
    \Psi(r,y) = - \frac{\mu}{\sqrt{\pi}}\mathrm{Re}\bigg( \frac{e^{-\zeta^2}}{\zeta^2}\bigg) \, .
\end{equation}

\section{Infinite Derivative Modification of the Kerr Metric}

In the local case, one can use the Kerr-Schild metric \eqref{KerrSchildMetric} with the potential $\Phi_0$ and null vectors $l_{\mu}$ given by \eqref{LocalPhi} and \eqref{Nullvector}, respectively in order to obtain a "standard" rotating black hole. Now we proceed as follows: We keep the metric $ds_0^2$ and the shear-free congruence which enters \eqref{KerrSchildMetric} unchanged (including their properties given in section \ref{Sec_KerrSchild}), but instead of the local potential $\Phi_0$ we substitute into the expression \eqref{KerrSchildMetric} the modified version of $\Phi$, which was obtained in the previous sections \eqref{nonlocalPhi}. This potential is a solution to the flat infinite derivative modification of the Laplace equation with the same point-like source in the complexified space as in the Kerr metric. As a result of this procedure, one obtains a metric which reproduces the Kerr metric asymptotically at large $r$.  In the limit when the non-locality parameter $\ell\to0$, it coincides with the latter. We call this metric the {\it nonlocal modification} of the Kerr metric, and it has the form
\begin{equation}\label{NLKerrMetric}
    ds^2=ds_0^2+\Phi (l_{\mu}dx^{\mu})(l_{\nu}dx^{\nu}) \, .
\end{equation}
As we have shown, this metric is finite at the ring, and in this sense, the ring singularity is smeared due to non-locality. 

It should be emphasized that one cannot guarantee that this metric is a solution of the exact nonlinear nonlocal generalization of the Einstein equations discussed in section \ref{IDGravity}. However, in the limit when $\ell\to 0$, it correctly reproduces the exact Kerr solution. Viewed this way, for small $\ell$, one may consider the modified metric as a nonlocal deformation of the metric from the local theory. We shall use this metric as a special type of non-singular rotating black hole model.
In the last sections of this Chapter, we discuss in more detail the properties of this model.

\section{Properties of the Metric}

Let us note that for the modified metric the quantity $\Sigma\Phi$ depends not only on the coordinate $r$ but also on the coordinate $y$ in the nonlocal metric. This represents a key departure from the Kerr metric and has several consequential implications 
\begin{itemize}
    \item In the local case, one can use the  Boyer-Lindquist coordinates in which all of the non-diagonal terms of the metric except $g_{t\phi}$ vanish. This is no longer valid for the nonlocal modification of the Kerr solution.
    \item The nonlocal version of the metric still has two Killing vectors $\mathbf{\xi}_{(t)} = \partial_t$ and $\mathbf{\xi}_{(\phi)} = \partial_{\phi}$, but these vectors do not satisfy the circularity conditions which are valid for the Kerr solution
    \begin{equation}
    \xi_{(\phi)[\alpha}\xi_{(t)\beta}\xi_{(t)\gamma;\delta ]} = \xi_{(t)[\alpha}\xi_{(\phi)\beta}\xi_{(\phi)\gamma;\delta ]} = 0 \, .
    \end{equation}
    \item Because the circularity conditions are not satisfied,   the surface $V=0$, where 
    \begin{equation}
        V = (\mathbf{\xi}_{(t)}\cdot\mathbf{\xi}_{(\phi)})^2 - \mathbf{\xi}_{(t)}^2\mathbf{\xi}_{(\phi)}^2 \, ,
    \end{equation}
    is not the event horizon anymore
\end{itemize}

Let us discuss the last point in more detail. One can check that the function $V$ vanishes when 
\begin{equation}
    \mathcal{V} \equiv \Delta_r^0 - \Sigma\Phi = 0\, ,
\end{equation}
is satisfied. We can calculate
\begin{equation}
    (\nabla\mathcal{V})^2 \equiv \mathcal{V}_{;\mu}\mathcal{V}^{;\mu} = \frac{1}{\Sigma}[\Delta_y(\Sigma\partial_y \Phi + 2y\Phi)^2 + V(\Sigma\partial_r\Phi + 2r(\Phi - 1))^2] \, .
\end{equation}
On the surface $\mathcal{S}_V$, which is defined by $V=0$, the second term inside the square bracket vanishes, and we are left with 
\begin{equation}
    (\nabla\mathcal{V})^2|_{\mathcal{S}_V} = \frac{1}{\Sigma}\Delta_y[\partial_y(\Sigma\Phi)]^2 \, . 
\end{equation}
If $\partial_y(\Sigma\Phi)\neq0$ and $|y|<a$ then $(\nabla\mathcal{V})^2>0$, which means that, in general, the surface $\mathcal{S}_V$ outside of the symmetry axis is timelike and hence it cannot be an event horizon. 

The ergosurface surface $\mathcal{S}_H$ of the Kerr metric is defined by  the equation $g_{tt} = \mathbf{\xi}_{(t)}^2 = 0$. In  Kerr-Schild coordinates \eqref{NLKerrMetric}, this surface is defined by the relation 
\begin{equation}
    \Phi = 1 \, .
\end{equation}
This is a surface with infinite red-shift. Outside of it, a particle can be at rest with respect to infinity such that its 4-velocity 
\begin{equation}
    u^{\mu} = \frac{\xi^{\mu}_{(t)}}{|\mathbf{\xi}^2_{(t)}|^{1/2}} \, ,
\end{equation}
is timelike.

The ergosurface is the outer boundary of the ergosphere—the region where particles can move either inward or outward in the radial direction, but are necessarily dragged into rotation around the black hole. In the Kerr geometry \eqref{KerrSchildMetric}, the inner boundary of the ergosphere coincides exactly with the event horizon. This property, however, is no longer preserved in the modified nonlocal Kerr geometry \eqref{NLKerrMetric}, where a separation between the ergosurface and the event horizon emerges.

In the ergoregion, a particle can move along a circular orbit such that its 4-velocity is proportional to a linear combination of the Killing vectors
\begin{equation}
    \eta^{\mu} = \xi_{(t)}^{\mu} + \omega \xi_{(\phi)}^{\mu} \, ,
\end{equation}
where $\omega$ is a constant describing angular velocity. The vector $\mathbf{\eta}$ is timelike when $\omega \in (\omega_-,\omega_+)$, where 
\begin{equation}
    \omega_{\pm} = \frac{-\mathbf{\xi}_{(t)}\cdot\mathbf{\xi}_{(\phi)} \pm \sqrt{V}}{\mathbf{\xi}^2_{(\phi)}} \, .
\end{equation}
If $\omega_+ = \omega_{-}$ the vector $\mathbf{\eta}$ is null. At $\mathcal{S}_V$ one has
\begin{equation}
    \Omega = \omega_{+}= \omega_{-}= -\frac{\mathbf{\xi}_{(t)}\cdot\mathbf{\xi}_{(\phi)}}{\mathbf{\xi}^2_{(\phi)}} \, .
\end{equation}

The surface defined by $V=0$ is the inner boundary of the ergosphere. In the case of the Kerr metric, the surface $\mathcal{S}_V$ coincides with the event horizon, it is null and it plays the role of a one way membrane. For our case in which the potential $\Phi$ is a more general function the situation is quite different: the surface $\mathcal{S}_V$ is timelike so it can be penetrated by outgoing particles and light rays. 
The position of the inner boundary $r = r_V(y)$ of the ergoregion, where $V=0$, is defined by 
\begin{equation}
    r^2 + a^2 - 2Mr = \Sigma\Psi \, ,
\end{equation}
with $\Psi$ defined by \eqref{Psi_eq}.
Since our correction $\Psi$ is a small term, the surface $\mathcal{S}_V$ will be slightly deviated with respect to the unperturbed Kerr horizon at $r=r_H$. Let us define
\begin{equation}
    h_V(y) = r_V(y) - r_H \, .
\end{equation}
Then we have explicitly that
\begin{equation}
    \hat{h}_V(y) \equiv \frac{1}{m}h_V(y) = \bigg[ \frac{\Sigma}{2b}\Psi\bigg]_{r=r_H} \, ,
\end{equation}
where $b=\sqrt{m^2 - a^2}$.
Since we are focusing on the $\mathrm{GF}_1$ model, we use the expression \eqref{Psi_eq} and find that
\begin{align}\label{eq_f_psinl}
    \hat{h}_V(y) &= -f(x) \, , \\
    f(x) &= \frac{\mu}{2\hat{b}}\bigg((1+\hat{b})^2 + (1-\hat{b})^2x^2 \bigg)\mathrm{Re}\bigg[ \frac{\mathrm{erfc(\zeta)}}{\zeta}\bigg] \, .
\end{align}
Where we defined 
\begin{equation}\label{diemnsionless_variables}
    x= \frac{y}{a} \, , \quad \hat{b} = \frac{b}{m} \, , \quad \mu = \frac{m}{\ell} \, .
\end{equation}

\subsection{Shift of the Event Horizon}

The event horizon of a stationary black hole coincides with the outer trapped surface. Introduced by Penrose to characterize the point of no return in gravitational collapse \cite{Penrose:1964wq}, trapped surfaces are closed spacelike surfaces such that their area decreases along any possible future direction. Consider a 2D surface $S$, and define the scalar
\begin{equation}
 \kappa = -g^{ab}H_aH_b|_S
\end{equation}
Here, $a$ and $b$ denote coordinates on this surface, and $H_a$ is the mean curvature one form. A surface $S$ is deemed trapped if $\kappa$ is positive; for black hole metrics, the event horizon resides where $\kappa = 0$, which is a special type of surface called \textit{marginally trapped surface}. 

In this section, we will follow the results of \cite{trapped}, where a formalism to find such surfaces is developed, in order to find the event horizon of our modified infinite derivative Kerr metric.
We assume that in the vicinity of the horizon the potential $\Phi$ will differ from its local, unperturbed value $\Phi_0$ only slightly. This means that $\Psi$ is small, and that the displacement of the horizon $h(y)$ for the infinite derivative modification of the Kerr metric $r_{H,\ell}$ with respect to the Kerr horizon $r_H$ is also a small quantity. We write
\begin{equation}
    r = r_{H,\ell} \equiv r_H+h(y) \, ,
\end{equation}
with $h(y)$ small. For now, we keep $\Psi$ arbitrary, only assuming that it is an even function of $y$.

In order to derive an equation describing the shift of the event horizon relative to the Kerr metric, we begin by introducing the coordinate
\begin{equation}\label{XX}
x = r - r_H - h(y) \, ,
\end{equation}
and consider a family of 2D surfaces $\mathcal{S}$ defined by
\begin{equation}\label{SURF}
t = t_0 \, , \quad x = x_0 \, ,
\end{equation}
where $t_0$ and $x_0$ are constant parameters. The 2D surface $\mathcal{S}_H$ defined by $x = 0$ corresponds to the intersection of the event horizon $\mathcal{H}$ with the 3D surface $t = t_0$. This implies that $\mathcal{S}_H$ is a marginally trapped surface. Consequently, the task of locating the event horizon reduces to determining the function $h(y)$ that defines $\mathcal{S}_H$.

To proceed, we transform to coordinates $(t, x, y, \phi)$ using the relations
\begin{equation}
dr = dx + \upsilon(y) dy \, , \quad \upsilon(y) = \dfrac{dh}{dy} \, ,
\end{equation}
and express the metric \eqref{NLKerrMetric} in the form
\begin{equation}\label{MET}
ds^2 = g_{ab} dx^a dx^b + 2g_{aA} dx^a dx^A + g_{AB} dx^A dx^B \, ,
\end{equation}
where the indices $a, b$ range over $\{0,1\}$ and $A, B$ over $\{2,3\}$. We define the coordinates explicitly as
\begin{equation}\label{CC}
x^0 = t \, , \quad x^1 = x \, , \quad x^2 = y \, , \quad x^3 = \phi \, .
\end{equation}
Fixing the coordinates $x^a$ specifies a 2D surface $\mathcal{S}$, with $x^A$ coordinates on it. The explicit form of the metric \eqref{MET} in these coordinates is
\begin{equation}\label{METR_X}
    \begin{split}
        g_{a b}dx^adx^b &= -(1 + \Phi)dt^2 + \frac{2\Phi \Sigma }{\Delta^0_r}dtdx  + \frac{\Sigma(\Phi\Sigma + \Delta^0_r)}{(\Delta^0_r)^2}dx^2 \, , \\
        g_{a A}dx^adx^A &= \frac{\Phi\Sigma \upsilon}{\Delta^0_r}dtdy + \frac{\Phi \Delta_y}{a}dtd\phi + \frac{\Sigma \upsilon(\Sigma \Phi + \Delta^0_r)}{(\Delta^0_r)^2}dxdy + \frac{\Phi \Delta_y \Sigma}{a \Delta^0_r}dxd\phi \, ,\\
        g_{A B}dx^Adx^B &= \frac{\Sigma((\Delta^0_r)^2 + \upsilon^2\Phi\Sigma\Delta_y + \upsilon^2 \Delta_y\Delta^0_r)}{\Delta_y(\Delta^0_r)^2}dy^2 + \frac{2 \Phi \Sigma \Delta_y \upsilon}{a\Delta^0_r}dyd\phi  \\
        & + \frac{\Delta_y(\Delta_y\Phi + \Delta^0_r)}{a^2}d\phi^2 \, .
    \end{split}
\end{equation}

Let us denote by $\gamma_{AB}$ the induced 2D metric on $\mathcal{S}$ and by $\gamma^{AB}$ its inverse. Following the formalism of \cite{trapped}, we introduce the following auxiliary quantities:
\begin{equation}\label{SEN}
\begin{split}
G &= \sqrt{\det g_{AB}} \equiv e^U \, ,\\
\vec{g}_a &= g_{aA} dx^A \, ,\\
\text{div}\, \vec{g}_a &= \dfrac{1}{G} \left( G \gamma^{AB} g_{aA} \right)_{,B} \, ,\\
H_\mu &= \delta_\mu^a \left( U_{,a} - \text{div}\, \vec{g}_a \right) \, .
\end{split}
\end{equation}

We now make use of the assumption that the potential $\Phi$ deviates only slightly from its local (Kerr) counterpart $\Phi_0$, and we write
\begin{equation}\label{HOR}
\Phi(r,y) = \Phi_0 + \lambda \Psi(r,y) \, ,
\end{equation}
introducing a dimensionless parameter $\lambda$ that quantifies the smallness of the deviation. Our analysis is restricted to terms up to first order in $\lambda$; at the end of the calculations, we set $\lambda = 1$.

Under this approximation, the $g_{AB}$ part of the metric \eqref{MET} takes the form
\begin{equation}\label{KSP0}
\begin{split}
g_{AB} dx^A dx^B &= \frac{\Sigma}{\Delta_y} dy^2 + \frac{2 \Phi \Sigma \Delta_y \upsilon}{a \Delta^0_r} dy d\phi + \frac{\Delta_y \Upsilon}{a^2} d\phi^2 \, ,\\
\Upsilon &= \Delta_y \Phi + \Delta^0_r \, ,
\end{split}
\end{equation}
from which it follows that
\begin{equation}
G \equiv \sqrt{\det g_{AB}} = \dfrac{\sqrt{\Sigma \Upsilon}}{a} \, .
\end{equation}

It is important to note that the coefficients in the metrics \eqref{NLKerrMetric} and \eqref{MET} are functions of $(r, y)$. When re-expressed in $(x, y)$ coordinates, the corresponding partial derivatives must be computed using the chain rule:
\begin{equation}\label{X_DER}
\begin{split}
\dfrac{\partial B(r,y)}{\partial x}\bigg|_y &= \dfrac{\partial B(r,y)}{\partial r}\bigg|_y \, ,\\
\dfrac{\partial B(r,y)}{\partial y}\bigg|_x &= \dfrac{\partial B(r,y)}{\partial y}\bigg|_r + \upsilon \dfrac{\partial B(r,y)}{\partial r}\bigg|_y \, .
\end{split}
\end{equation}

The $t$-component of $U_{,a}$ vanishes, while the $x$-component is given by
\begin{equation}\label{Ux}
U_{,x} = \frac{\partial_r(\Sigma \Upsilon)}{2 \Sigma \Upsilon} \, .
\end{equation}
For the divergence of the vector field $\vec{g}_a$, one finds:
\begin{equation}\label{KSPt}
\text{div}\, \vec{g}_t = \frac{1}{2\Sigma \Upsilon^2} \left[ \Delta_y \Phi \Upsilon (2\Sigma \partial_y \upsilon + \upsilon \partial_y \Sigma) + \Sigma \upsilon (\Upsilon + \Delta^0_r) \partial_y \Upsilon \right] \, ,
\end{equation}
\begin{equation}\label{KSPx}
\begin{split}
\text{div}\, \vec{g}_x &= \frac{1}{2\Sigma \Delta^0_r \Upsilon^2} \Big[ \upsilon \Delta_y \Upsilon (\Upsilon + 3\Sigma \Phi) \partial_y \Sigma + \Sigma \upsilon \Delta_y \left( \Delta^0_r (\Upsilon + \Sigma) \right) \partial_y \Phi \\
&\quad + \Sigma \upsilon \left( \Sigma \Phi (\Upsilon + \Delta^0_r) + \Upsilon (2\Upsilon + \Delta_y \Phi) \right) \partial_y \Delta_y + 2\Sigma \Delta_y \Upsilon (\Upsilon + \Sigma \Phi) \partial_y \upsilon \Big] \, .
\end{split}
\end{equation}

Substituting these expressions into $H_\mu$, as defined in \eqref{SEN}, we then evaluate the surface gravity $\kappa$ using the \texttt{GRtensor} package in \texttt{Maple}. In accordance with our approximation, the inverse metric $g^{ab}$ is truncated to include only zeroth- and first-order terms in $\lambda$.

Following the perturbative expansion, we again substitute \eqref{HOR} into the resulting expression for $\kappa$, retaining only the leading-order contributions in $\lambda$. In particular, this allows us to replace $\Psi(r,y)$ by $\Psi(r_h, y)$, since $\Psi$ is already first order in $\lambda$.

As expected, the contribution to $\kappa$ at order $\lambda^0$ vanishes, confirming that $r = r_h$ corresponds to the event horizon of the unperturbed Kerr metric. Imposing the condition $\kappa = 0$ at first order in $\lambda$ yields a differential equation for the function $h(y)$, which characterizes the displacement of the event horizon in the perturbed metric
\begin{equation}\label{ode_shift}
\begin{split}
    &\frac{d}{dy}\bigg[ (a^2 - y^2)\frac{dh}{dy}\bigg] - (\alpha + \tilde{\beta}y^2)h^2 = \varpi\Psi \, , \\
    &\alpha = \frac{\beta}{b}(m(4m^2 + 7mb+4b^2) + b^3) \, , \\
    & \tilde{\beta} = \frac{b^2}{4m^2r_H^2} \, , \quad \varpi = - \frac{1}{2b}(r_H^2+y^2)(\alpha + \tilde{\beta}y^2) \, .
\end{split}
\end{equation}

In order to find a solution, it is convenient to write equation \eqref{ode_shift} in dimensionless form by using $\hat{h} = h/m$ and \eqref{diemnsionless_variables}
\begin{equation}\label{ode_shift_dimless}
\begin{split}
   & \frac{d}{dx}\bigg[ (1-x^2)\frac{d\hat{h}}{dx}\bigg] - (\alpha + \beta x^2) = F(x) \, , \\
   &\beta = \tilde{\beta}a^2 = \frac{\hat{b}^2(1-\hat{b}^2)}{4(1+ \hat{b})^2} \, , \\
   & \alpha = \frac{\hat{b}}{4(1+\hat{b})^2}(4 + 7\hat{b} + 4\hat{b}^2 + \hat{b^3}) \, , \\
   &F = -\frac{1}{2\hat{b}}((1+\hat{b})^2 + (1-\hat{b^2})x^2)(\alpha + \beta x^2)\Psi \, .
\end{split}
\end{equation}
Here $\hat{h}$ is an even function of $x$ and therefore it satisfies 
\begin{equation}\label{bdry_h}
    \frac{d\hat{h}}{dx}\bigg|_{x=0}=0 \, .
\end{equation}
Both $\hat{h}(x)$ and $F(x)$ are regular at the symmetry axis, defined in these coordinates by $x=\pm1$. Near this axis, they can be expanded as
\begin{equation}
\begin{split}
    \hat{h}(x) &= \hat{h}_0 + \hat{h}_1(1-x^2) + O((1-x^2)^2) \, , \\
    F(x) &= F_0 + F_1(1-x^2) + O((1-x^2)^2) \, .
\end{split}
\end{equation}
Using these expansions in equation \eqref{ode_shift_dimless} we can obtain the following relation
\begin{equation}\label{bdry_ode_dimless}
    \bigg[ \frac{d\hat{h}}{dx} + \frac{1}{4}(\alpha + \beta)\hat{h} - \frac{1}{4}F\bigg]_{x=\pm1} = 0 \, .
\end{equation}
Equation \eqref{ode_shift_dimless} equipped with the boundary conditions \eqref{bdry_h} and \eqref{bdry_ode_dimless} is a well posed boundary value problem that can be then solved numerically. 
First, we will show that for $F=0$ the homogeneous equation \eqref{ode_shift_dimless} does not have a regular solution. This equation is invariant under $\hat{h}(x) \to - \hat{h}(x)$, therefore it is sufficient to consider only the $\hat{h}(0)>0$ case. Using the initial condition \eqref{bdry_h} we have
\begin{equation}
    \frac{d\hat{h}}{dx} = \frac{1}{1-x^2}\int_0^x(\alpha + \beta x^2)\hat{h}(x)dx \, .
\end{equation}
From this relation, we can conclude that $\hat{h}(x)$ is a positive monotonically growing function of $x$, and as a result of this $d\hat{h}/dx$ grows infinitely at $x=1$. Additionally, let us note that this homogeneous equation is the equation for the oblate spheroidal angle functions \cite{Flammer2014}. Given a certain value of $\beta$, regular solutions exist only for special values of $\alpha$, which are the eigenvalues of this problem. For our adopted form of the coefficients $\alpha$ and $\beta$, this homogeneous equation only has the trivial regular solution $\hat{h}(y) = 0$.

\subsubsection{Numerical Solutions}

\begin{figure}[!hbt]
    \centering
      \includegraphics[width=0.7\textwidth]{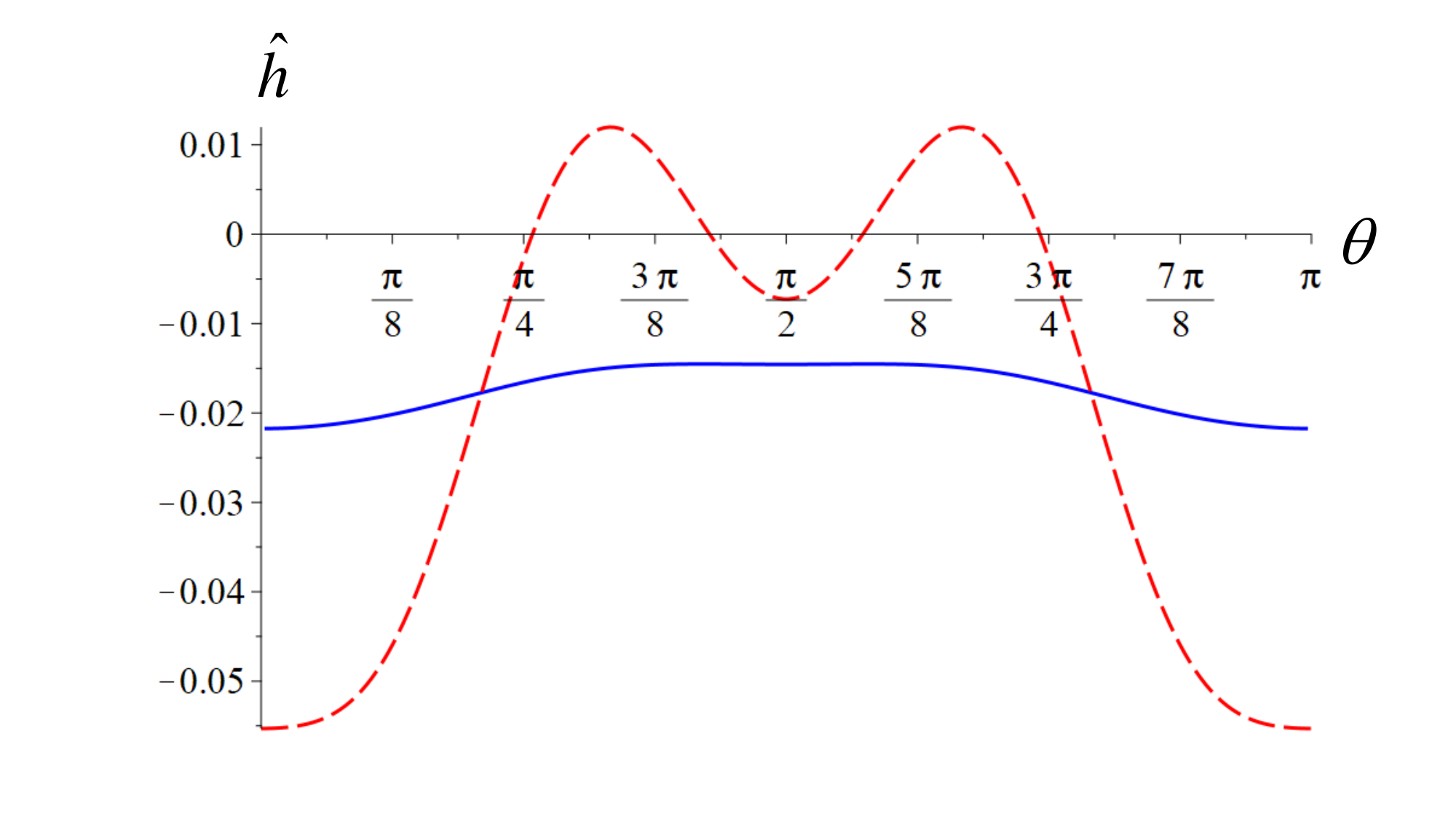}
\vspace{-0.8cm}
    \caption{\label{F7}  Plots of $\hat{h}$ (solid line) and $\hat{h}_V(\theta)$ (dash line) as a function of the angle $\theta$ for parameters $\mu=3$ and $\hat{b}=0.3$ ($a=0.9$). }
\end{figure}

Now, in order to find a numerical solution, it will be convenient to use a function $\hat{h}(\theta)$ where $x=\cos\theta$ for $\theta \in [0,2\pi]$. We may rewrite \eqref{ode_shift_dimless} as
\begin{equation}
    \frac{d^2\hat{h}}{d\theta^2} + \cot\theta \frac{d\hat{h}}{d\theta} - (\alpha + \beta\cos^2\theta)\hat{h} = F(\cos^2\theta) \, .
\end{equation}
We want a solution which satisfies the condition
\begin{equation}\label{bdry_anglecoords}
    \frac{d\hat{h}}{d\theta}\bigg|_{\theta = \pi/2} = 0 \, ,
\end{equation}
and that is regular at $\theta = 0$ and $\theta = \pi$.

\begin{figure}[!hbt]
    \centering
      \includegraphics[width=0.7\textwidth]{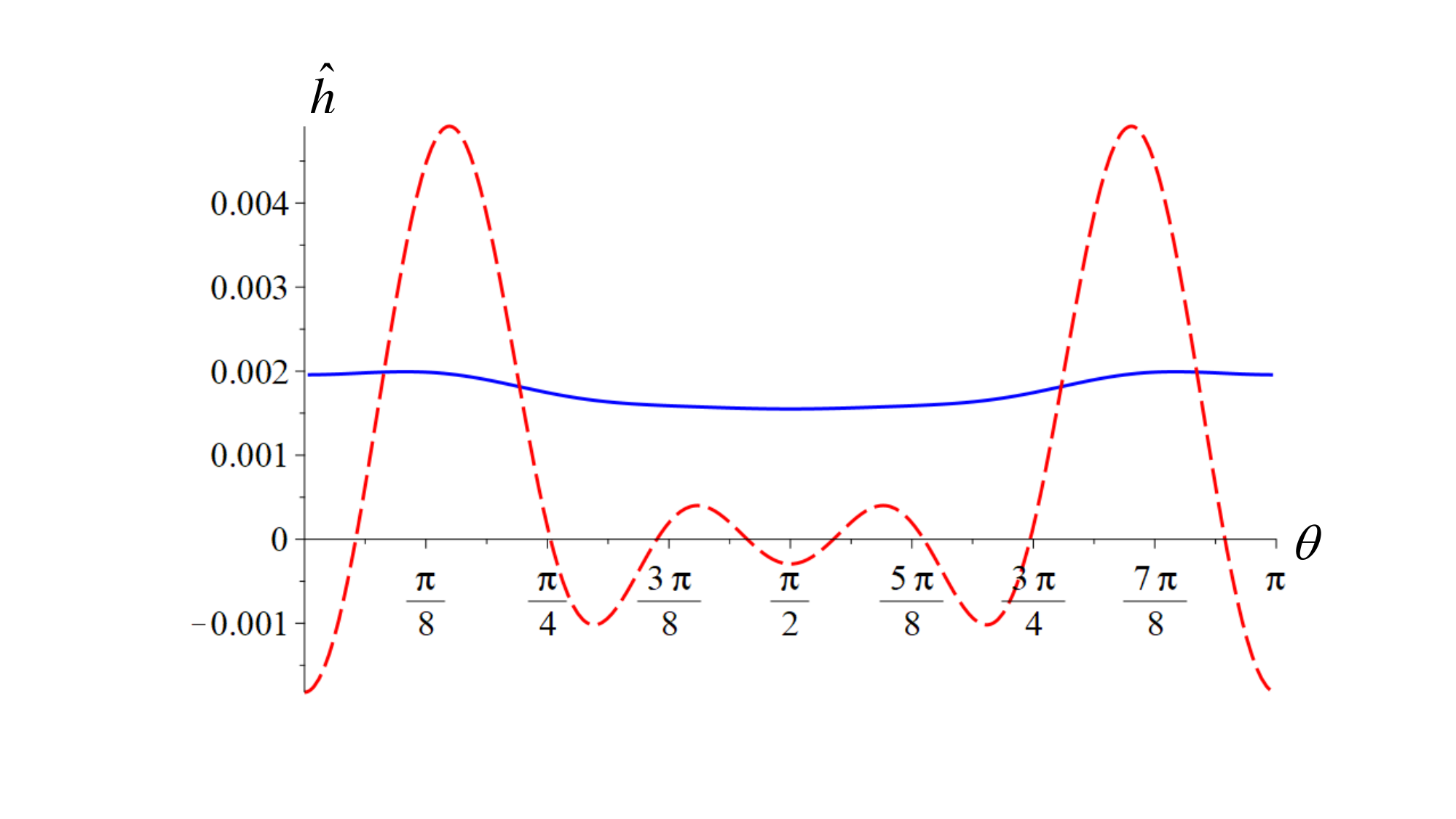}
 \vspace{-0.8cm}
   \caption{\label{F8} Plots of $\hat{h}$ (solid line) and $\hat{h}_V(\theta)$ (dash line) as a function of the angle $\theta$ for parameters $\mu=4$ and $\hat{b}=0.3$ ($a=0.9$). }
\end{figure}

\begin{figure}[!hbt]
    \centering
      \includegraphics[width=0.7\textwidth]{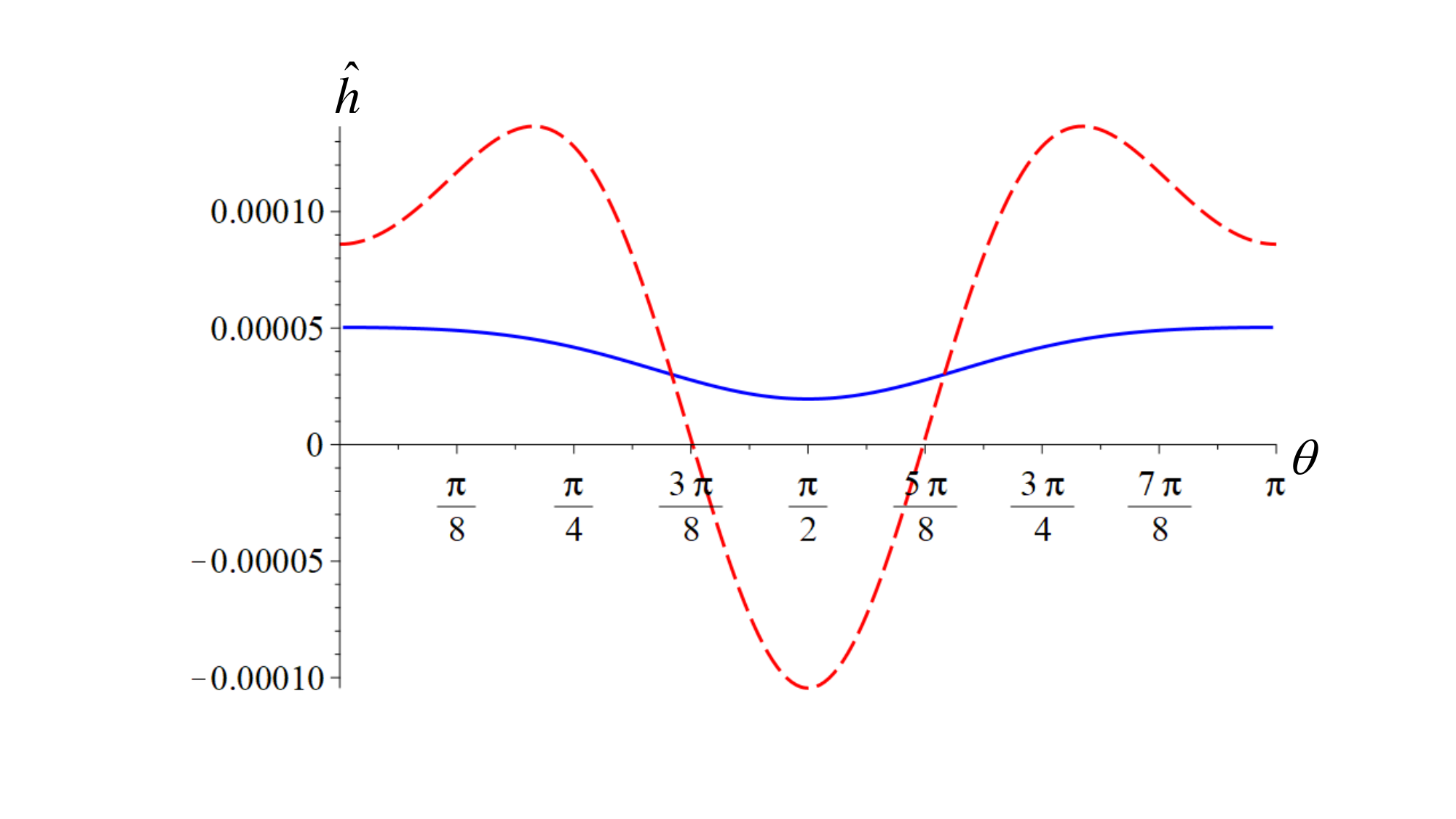}
\vspace{-0.8cm}
   \caption{\label{F9} Plots of $\hat{h}$ (solid line) and $\hat{h}_V(\theta)$ (dash line) as a function of the angle $\theta$ for parameters $\mu=3$ and $\hat{b}=0.9$ ($a=0.44$). }
\end{figure}

At this point, we may choose $\Psi$ to take the form \eqref{Psi_eq}. This means that the function $F(x)$ in equation \eqref{ode_shift_dimless} takes the form 
\begin{equation}
    F(x) = (\alpha + \beta x^2)f(x) \, ,
\end{equation}
where $x=\cos\theta$ and $f(x)$ is given by \eqref{eq_f_psinl}. In order to find a regular solution with the boundary condition \eqref{bdry_anglecoords} we use a specially designed numerical solver\footnote{
This code using a pseudo-spectral method and Gauss collocation grid, was developed and provided to us by Professor A. Frolov, to whom we are grateful.
}
In figures \ref{F7}, \ref{F8} and \ref{F9} we show the plots of $\hat{h}(\theta)$ and $h_V(\theta)$ for some selected values of the parameters $\mu$ and $\hat{b}$.

\section{Conclusions}

In this Chapter, we have constructed a model of a regular rotating black hole. The starting point was the Kerr-Schild form of the metric, which we modified by altering the potential that enters this representation. Since this potential satisfies the Laplace equation for a source located in complex spacetime, we replaced the standard Laplacian operator $\Delta$ with its infinite-derivative counterpart, $f(\Delta)\Delta$. Specifically, we employed the form factor $f(\Delta) = \exp(-\ell^2 \Delta)$, as it allows for an explicit analytic expression for the potential $\Phi$.
A notable strength of this model is its regularity at the horizon. This contrasts with earlier approaches that involved linearizing the Kerr metric by expanding in powers of the mass parameter $m$ and retaining only the zeroth and first-order terms. Such approximations typically result in metrics that are singular and lack a proper horizon structure.
An interesting geometric feature of the Kerr-Schild form of the Kerr metric is the existence of a coordinate transformation that brings the metric to a form with a single non-zero off-diagonal component, $g_{t\phi}$. This simplification does not apply to the model developed here, marking a significant departure from other models found in the literature \cite{Buoninfante_2018Ring,Torres2022,Gurses:1975vu,Babichev:2020qpr,Zhou:2023lwc,Baines2023}. Furthermore, in contrast to the local Kerr metric, where the horizon coincides with the surface $V = 0$, our model exhibits a deviation from this behavior. The correction terms that characterize this deviation were derived and computed numerically.
It is also important to emphasize that in the limit $\ell \to 0$, corresponding to vanishing nonlocality, the standard Kerr solution is recovered with the potential $\Phi = 2mr/(r^2 + y^2)$. All nonlocal corrections are governed by the dimensionless ratio $\ell/m$.
While this model is not an exact solution to the full nonlocal gravitational field equations, it is a promising approximation. We may hope that it captures key features of the as-yet unknown exact solution describing a rotating black hole in an infinite derivative theory of gravity.

	\chapter{Regular Black Hole Models From Higher Curvature Modifications}
\label{Ch5}

\section{Introduction}

As discussed in the Introduction, another possible route to modifying Einstein's theory of gravity is to include higher-order curvature terms in the gravitational action. In general, such modifications lead to field equations containing derivatives of the metric higher than second order. Consider the following action:
\begin{equation}
W = \dfrac{1}{\kappa} \int d^D x \sqrt{-g} , L(g^{\mu\nu}, R_{\alpha\beta\gamma\delta}), .
\end{equation}
Varying this action with respect to the metric yields the field equations:
\begin{equation}
\begin{split}
P^{\alpha\beta\gamma\delta} R_{\lambda\beta\gamma\delta} &- \dfrac{1}{2} \delta_{\lambda}^{\alpha} L - 2 \nabla^{\beta} \nabla^{\gamma} P^{\alpha}_{\ \ \beta\gamma\lambda} = 0   \, , \\
&P^{\alpha\beta\gamma\delta} = \dfrac{\partial L}{\partial R^{\alpha\beta\gamma\delta}} \, .
\end{split}
\end{equation}
It is evident that the last term on the left-hand side involves fourth-order derivatives of the metric. There exists a notable class of nonlinear curvature models in which these higher-derivative terms vanish. This occurs when the following condition is satisfied:
\begin{equation} \label{PPP}
\nabla^{\gamma} P^{\alpha\beta\gamma\delta} = 0 , .
\end{equation}
Gravitational theories that satisfy this condition are known as Lovelock gravity, a well-studied higher-dimensional generalization of Einstein’s theory of gravity. It is built from special combinations of curvature tensors, called Lovelock terms, that extend the Einstein–Hilbert action while still producing field equations containing only second derivatives of the metric. In four dimensions it reduces to General Relativity (plus a cosmological constant), but in higher dimensions it includes extra curvature contributions such as the Gauss–Bonnet term (see, e.g., \cite{lovelock1971,PADMANABHAN2013115,Charmousis2009,Pavluchenko2024,Nikolaev2025,Brustein2013}).

More recently, a new class of models has emerged in which a weaker version of condition \eqref{PPP} is imposed, specifically, that it holds only for static, spherically symmetric spacetimes. These theories are referred to as quasitopological gravity, and they have been formulated primarily in spacetime dimensions $D \ge 5$.

A particularly remarkable feature of these models is their admission of many regular black hole solutions, including well-known examples such as the Hayward metric. In this Chapter, following the results of \cite{rbh_pfkz}, we investigate regular black hole solutions arising from gravitational theories with nonlinear curvature terms. We begin by reviewing the properties of Hayward black holes and then proceed to describe the method of spherical reduction, which simplifies the problem by focusing on spherically symmetric geometries.

To illustrate the utility of this reduction procedure, we begin by showing how the Schwarzschild–Tangherlini solution arises from the reduced Einstein–Hilbert action. We then examine spherically symmetric solutions in Einstein–Born–Infeld theory, and finally explore how the reduced Einstein-Hilbert action is modified within the framework of quasitopological gravity. In fact, the resulting reduced action can be formally applied in four-dimensional spacetimes as well. In the final sections, we present a further generalization of the quasitopological reduced action that incorporates additional curvature invariants.

The principal outcome of this Chapter is the identification of a broad new class of spherically symmetric, regular black hole solutions that obey Markov’s limiting curvature condition, thereby eliminating singularities.

\subsection{The Hayward Metric and the Limiting Curvature Conjecture }\label{sec:Hayward}

The concept of limiting curvature has a long history, initially proposed by Markov \cite{limcur_markov}. The idea is the existence of a limiting density for all forms of matter, such that there exists a density at which the differences in the microscopic physical properties between various types of matter disappear. This could also be conjectured as the existence of some length scale parameter $\ell$ which regulates the curvature, and it appears, for example, in a black hole metric. This type of regulator is commonly included in regular black hole metric models, as already discussed in the Introduction. An important example is the Hayward 4D metric, given in equation \eqref{HAYWARD} and reproduced here for convinience \cite{Hayward}:
\begin{equation}\label{haymet}
\begin{split}
    ds^2 &= -f(r)dt^2 + f(r)^{-1}dr^2 + r^2 d\Omega^2\, ,\\
    f(r) &= 1 - \frac{2mr^2}{r^3 + 2\ell^2m}\, ,\quad
    d\Omega^2=d\theta^2 + \sin^2\theta d\phi^2, \quad 
\end{split}
\end{equation}
Higher-dimensional and other generalizations of the Hayward metric are discussed in \cite{Frolov:2016pav}. At far distances $r\to\infty$ the metric \eqref{haymet} has the following asymptotic
\begin{equation}
    f = 1 - \frac{2m}{r} + \mathcal{O}(r^{-4}) \quad \text{as} \quad r \to \infty,
\end{equation}
and hence reproduces the Schwarzschild solution. Near the center at $r=0$, it approaches the deSitter metric
\begin{equation}
    f = 1 - \frac{r^2}{\ell^2} + \mathcal{O}(r^5) \quad \text{as} \quad r \to 0.
\end{equation}
One often says that it has a de Sitter-like core.

\begin{figure}[!hbt]
    \centering
    \includegraphics[width=0.7\textwidth]{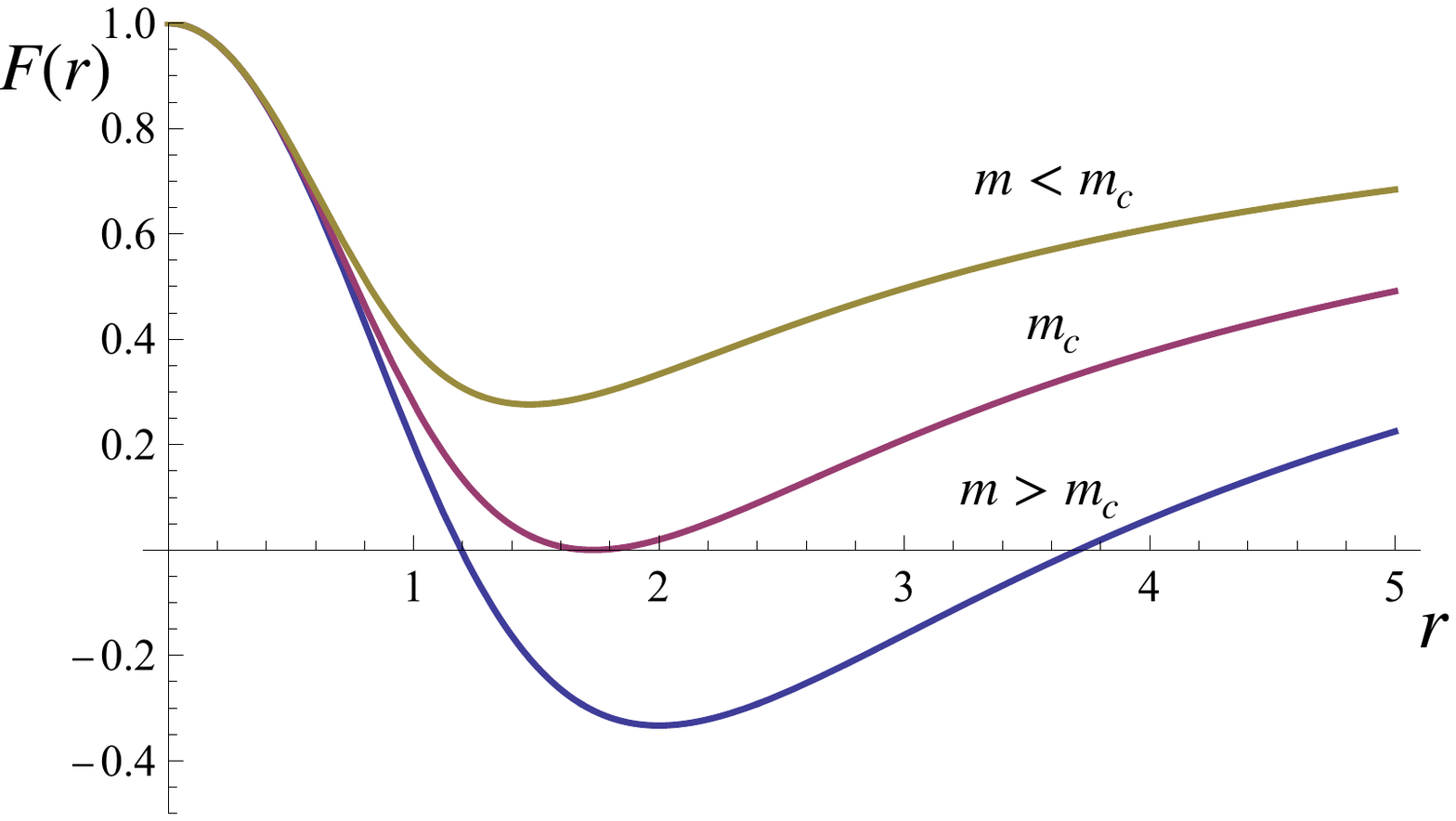}
    \vspace{-1cm}
    \caption{Metric function $F(r)$ of the Hayward solution for various masses with fixed $\ell = 1$. For masses below the critical value, no horizon forms. For larger masses, inner and outer horizons appear.}\label{PLTHAY}
\end{figure}

There also exists a critical mass $m_c = \frac{3\sqrt{3}}{4} \ell$ above which the spacetime has two horizons, and below which it does not have any horizon. This behavior is illustrated in Figure~\eqref{PLTHAY}.

For $m>m_c$ he outer horizon coincides with the event horizon. A new feature that distinguishes this metric from the Schwarzschild black hole is the presence of a second, inner horizon. A characteristic property of the Hayward and similar metrics describing regular black holes is that $f(r=0)=1$. It is easy to check that under this condition, the function $f(r)$ should have an even number of zeros $r=r_i$. In this sense, the Hayward metric, which has only two horizons, is rather simple.

Let us introduce the following coordinate
\begin{equation} 
y = r/A\, , \quad A = (2m\ell^2)^{1/3}\, .
\end{equation} 
Using this new coordinate, one can write the metric \eqref{haymet} in the form
\begin{equation} 
ds^2=A^2\big[
-\tilde{f}(y)d(t/A)^2+\dfrac{dy^2}{\tilde{f}}
+y^2 d\Omega^2\big]\, .
\end{equation}
Calculation of  the Ricci scalar $R$ and the curvature invariants $\mathcal{S}^2 = S_{\mu\nu}S^{\mu\nu}$ and $\mathcal{C}^2 = C_{\mu\nu\alpha\beta}C^{\mu\nu\alpha\beta}$, where $S_{\mu\nu} = R_{\mu\nu} - \frac{1}{4}g_{\mu\nu}R$, for this metric gives
\begin{equation}
    R = -\frac{6}{\ell^2} \frac{y^3 - 2}{(y^3 + 1)^3}, \quad \mathcal{S} = \frac{9}{\ell^2} \frac{y^3}{(y^3 + 1)^3}, \quad \mathcal{C} = \frac{\sqrt{12}}{\ell^2} \frac{y^3(y^3 - 2)}{(y^3 + 1)^3}.
\end{equation}
It can be seen that in terms of the coordinate $y$ these invariants are independent of the mass parameter $m$ and contain only rational functions of $y$ and the scale factor $\ell^{-2}$. The denominators in these expressions are always positive. The scalar invariants vanish at infinity and remain finite at $y=0$. At this point, one has $\mathcal{S} =\mathcal{C} =0$ and $R=12/\ell^2$, so that in this sense, the metric is regular. All the invariants are bounded by some value $C/\ell^2$, where $C$ is a dimensionless constant depending on the chosen invariant. This means that the metric \eqref{haymet} satisfies Markov's condition. 

The properties discussed above make the Hayward metric a particularly attractive candidate for modeling regular black holes. However, it was introduced in an ad hoc manner and does not arise as a solution to any known modified gravity field equations. It is worth noting that numerous other models exhibiting similar features have been proposed in the literature (see, e.g., \cite{Hayward,FrolovMetric,Carballo-Rubio_2020,carballo:geodesicallycompleteBHs,Fragomeno:2024tlh,nonsingularparadigmblackhole,Dymnikova:1992ux,Bardeen1968,PRD:Borde,Mars:1996khm,PRD:Lemos,Simpson_2019,ansoldi2008,visser:RQBH}).

However, ad hoc metric constructions face several challenges:
\begin{itemize}
    \item Some models require finely tuned or unphysical initial conditions \cite{Horowitz1995}
    \item Describing black hole interiors may necessitate alternative formulations \cite{Ashtekar_2006}
    \item Some of these models of regular black holes are coupled to exotic matter
\cite{Santos2024,BAMBI2013329,Rodrigues_2018,OVALLE2023138085}
\end{itemize}

Recently, however, the Hayward metric was derived for $D\ge 5$ spacetime dimensions using the so-called Quasitopological gravity action \cite{QT_BH}, along with several other spherically symmetric black hole metrics. This becomes feasible when an infinite number of higher-order curvature terms are considered. Such an approach is particularly interesting because it connects the idea of a universal bound on curvature to an extension of General Relativity in which additional terms are coupled via a fundamental length scale, $\ell$.

\section{Spherical Reduction}

Finding solutions in a theory with an infinite number of curvature power terms is, of course, a priori a very difficult—if not impossible—task. However, this is where the restriction to spherically symmetric solutions becomes especially useful: one can employ symmetry principles to reduce the full gravitational field equations to a much simpler form, as we will see in this section.

\subsection{Symmetry Reduction}

In four-dimensional spacetime, the metric contains 10 components, each a function of four coordinates. This complexity, combined with the nonlinearity of the Einstein equations, makes finding a general solution virtually impossible. Typically, one simplifies the problem by imposing restrictions on the form of the solution. A common approach is to consider solutions that exhibit certain symmetries.

There are two main ways to incorporate symmetry when searching for solutions to the Einstein equations. One can first derive the Einstein equations from the Einstein-Hilbert action and then impose the symmetry ansatz on the resulting equations. Alternatively, one can directly substitute the ansatz into the action and obtain a reduced set of equations for the metric by varying this reduced action.

\begin{figure}[!hbt]
    \centering
    \includegraphics[width=0.7\textwidth]{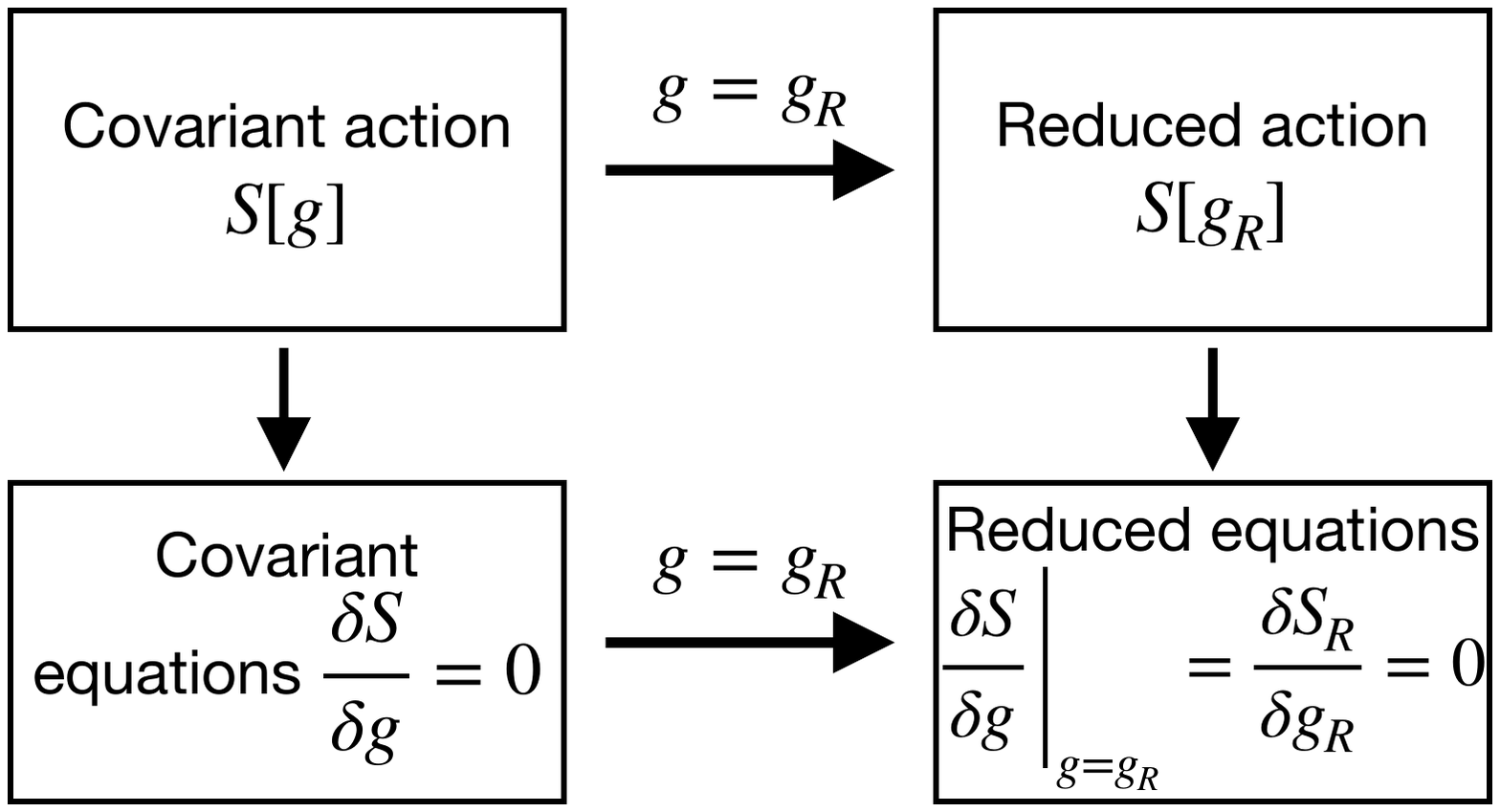}
    \vspace{-1cm}
    \caption{Schematic representation of the symmetry reduction procedure for a covariant action. Here we use the notation $g$ to denote an arbitrary metric, while  $g_R$ represents an ansatz displaying some type of symmetry}\label{SR}
\end{figure}

It is important to note that these two procedures do not necessarily yield the same results. The reason is straightforward: in the second approach, the variation is performed over a reduced set of variables, and there is no guarantee that this is equivalent to varying the full action with respect to a larger set of parameters. The challenge of finding solutions via a symmetry-reduced action is not unique to gravity, as it arises in general Lagrangian field theories. While the correspondence between solutions of the reduced and full Lagrangians is not always guaranteed, it does hold under certain symmetry conditions, as stated by Palais’ Principle of Symmetric Criticality (PSC) \cite{PSC}. This principle asserts that “for any group-invariant Lagrangian, the equations obtained by restriction of the Euler-Lagrange equations to group-invariant fields are equivalent to the Euler-Lagrange equations of a canonically defined, symmetry-reduced Lagrangian” \cite{PSC-GR}. In particular, the validity of PSC has been demonstrated for the spherical reduction of Einstein gravity (see e.g. \cite{PSC-GIS,PSC-GVTY,rbh_pfkz}).

\subsection{Spherical Symmetry in Einstein Gravity}

The spherical reduction of the Einstein equations is discussed in the Appendix~\ref{app:SphReduction}. Here we only summarize the results which will be used later. 

First, let us note that there are several distinct``levels” at which such a dimensional reduction can be performed. At the first stage, one may consider a reduced metric of the form
\begin{equation} \label{DIL}
ds^2=g_{ij}dx^i dx^j +\rho^2(x^i) d\Omega^2_{D-2}\, ,
\end{equation} 
where $d\Omega^2_{D-2}$ is the metric on $(D-2)$ dimensional round sphere. The reduced action for this metric depends on $g_{ij}(x^i)$ and $\rho(x^i)$, and it is, in fact, the action for a 2D dilaton gravity.

At the next``level" one introduces an additional restriction by requiring that the metric is static and uses the following ansatz for the metric
\begin{equation}\label{sssastz}
    ds^2 = -N(r)^2f(r)dt^2 + f(r)^{-1}dr^2 + \rho(r)^2d\Omega^2 \, .
\end{equation}
Where the $N$, $f$ and $\rho$ are arbitrary functions of the coordinate $r$. 

If $d\rho/dr>0$, one can impose an additional (gauge) condition by choosing $\rho=r$, so that the metric takes the form
\begin{equation}\label{STAT}
    ds^2 = -N(r)^2f(r)dt^2 + f(r)^{-1}dr^2 + r^2d\Omega^2 \, .
\end{equation}
The corresponding reduced action is now a functional that depends on the two functions $N(r)$ and $f(r)$. 

According to the established Palais’ Principle of Symmetric Criticality, the reduced actions at each ``level" of spherical reduction are sufficient to obtain the corresponding solutions. This sequence of reductions can be described as a top-down approach. A natural question is whether the inverse, or "bottom-up," procedure is also valid. Specifically, does there exist a covariant action that, upon spherical reduction, yields a chosen reduced action? This remains an interesting open question.

The reduced Einstein-Hilbert action for the static spherically symmetric \eqref{STAT} is (see Appendix~\ref{app:SphReduction}) 
 \begin{equation}
     S = B \int dt dr N'r^{D-1}p \, ,
 \end{equation}
 where $B=  (D-2)\Omega_{D-2}/2\mathcal{\kappa} $ and remembering that the prime symbol denotes a derivative with respect to $r$. Varying this action yields the following equations
 \begin{equation}
     N' = 0 \quad \mathrm{and} \quad [r^{D-3}(1-f)]'=0 \, .
 \end{equation}
 Where the equation on the left corresponds to a variation over $f(r)$, while the equation on the right corresponds to a variation over $N(r)$. The first of these equations shows that $N=\mathrm{const}$. By rescaling time, $t$ can be set equal to $1$.
For this choice, the second equation can be easily integrated and gives
\begin{equation}\label{schwtangh}
f=1-\dfrac{\mu}{r^{D-3}}\, , \quad  \mu = m/B.
\end{equation} 
Hence, the corresponding metric \eqref{STAT} is nothing but the Schwarzschild-Tangherlini solution of the higher-dimensional Einstein equations.

\subsection{Spherically Reduced Action for Quasitopological Gravity}

Next, we discuss the main results of \cite{QT_BH}, in which the authors used these tools to obtain the radial static spherically reduced action for the Quasitopological gravity models, which takes the form 
\begin{equation}\label{ASQT}
    S_{\mathrm{SQT}} = -B\int dtdr r^{D-1} {N}' h(p) \, ,
\end{equation} 
where $h(p)$ is a function of the scalar invariant $p$ 
\begin{equation} \label{pbasic}
p=\dfrac{1-f}{r^2} \, .
\end{equation} 

In order to obtain this action, one considers specially constructed scalar curvature invariants of the power $n$ constructed so that their variation over spherically symmetric spacetime does not contain higher than second order derivatives \cite{Hennigar:2017ego,Bueno:2020gq, Bueno:2022res}. In general, the function $h(p)$ can be written in the form of a series
\begin{equation} \label{hhha}
h(p)=\sum_{n=1}^{\infty}\alpha_np^n\, ,
\end{equation} 
where  $p^n$ stands for the term of the covariant action corresponding to curvature invariants of power $n$. It was shown that the ambiguity in specifying these contributions involves an arbitrary constant $\alpha_n$.  As we shall see later, the choice $\alpha_n = \ell^{2(n-1)}$ corresponds to solutions that take the form of the Hayward metric. In this case, equation \eqref{hhha} takes the form 
\begin{equation}\label{hhhh}
    h(p) = \frac{1}{\ell^2}\sum_{n=1}^{\infty} (\ell^2 p)^n \, .
\end{equation} 
The length parameter $\ell$, which enters \eqref{hhhh}, is required since both functions $p$ and $h(p)$ have dimensions of $length^{-2}$. Let us stress that all these results were obtained in spacetimes of dimensions five and higher.

It is easy to see that the variation of the reduced action \eqref{ASQT} with respect to $f(r)$ gives the equation $N'=0$. This allows one to set $N=1$ after variations. The equation obtained by the variation over $N(r)$ gives
\begin{equation} 
(r^{D-1}h(p))'=0\, .
\end{equation} 
The solution of this equation is
\begin{equation} \label{hhpp}
h(p)=\dfrac{\mu}{r^{D-1}}\, ,
\end{equation} 
where $\mu$ is the constant of integration, defined by \eqref{schwtangh}. For a chosen function $h(p)$, equation \eqref{hhpp} is an algebraic equation which allows one to find $p=p(r)$, and hence to reconstruct the corresponding metric function $f$
\begin{equation} \label{ffpp}
f=1-r^2 p(r)\, .
\end{equation} 

Using this method, one can obtain the desired class of solutions. Black hole solutions are distinguished by the requirement that the function $f(r)$ contains at least one zero. An important result obtained in \cite{QT_BH} is that if the number of non-vanishing terms in the series \eqref{hhhh} is finite, all corresponding black hole solutions exhibit a singularity at $r = 0$. In contrast, if the series \eqref{hhhh} contains an infinite number of terms, it becomes possible to construct regular black hole solutions.

In particular, equation \eqref{hhhh} can be written as
\begin{equation} 
h(p)=\dfrac{p}{1-\ell^2 p}\, .
\end{equation} 
Then, we can solve the equation \eqref{hhpp}. One gets
\begin{equation} \label{ppsi}
p=\dfrac{\psi}{1+\ell^2 \psi}\, .
\end{equation} 
Using this relation, together with \eqref{ffpp} one finds
\begin{equation}\label{DDHA}
f=1-\dfrac{\mu r^2}{r^{D-1}+\ell^2 \mu}\, .
\end{equation} 
This solution is the $D$-dimensional version of the Hayward metric.

\subsection{Nonlinear Electrodynamics Interacting With Einstein Gravity: Spherical Reduction}

In order to further illustrate the power of the spherical reduction approach, we finally consider spherically symmetric solutions of the Einstein equations coupled to the equations of a nonlinear electrodynamics model.

The Einstein-Hilbert action, combined with the action for such a nonlinear electrodynamics model, takes the form
\begin{equation}\label{EHA}
\begin{split}
S[\,g,\,A]&=S_A+S_g\, ,\\
S_A=&- \int d^D x \, \sqrt{-g} P(\mathcal{F})\, \\
S_g&=\frac{1}{2\varkappa}\,\int d^D x \,\sqrt{-g}\, R\, ,\\
\mathcal{F}&=-\frac{1}{2}F_{\mu\nu}F^{\mu\nu} \, ,\quad 
F_{\mu\nu}=A_{\mu;\nu}-A_{\nu;\mu}\, .
\end{split}
\end{equation}
We keep the number $D$ of spacetime dimensions arbitrary.
Here $P$ is an arbitrary function, which for small $\mathcal{F}$ has the following asymptotic form
\begin{equation}\label{PPPP}
P(\mathcal{F})=\frac{1}{2}\mathcal{F}+...\, .
\end{equation}
We do not include the other quadratic in the field strength invariants since they vanish for a static spherically symmetric field $F_{\mu\nu}$. The condition \eqref{PPPP} guarantees that in the weak field limit, the non-linear electrodynamics model reduces to the standard Maxwell theory.

A static spherically symmetric ansatz for the fields in $D$-dimensional spacetime is
\begin{equation}
\begin{split}
ds^2&=-N^2 f dt^2+\dfrac{dr^2}{f}+r^2 d\Omega_{D-2}^2\, ,\\
A_{\mu}&=\varphi \delta_{\mu}^t\, ,
\end{split}
\end{equation}
where $N$, $f$ and $\varphi$ are arbitrary functions of $r$. One has
\begin{equation}
\mathcal{F}=\frac{(\varphi')^2}{N^2}\, .
\end{equation}
Using these relations, one obtains the following expression for the spherically reduced action 
\begin{equation}
S_{SR}[N,f,\varphi]=B\int dt dr \big[ N'r^{D-1}p+ 2 N \varkappa r^{D-2}P(\Phi^2/N^2)\big]\, .
\end{equation}
Here $\Phi=\varphi'$. The variation with respect to all 3 variables of the reduced action gives the following set of reduced field equations 
\begin{equation}\label{SEQ}
    \begin{split}
&N'=0\, ,\\
&(r^{D-1}p)'+ 2\varkappa r^{D-2}P - 4\varkappa r^{D-2}\frac{\Phi^2}{N^2}\frac{dP}{d\mathcal{F}}  = 0\, \\
&\big(2\varkappa r^{D-2}\dfrac{\Phi}{N} \dfrac{dP}{d\mathcal{F}}\big)'=0\, .
\end{split}
\end{equation}
Using the ambiguity in the choice of $t$ we set $N=1$. For this choice, the integration of the last equation gives
\begin{equation}\label{eq_ED3}
2r^{D-2}\Phi \dfrac{dP}{d\mathcal{F}}=Q\, .
\end{equation}
Where $Q$ is an integration constant. This equation can be used to find $\Phi$ as a function of $r$. For $N=1$ and using \eqref{eq_ED3}, the second equation in \eqref{SEQ} becomes
\begin{equation} \label{rrff}
(r^{D-1}p)'+2\varkappa r^{D-2}P - 2\varkappa Q\Phi = 0 \, .
\end{equation} 
Substituting the obtained solution for $\Phi(r)$ into \eqref{rrff} and integrating the obtained equation, one can find $p=p(r)$. This allows one to obtain $f(r)$ by using \eqref{ffpp}. Let us consider the Born–Infeld model, for which
\begin{equation}
P(\mathcal{F}) = b^2\left(1 - \sqrt{1 - \frac{\mathcal{F}}{b^2}} \right)
= b^2\left(1 - \sqrt{1 - \frac{\Phi^2}{b^2}} \right) \, .
\end{equation}
In our gauge choice, $\mathcal{F} = \Phi^2$. Taking the derivative, we find
\begin{equation}
\frac{dP}{d\mathcal{F}} = \frac{1}{2} \left( 1 - \frac{\Phi^2}{b^2} \right)^{-1/2} \, .
\end{equation}
Substituting this into \eqref{eq_ED3} allows us to solve for $\Phi$, which gives
\begin{equation}\label{eq_PhiBorn}
\Phi = \frac{Q}{\sqrt{r^{2D-4} + \frac{Q^2}{b^2}}} \, .
\end{equation}
Using this result, the function $P$ can be written as
\begin{equation}\label{eq_PBorn}
P = b^2 \left(1-\frac{r^{D-2}}{\sqrt{r^{2D-4} + Q^2/b^2}}\right) \, .
\end{equation}
Equations \eqref{eq_PhiBorn} and \eqref{eq_PBorn} may then be substituted into \eqref{rrff}, yielding
\begin{equation}
(r^{D-1}p)' + 2\varkappa b^2 \left(r^{D-2} - \sqrt{r^{2D-4} + Q^2/b^2}\right) =0 \, .
\end{equation}
This equation can be solved for $p$, which in turn determines the metric function $f$ via \eqref{ffpp},
\begin{equation}
\begin{split}\label{metric_EBI}
f &= 1 - \frac{\mu}{r^{D-3}} + \frac{2b^2\varkappa}{D-1}\bigg[r^2 - \frac{\sqrt{r^{2D-4} + Q^2/b^2}}{r^{D-4}} + \\ 
& \frac{D-2}{D-3} \frac{Q^2}{b^2}\frac{1}{r^{2D-6}} {}_2F_1 \bigg( \frac{1}{2}, \frac{D-3}{2D-4};\frac{7-3D}{4-2D};-\frac{Q^2}{b^2}\frac{1}{r^{2D-4}} \bigg) \bigg] \, .
\end{split}
\end{equation}
where ${}_2F_1$ denotes the Gauss hypergeometric function \cite{NIST2010}.

The expression above reduces to the Reissner–Nordström solution in the limit $b \to \infty$ and $D=4$, which is the same limit in which Maxwell’s theory is recovered from the Born–Infeld model:
\begin{equation}
f|_{b\to \infty} = 1 - \frac{\mu}{r} + \frac{\kappa Q^2}{r^2} + O(1/b^2) \, .
\end{equation}

The asymptotic behavior of the solution for $r\to\infty$ in $D=4$ is given by
\begin{equation}
f(r \to \infty) = 1 - \frac{\mu}{r} + \frac{\kappa Q^2}{r^2} + O(1/r^4) \, .
\end{equation}

This expression coincides with the weakly nonlinear regime, where Born–Infeld corrections manifest only as small deviations from the Reissner–Nordström solution. These nonlinear effects become significant only in the high-curvature region near the origin.

\section{Generalization of the Reduced Quasitopological Gravity Action}\label{sec2Ddil}

We now turn to other regular black hole solutions and their generalizations, using those arising from quasitopological gravity as a starting point. Our goal is to extend these solutions in a broader context, exploring modifications that preserve regularity while incorporating additional physical features. Before proceeding, we first define the key geometric quantities that will be used throughout this construction.

\subsection{Basic Curvature Invariants}

We will begin by discussing the properties of static spherically symmetric metrics in more detail. As mentioned in the Appendix~\ref{app:SphReduction}, there exist four independent curvature invariants for such metrics, which we denote by $p,q,u,v$. The non-zero components of the $D$-dimensional Riemann curvature tensor have the form
\begin{equation}\label{RRRR}
\begin{split}
    & R_{jl}^{\ \ ik} = v\delta^{ik}_{jl} \, , \quad  R_{ta}^{\ \ tb} = u\delta^{b}_{a} \, , \\
    &  R_{ra}^{\ \ rb} = q\delta^{b}_{a} \, , \quad   R_{ab}^{\ \ cd} = p\delta^{cd}_{ab} \, .
\end{split}
\end{equation}
where $i,j=0,1$ and $a,b=2,\dots,(D-1)$. For the metric \eqref{STAT}, these quantities have the following form
\begin{equation}\label{pquv0}
\begin{split}
    & p = \frac{1- f}{r^2}\, , \quad v = -\frac{1}{2}\left(f''+ 3f'\frac{N'}{N} + 2f\frac{N''}{N} \right) \, , \\
    & q = - \frac{f'}{2r}  \, , \quad u = -\left( \frac{f'}{2r} + f\frac{N'}{r N}\right) \, .
\end{split}
\end{equation}
Here and later on, we denote by a prime the derivative with respect to $r$. We call functions $p$, $q$, $u$ and $v$ basic curvature invariants. Other local scalar invariants constructed from the curvature can be expressed as a function of these basic curvature invariants \footnote{ As shown by Narlikar and Karmarkar \cite{NarlikarKarmarkar1949}, for spherically symmetric metrics that depend on both spatial and temporal coordinates, a set of four invariants is sufficient to generate all other algebraic curvature invariants. The metric \eqref{STAT} falls within a broader class of so-called warped product metrics. A detailed discussion of the complete set of curvature invariants sufficient for constructing all algebraic invariants of the Riemann tensor, along with further references, can be found in \cite{SANTOSUOSSO1998381}.}.
For example, the Ricci scalar has the form
\begin{equation}\label{R}
R= \big[v+(D-2)q\big]+\big[v+(D-2)u\big] +(D-2)\big[(D-3)p+q+u\big] .
\end{equation}
Expressions for the other scalar invariants are given in Appendix~\ref{app:SphReduction}.

One may observe that the parameter $p$, which appears in the reduced action of the quasitopological theory \eqref{pbasic}, corresponds to the first of the basic curvature invariants characterizing a static spherically symmetric spacetime. This observation suggests a broader generalization: the action can be extended by allowing the function $h$ to depend on all basic curvature invariants of the spacetime. In this way, we consider the following form of the reduced action:
\begin{equation}\label{GEN}
    S = -B\int dr dt {N}' r^{D-1} h(p,q,u,v) \, .
\end{equation} 

\subsection{Covariant Action From a Bottom-Up Approach}

A natural question is: can one reconstruct a covariant $D$-dimensional action whose spherical reduction yields the action \eqref{GEN}? The answer is yes, provided that the function $h$ depends solely on the invariant $p$. In other words, this reconstruction is possible if one begins with the reduced action derived from quasitopological gravity, as given in \eqref{ASQT}. To demonstrate this, one starts from the action \eqref{ASQT}, performs an integration by parts, and discards total derivative (surface) terms, which do not affect the equations of motion. Following this procedure, the reduced action takes the form:
\begin{equation}\label{ACT1}
S=B\int dt dr\, N\big[r^{D-1}h( p)\big]'  .
\end{equation}
The function $N$ can be expressed in terms of the $D$-dimensional metric density as $\sqrt{-g}=N r^{D-2}\sqrt{\hat{g}}$, where $\hat{g}=\det(\hat{g}_{ab})$ corresponds to the metric $d\Omega_{D-2}^2$ of the $(D-2)$-dimensional unit sphere, given in \eqref{haymet}.

First, let us consider  the following combination
\begin{equation}
 \frac{1}{r^{D-2}}\frac{\partial}{\partial r}\big[r^{D-1}h( p)\big]=(D-1)h + r\frac{\partial}{\partial r}h( p) =(D-1)h +r p'\frac{\partial}{\partial p}h( p)  .
 \end{equation}
Using (\ref{pquv0}) one obtains the identity
\begin{equation}
rp'=2(q-p) \, .
\end{equation}
Then one can see that
\begin{align}\nonumber
\sqrt{-g}\Big[(D-1)h +2(q-p)\frac{\partial h( p)}{\partial p}\Big]
=\sqrt{\hat{g}}N\big(r^{D-1}h( p)\big)' .
\end{align}
The left-hand side of this expression has a completely $D$-dimensional covariant form, and can be written explicitly in terms of curvature invariants $R$, $C_{\alpha\beta\mu\nu} C^{\alpha\beta\mu\nu}$, $R_{\mu}^{\nu} R_{\nu}^{\mu}$, and $R_{\mu}^{\nu} R_{\nu}^{\sigma}R_{\sigma}^{\mu}$ using their explicit equations, which can be found in Appendix~\ref{app:SphReduction}.

It means that for any given function $h(p)$, the covariant action
\begin{align}\label{EHAcov}
S[\,g\,]=&\frac{1}{2\varkappa}\,\int d^D x \,\sqrt{-g}\Big[(D-1)h+2(q-p)\frac{\partial}{\partial p}h \Big] ,
\end{align}
after spherical reduction reproduces the reduced action \eqref{ASQT} from quasitopological gravity.

In more general cases, e.g., with $h(p,q)$, the problem of reconstructing a covariant action becomes significantly more challenging. At present, we cannot guarantee the existence of a covariant action that reproduces our ansatz \eqref{GEN} upon spherical reduction. This remains an interesting open problem and merits further study.

 Furthermore, let us emphasize that one can rewrite the adopted ansatz for the metric in terms of the curvature invariants scalars. However, if one use such a covariant action for getting equation beyond spherical symmetry ansatz, in a general case one obtains fourth order in derivatives equations which might have pathological solutions.
 
\subsection{Generalized Action Equations}

The main focus of this Chapter is to construct regular black hole metrics; to that end, we will use the action \eqref{GEN} to extract solutions. Let us begin by introducing the shorthand notation
\begin{equation} 
W=N'r^{D-1}\, .
\end{equation}
It is safe to assume that $h$ depends only on $f$ and its derivatives up to second order, as long as there are no derivatives of the basic curvature invariants $q,v,u$ in the argument of the function $h$. Then the variation of the action \eqref{GEN} over $f$ yields
 \begin{equation}\label{EOMS2DF}
     \frac{\partial h}{\partial f}W - \left( \frac{\partial h}{\partial f'} W\right)' + \left( \frac{\partial h}{\partial f''}W\right)'' = 0 \, ,
 \end{equation}
Under the same assumption, one can safely neglect higher-than-second derivatives of the function $N$ in the action. Varying the action with respect to $N$, we obtain
\begin{equation}\label{EOMS2DH}
    (r^{D-1}h)' - \frac{\partial h}{\partial N}W + \left( \frac{\partial h}{\partial N'} W\right)' - \left( \frac{\partial h}{\partial N''}W \right)'' = 0 \, .
\end{equation}
Both equations \eqref{EOMS2DF} and \eqref{EOMS2DH} require appropriate boundary conditions. However, these equations are generally nonlinear; as a consequence, the proper choice of such conditions is highly non-trivial. Instead, we will opt for fixing some of the functions of our model, namely, notice that equation \eqref{EOMS2DF} always has the solution $W=0$, which leads to
\begin{equation}
   N' = 0  \, .
\end{equation}
By analogy with the standard gauge choices in GR and quasitopological gravity, we set $N=1$. This choice reduces \eqref{EOMS2DH} to 
\begin{equation}
    \frac{d}{dr}(r^{D-1}h) = 0 \, ,
\end{equation}
which can be readily integrated and obtain 
\begin{equation}\label{chareq_h}
    h = \frac{\tilde{\mu}}{r^{D-1}} \, ,
\end{equation}
where $\tilde{\mu}$ is an integration constant. 

For the remainder of this Chapter, we will focus on solutions of this form. The constant $\tilde{\mu}$ dominates at the asymptotic limit of the solution at large distance, and it is proportional to the mass parameter $\mu$ of the solution $\tilde{\mu} = Z\mu$. The coefficient $Z$ will depend on the parameters of the theory. The mass parameter $\mu$ is related to the mass of the black hole $m$ by
\begin{equation}
    \mu = \frac{2\mathcal{\varkappa}m}{(D-2)\Omega_{D-2}} = \frac{m}{B} \, .
\end{equation}
Where $\mathcal{\varkappa}$ is the $D$-dimensional gravitational coupling (see Appendix~\ref{app:SphReduction}). Then, variations over the action \eqref{GEN} lead to the following equations 
\begin{equation}
    N=1, \quad h = \frac{\tilde{\mu}}{r^{D-1}} \, .
\end{equation}

\subsection{Scaling Property}
There is an interesting property of this solution that we will discuss next. The basic curvature invariants $p$, $q$, $u$, and $v$ each have dimensions of $1/\text{[Length]}^2$. This follows directly from their definition in equation \eqref{RRRR} and their explicit forms given in equation \eqref{pquv0}. The action \eqref{ACT1}, when evaluated in the case $h = p$, reproduces the reduced Einstein–Hilbert action. Consequently, the function $h$ must also carry dimensions of $1/\text{[Length]}^2$. 

In the following, we consider a function $h$ that is non-linear in the basic curvature invariants \eqref{RRRR}. For instance, take $h = p + \beta p^2$. In order for this expression to be dimensionally consistent, the coefficient $\beta$ must have dimensions of $\text{[Length]}^2$. This implies that the general non-linear theory must involve a fundamental length scale, which we denote, again, by $\ell$.\footnote{If the theory contains multiple length parameters, we designate one of them as $\ell$. The ratios of the other lengths to $\ell$ then enter the model as dimensionless parameters.}

In quantum gravity, a natural choice for such a fundamental length is the Planck length $\ell_{\mathrm{Pl}}$. However, we do not assume $\ell = \ell_{\mathrm{Pl}}$ and instead keep $\ell$ arbitrary. In the non-linear case, $\ell$ serves as a characteristic scale at which non-linear effects become significant.
 
It will be useful to write the invariants in the radial gauge $\rho = r$ with $N=1$, since we can relate them as
\begin{equation}\label{BIRel}
\begin{split}
    &q = u =p +\frac{1}{2}r\frac{\partial}{\partial r}p \, , \\
    &v = p + 2 r \frac{\partial }{\partial r}p + \frac{1}{2} r^2 \frac{\partial^2}{\partial r^2} p \, .
\end{split}
\end{equation}
This means that the function $h$ as defined in equation \eqref{chareq_h} can be written in terms of $p$, $p'$ and $p''$
\begin{equation}\label{eqH}
    h = h(r,p,p',p'') = \frac{\tilde{\mu}}{r^{D-1}} \, ,
\end{equation}
which makes equation \eqref{eqH} an ordinary partial differential equation for $p$ whose solution is a function of r and the two parameters $\tilde{\mu}$ and $\ell$
\begin{equation}
    p = p(r,\tilde{\mu},\ell) \, .
\end{equation}
We may cast this solution in terms of some dimensionless function $P$. According to the Buckingham $\pi$ theorem, since our $D$-dimensional solution depends on $3$ physical variables, then
\begin{equation}
    p = \ell^{-2}P(r/\ell,\tilde{\mu}/\ell^{D-3})\, .
\end{equation}
Furthermore, we may cast the entire solution in terms of one dimensionless coordinate $y$ such that $p=p(y)$. This is possible thanks to the relations \eqref{BIRel}. In order to see this, we first define
\begin{equation}
    y = \frac{r^{D-1}}{\ell^2\tilde{\mu}} \, .
\end{equation}
Then equations \eqref{BIRel} become 
\begin{align}
    &q = u = p +\frac{D-1}{2} y \frac{\partial}{\partial y} p \, , \\
    & v = p + \frac{(D-1)(D+2)}{2}y \frac{\partial}{\partial y}p + \frac{(D-1)^2}{2} y^2 \frac{\partial}{\partial y^2} p \, .
\end{align}
And equation \eqref{eqH} may be multiplied on both sides by $\ell^2$ so that 
\begin{equation}\label{eqHY}
    \hat{h}(y,\hat{p},d\hat{p}/dy,d^2\hat{p}/dy^2) = \frac{1}{y} \, ,
\end{equation}
where 
\begin{equation}
    \hat{p} = \ell^2p , \quad \hat{h} = \ell^2 h\, .
\end{equation}
Which means that the solution for \eqref{eqHY} has solutions which depend only on the dimensionless coordinate $y$, as it depends on the curvature invariant $p$ and $y$ while remaining completely independent of the parameters $\ell$ and $\tilde{\mu}$. We call this the scaling property of our model and refer to equation \eqref{eqHY} as the master equation. It will be the topic of the following section to find physically sensible solutions for this equation. 

\section{Regular Black Hole Solutions}

\subsection{Master Equation}

In general, equation \eqref{eqHY} is some non-linear ODE. To simplify this equation, we impose a few conditions on the function $h$ 
\begin{itemize}
    \item $h$ is a function $h(\psi)$ of the linear combination of the basic curvature invariants $\psi = ap + bq + cv$
    \item The function has the form 
    \begin{equation}
        h(\psi) = \frac{\psi}{1-\ell^2\psi} \, .
    \end{equation}
    \item $a$ is positive
\end{itemize}
The second condition reduces equation \eqref{eqHY} to the rather simple form
\begin{equation}\label{eqC2}
    \ell^2\psi = \frac{1}{1+y} \, .
\end{equation}
The first condition guarantees that equation \eqref{eqC2} has a form which is linear in the basic curvature invariants
\begin{equation}\label{eqC1}
     ap + bq + cv = \frac{1}{\ell^2} \frac{1}{1+y} \, .
\end{equation}
The third condition is for convenience, as it lets us normalize everything as
\begin{equation}
    \hat{b} = b/a \, ,\quad \hat{c} = c/a\, , \quad \hat{\ell} = \sqrt{a}\ell\, ,\quad \hat{\mu} = \tilde{\mu} /a \, .
\end{equation}
So that the equation \eqref{eqC1} becomes
\begin{equation}\label{eqhatP}
    p + \hat{b}q + \hat{c}v = \frac{1}{\hat{\ell}^2}\frac{1}{1+y} \, .
\end{equation}
Notice that this normalization does not affect $y$ since
\begin{equation}
    y = \frac{r^{D-1}}{\ell^2 \mu} = \frac{r^{D-1}}{\hat{\ell}^2\hat{\mu}} \, .
\end{equation}
Then we can set $a=1$ for the remainder of the Chapter and study equation \eqref{eqhatP}. We will, as well, omit the hats on top of the variables for simplicity
\begin{equation}\
    p+bq+cv = \frac{1}{\ell^2}\frac{1}{1+y} \, .
\end{equation}
Finally, we may write the equation in a more compact notation 
\begin{equation}\label{MEQP}
    \mathcal{D}p = \frac{1}{\ell^2}\frac{1}{1+y} \, ,
\end{equation}
where we have defined 
\begin{equation}
    \mathcal{D} = Ay^2\frac{d^2}{dy^2} + By\frac{d}{dy} + C \, ,
\end{equation}
and
\begin{equation}
    A = \frac{1}{2}(D-1)^2c \, ,\quad B =\frac{1}{2}(D-1)(b + [D+2]c) \, , \quad C = 1+b+c \, .
\end{equation}
We will refer to equation \eqref{MEQP} as the master equation and look for its solutions. 

\subsection{Solutions of the Master Equation}

\begin{figure}[!hbt]
    \centering
\includegraphics[width=0.7\textwidth]{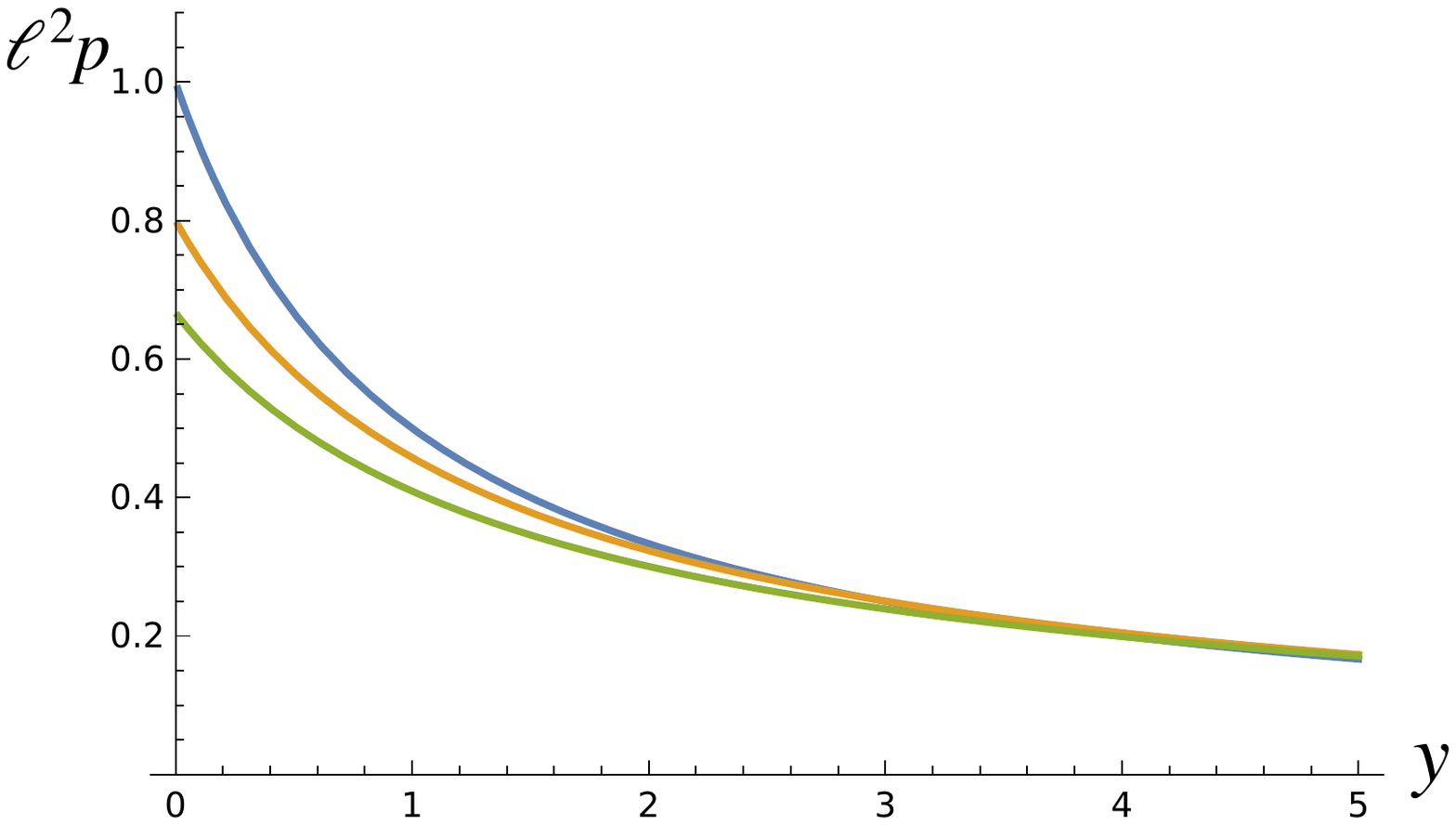}
\vspace{-1cm}
    \caption{Basic invariant $p$ in $D=4$ dimensions with $c=0$. The cases $b = 0$ (blue), $b=1/4$ (orange) and $b=1/2$ (green) are shown.} \label{PLTpD4}
\end{figure}

\begin{figure}[!hbt]
    \centering
\includegraphics[width=0.7\textwidth]{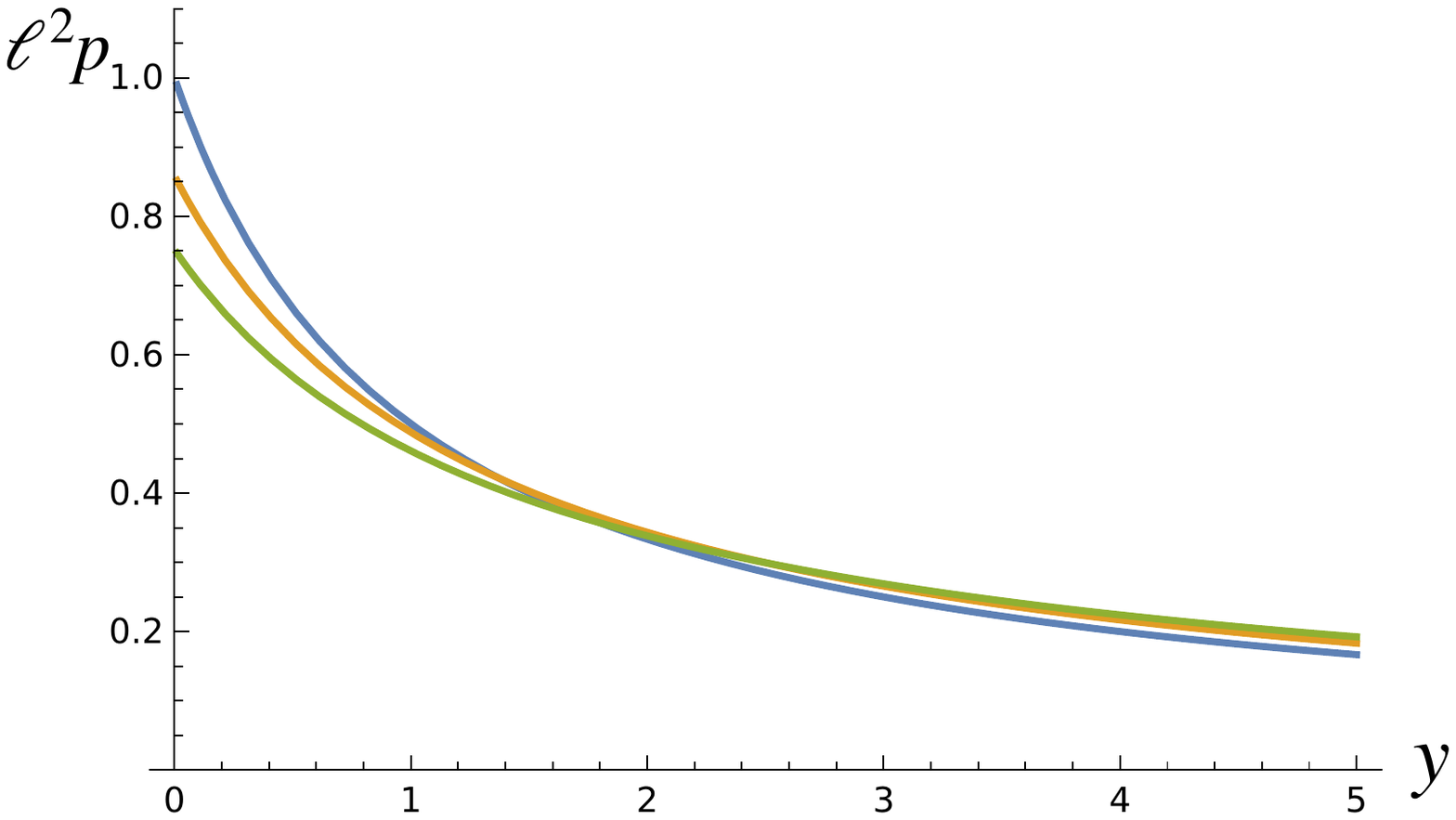}
\vspace{-1cm}
    \caption{Basic invariant $p$ in $D=5$ dimensions with $c=0$. The cases $b = 0$ (blue), $b=1/6$ (orange) and $b=1/3$ (green) are shown.} \label{PLTpD5}
\end{figure}

A solution of this equation that is smooth for $y \in [0,\infty)$, finite at $y=0$ and decreases as $1/y$ as $y$ approaches infinity will be called regular. 
In order to find solutions for this equation, we will first consider the homogeneous equation
\begin{equation}
    \mathcal{D}p = 0 \, .
\end{equation}
At this point, it is useful to denote 
\begin{equation}
    \Delta = \sqrt{(b-c)^2 - 8c} \, , \quad \lambda_{\pm} = \frac{b+3c \pm \Delta}{2(D-1)c} \, .
\end{equation}
and keep in mind the relations
\begin{align}\label{lmbdaREL}
    \lambda_+ - \lambda_- = \frac{\Delta}{(D-1)c} \, , \quad \lambda_+ + \lambda_- \frac{b+3c}{(D-1)c} \, ,\quad \lambda_+ \lambda_- = \frac{2(1+b+c)}{(D-1)^2 c} \, .
\end{align}
We require that $\Delta >0$, for which the general solution is 
\begin{equation}
    p = \frac{C_+}{y^{\lambda_+}} + \frac{C_-}{y^{\lambda_-}} \, .
\end{equation}
$C_{\pm}$ are the two integration constants. The asymptotic properties we impose on our solution are going to determine some constraints on these constants, namely if $C_{\pm}\neq 0$ the power $\lambda_{\pm}$ in the denominators has to be greater than or equal to $1$ such that the solutions decrease sufficiently fast at infinity. In this case, the solution still diverges as $y=0$. Therefore, the homogeneous equation does not have non-trivial regular solutions.
The inhomogeneous master equation \eqref{MEQP} has a solution of the form
\begin{equation}\label{solP}
    p = \mathcal{N} \bigg( \Phi_0(y,\lambda_-) - \Phi_0 \Phi_0(y,\lambda_+)  \bigg) \, ,
\end{equation}
where 
\begin{equation}
    \mathcal{N} = \frac{2}{(D-1)\ell^2\Delta} \, ,
\end{equation}
and 
\begin{equation}
    \Phi_0(y,\lambda) = \Phi(-y,1,\lambda) \, .
\end{equation}
Is the Lerch transcendent function. Figures \ref{PLTpD4} and \ref{PLTpD5} show a plot of the invariant $p$ in four and five dimensions for different parameters $b$ and $c$. 

\subsection{Parameters Constraints}

\begin{figure}[!hbt]
    \centering
\includegraphics[width=0.7\textwidth]{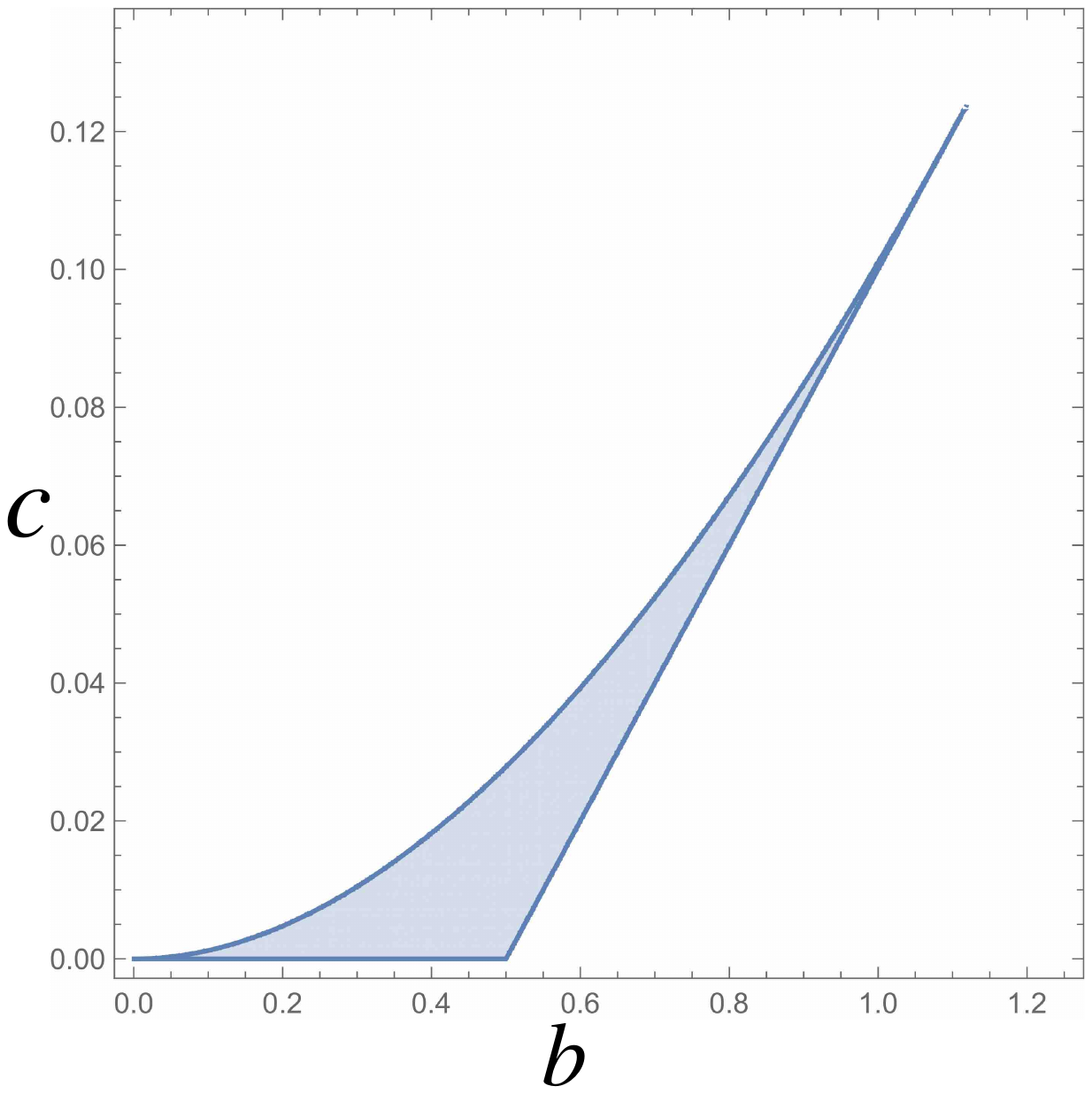}
    \caption{\label{DOMAIN_4} A two-dimensional plane of $(b,c)$ parameters.
    A shaded region in this plane shows allowed values of these parameters for $D=4$ and for the choice $k=2$.  }
\end{figure}

\begin{figure}[!hbt]
    \centering
\includegraphics[width=0.7\textwidth]{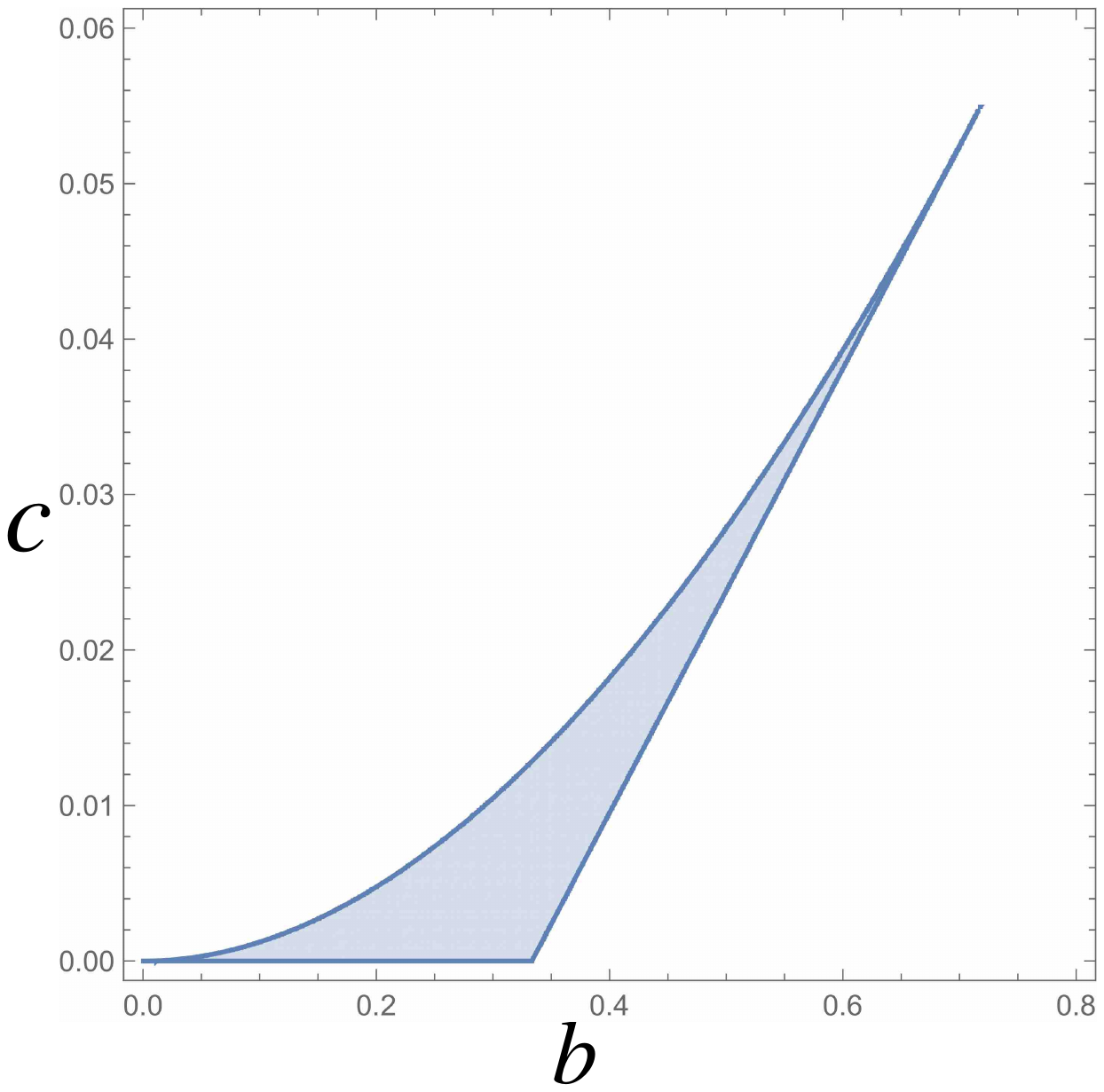}
    \caption{\label{DOMAIN_5}  A two-dimensional plane of $(b,c)$ parameters.
    A shaded region in this plane shows allowed values of these parameters for $D=5$ and for the choice $k=2$.  }
\end{figure}

The choice for the values of these parameters is justified by the assumption that the $\lambda_{\pm}$ are real and positive. If they are complex, then the solutions would oscillate at infinity. To guarantee that they are real, it suffices to impose the condition
\begin{equation}
    (b-c)^2 - 8c >0 \, .
\end{equation}
The condition that $\lambda_{\pm}$ are positive allows one to use the integral representation of the functions $\Phi_0(y,\lambda_{\pm})$, which makes it possible to guarantee that the solution has well-defined asymptotic behavior at infinity. 
The asymptotics of the solution at $y=0$ can be found with the aid of the definition of the Lerch function \cite{Bateman1953,NIST2010}, which gives
\begin{equation}
    p = \frac{1}{(1+b+c)\ell^2} + O(y) \, .
\end{equation}
Which means that the function $f$ which enters the metric has the form
\begin{equation}
    f = 1 - \frac{r^2}{(1+b+c)\ell^2 } + \ldots \, .
\end{equation}
Therefore, the resulting solution describes a regular black hole whose metric near the center corresponds to a deSitter-like core if $1+b+c>0$, and an anti-deSitter-like core if $1+b+c<0$. We will assume that our regular black hole models possess a deSitter core, hence focusing on the case 
\begin{equation}
    1+b+c >0 \, .
\end{equation}
Because we have imposed that $\lambda_{\pm} >0$, the third relation in \eqref{lmbdaREL} implies that $c>0$. We can conclude that 
\begin{equation}
    \lambda_+ > \lambda_- \, .
\end{equation}
At infinity, using the asymptotic form of the Lerch function found in \cite{Lerchasympt,FERREIRA2004210,Lagarias2016} one can show that $p$ takes the form
\begin{equation}\label{pinfty}
    p = \mathcal{N}\pi \left[ \frac{1}{\sin(\pi \lambda_-)y^{\lambda_-}} - \frac{1}{\sin(\pi \lambda_+)y^{\lambda_+}}\right] - \sum_{n=1}^{\infty}\frac{B_n}{y^n} \, ,
\end{equation}
where 
\begin{equation}
    B_n = \mathcal{N}(-1)^n \left[\frac{1}{\lambda_- - n} - \frac{1}{\lambda_+-n} \right] \, .
\end{equation}
After some algebra one may find that 
\begin{equation}
    B_1 = \frac{1}{Z\ell^2}\, , \quad Z = 1 - \frac{1}{2}(D-3)b + \frac{1}{2}(D-3)(D-2)c \, .
\end{equation}
It is important to reproduce the Schwarzschild-Tangherlini metric at infinity, therefore the first term between the brackets in equation \eqref{pinfty} should decrease faster than $1/y$. In order to meet this condition, we will impose 
\begin{equation}
    \lambda_- > k \, ,
\end{equation}
where $k$ is some arbitrary number larger than $1$. A similar condition for $\lambda_+$ holds since $\lambda_+>\lambda_-$. We can use this relation to write 
\begin{equation}\label{eqbndgamma}
    b-\gamma c>\Delta \, ,\quad \gamma = 2(D-1)k - 3 \, .
\end{equation}
Hence
\begin{equation}
    b>\gamma c \, .
\end{equation}
Since for $D \geq 3$ the parameter $\gamma$ is positive, $b$ should be positive as well. The final constraint we will derive for the parameters of the theory comes from rewriting the inequality in \eqref{eqbndgamma} 
\begin{equation}
    8 - 2(\gamma - 1)b + (\gamma^2 - 1)c >0 \, .
\end{equation}
If all of the conditions hold for $k=1$, then we have guaranteed that the first term in the bracket in \eqref{pinfty} decreases faster than $1/y$. The leading asymptotic term at infinity of $p$ is $B_1/y$, the function $f$ takes the form
\begin{equation}
    f = 1 - \frac{\tilde{\mu}}{Zr^{D-3}} + \ldots \, .
\end{equation}
In order to get the proper asymptotics that match the Schwarzschild-Tangherlini metric we define 
\begin{equation}
   \tilde{\mu} = Z\mu \, ,
\end{equation}
where $\mu$ is our mass parameter.
With that, we are finished imposing constraints in the parameters of the model. These are represented in the Figures \ref{DOMAIN_4} and \eqref{DOMAIN_5}, since the domain of allowed values depends on the number of dimensions $D$, we show them in $D=4$ and $D=5$ respectively. To summarize the information displayed in the figures 
\begin{itemize}
    \item $\lambda_->k$ and $c>0$ completely parametrize the allowed values of $b$ and $c$ for some $k>1$
    \item The parameter space is bounded at the bottom line by $c=0$
    \item The leftmost line is defined by $b = \sqrt{8c}+c$ and it arises from $\lambda_->k$
    \item The rightmost line is defined by $b=[(D-1)k-1]c + 2/[(D-1)k-2]$ and it arises from $\Delta \in \mathbb{R}$
\end{itemize}

\subsection{Special Cases}
Now that we have completely restricted the choice of physically meaningful parameters, we check some of the special choices they can take. First, let us consider the case when $c=0$, so $\psi = p+bq$. By taking the limit $c \to 0$ in our previous equations, one obtains   
\begin{equation}
    \Delta = b - \frac{(4+b)c}{b} \, , \quad \lambda_- = \frac{2(1+b)}{(D-1)b} \, ,\quad \lambda_+ \to \infty \, .
\end{equation}
By using the integral representation of $\Phi_0(y,\lambda)$ \cite{NIST2010}, one can show that it vanishes in the $\lambda \to \infty$ limit. Therefore 
\begin{equation}
    p = \mathcal{N}\Phi_0(y,\lambda_-) \, ,\quad \mathcal{N} = \frac{2}{(D-1)\ell^2b} \, .
\end{equation}
Near the origin $r=0$, the metric function $f$ takes the form 
\begin{equation}
    f = 1 - \frac{r^2}{(1+b)\ell^2} + \ldots \, .
\end{equation}
Since we have already imposed that $1+b>0$, the regular black hole has a deSitter-like core. For the other constraints, we have that
\begin{equation}
    4 - (\gamma - 1)b>0 \,, \quad \gamma =2(D-1)k-3 \, ,
\end{equation}
guarantees that the term $\sim 1/y^{\lambda_-}$ is the leading term in the expansion for $p$ at infinity. The mass parameter $\mu$ in this case is related to the integration constant $\tilde{\mu}$ via
\begin{equation}
    \tilde{\mu} - \bigg(1-\frac{1}{2}(D-3)\bigg) \mu \, .
\end{equation}
Another choice one can make is $b=c=0$, for which $\psi = p$. Then the equation \eqref{MEQP} reduces to a simple algebraic equation
\begin{equation}
    p = \frac{1}{\ell^2}\frac{1}{1+y} \, .
\end{equation}
In this case the solution becomes 
\begin{equation}
    p = \frac{\tilde{\mu}}{r^{D-1} + \ell^2\tilde{\mu}} \, .
\end{equation}
But now we also have that $\tilde{\mu} = \mu$ and we do not have to worry about any other constraint. This expression reproduces the Hayward metric if one solves to find the metric function $f$. 
Therefore, this special case reduces to Quasitopological gravity, and if one takes the limit $\ell \to 0$ then it recovers the Schwarzschild-Tangherlini metric.

\section{Bounds of Scalar Curvature Invariants}

\subsection{Solution Bounds}
The last property we are going to discuss is the boundedness of $p$, using the integral representation \cite{NIST2010} we can write 
\begin{equation}
    \frac{p}{\mathcal{N}} = \int_0^{\infty} \frac{e^{-\lambda_-x}(1 - e^{-(\lambda_+ - \lambda_-)})dx}{1+ye^{-x}} \, .
\end{equation}
The expression in parenthesis in the numerator is always positive if $\lambda_+ > \lambda_-$, this guarantees that the basic scalar invariant $p$ is positive as well. Furthermore, note that  $1+ye^{-x} \geq 1$, then the value of the denominator that most increases the value of the integral is 1. If we explicitly calculate this integral, we can find an upper bound
\begin{equation}
    0<p< \frac{1}{\ell^2(1+b+c)} \, .
\end{equation}
This property will become important when we calculate the curvature invariants of our model. 
In the next section, we will study the properties of our solution and construct a regular black hole with it in the next section. 

\subsection{Curvature Invariants}
\begin{figure}[!hbt]
    \centering
\includegraphics[width=0.7\textwidth]{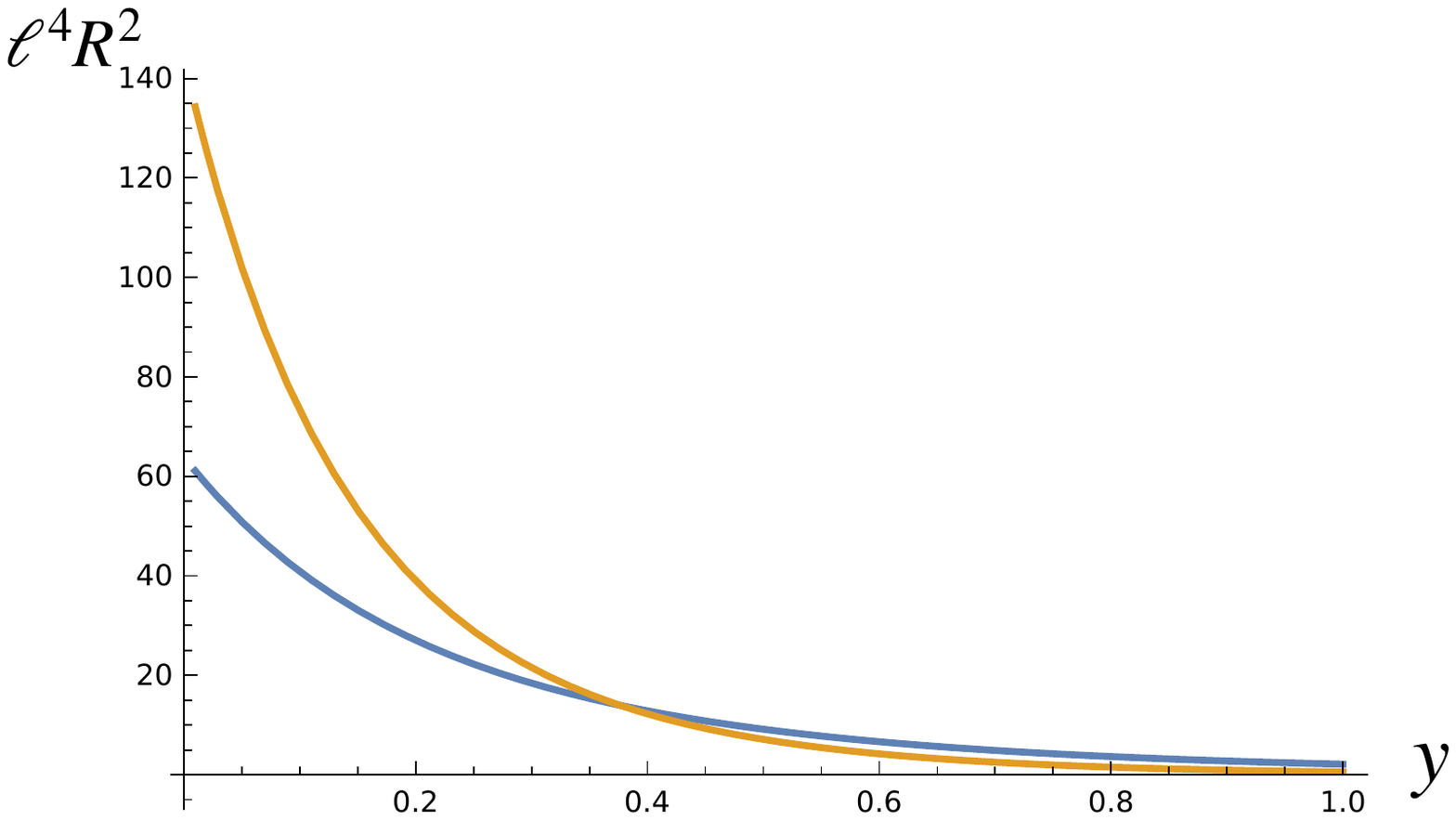}
\vspace{-1cm}
\caption{Dimensionless curvature invariant $\ell^{4}R^{2}$ as a function of $y$ in dimension $D=4$ with $c=0$ and $b=0$ (orange), $b=b_m$ (blue).}\label{PLTR2}
\end{figure}

\begin{figure}[!hbt]
    \centering
\includegraphics[width=0.7\textwidth]{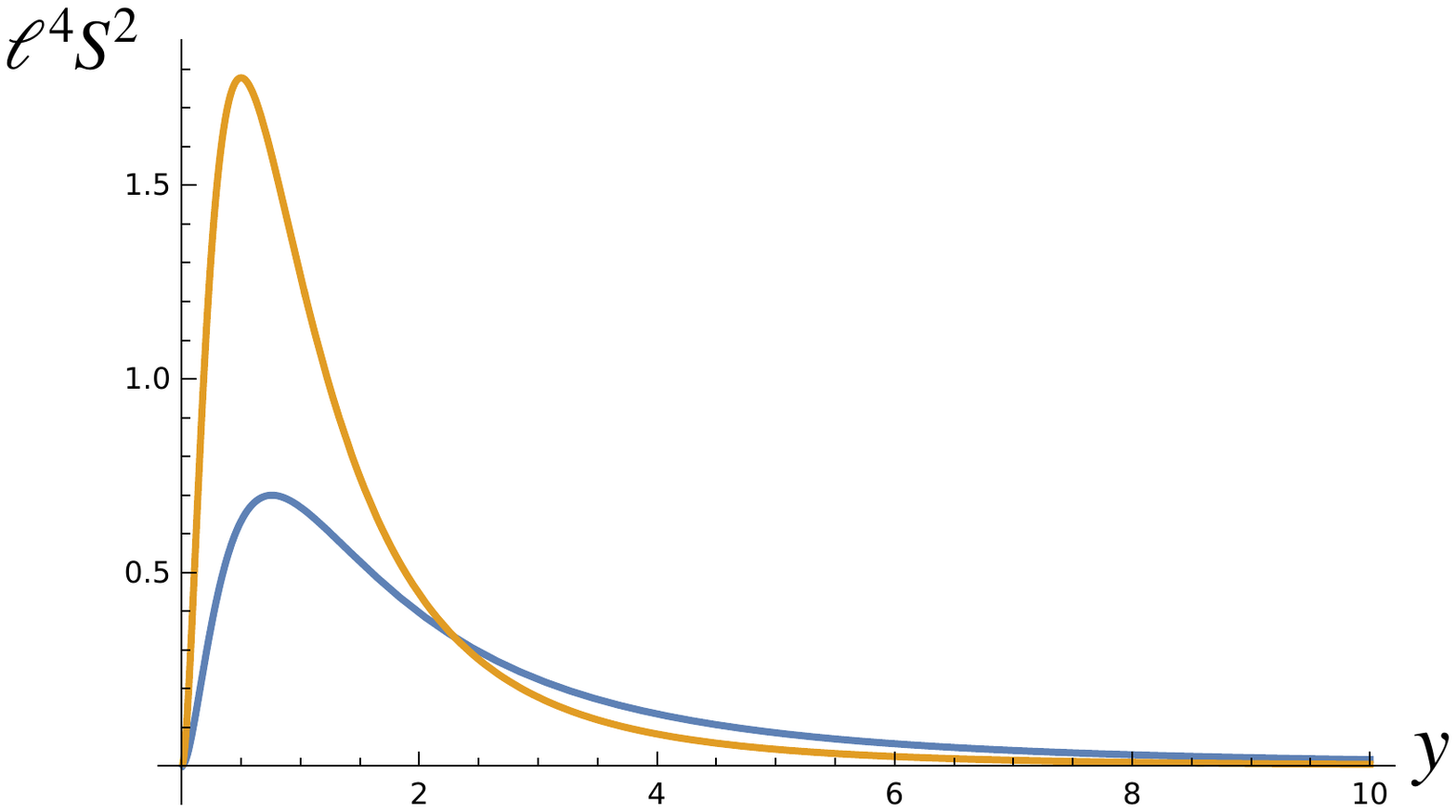}
\vspace{-1cm}
 \caption{Dimensionless curvature invariant $\ell^{4}S_{\mu\nu}^{2}$ as a function of $y$ in dimension $D=4$ with $c=0$ and $b=0$ (orange), $b=b_m$ (blue).}\label{PLTS2}
\end{figure}

\begin{figure}[!hbt]
    \centering
\includegraphics[width=0.7\textwidth]{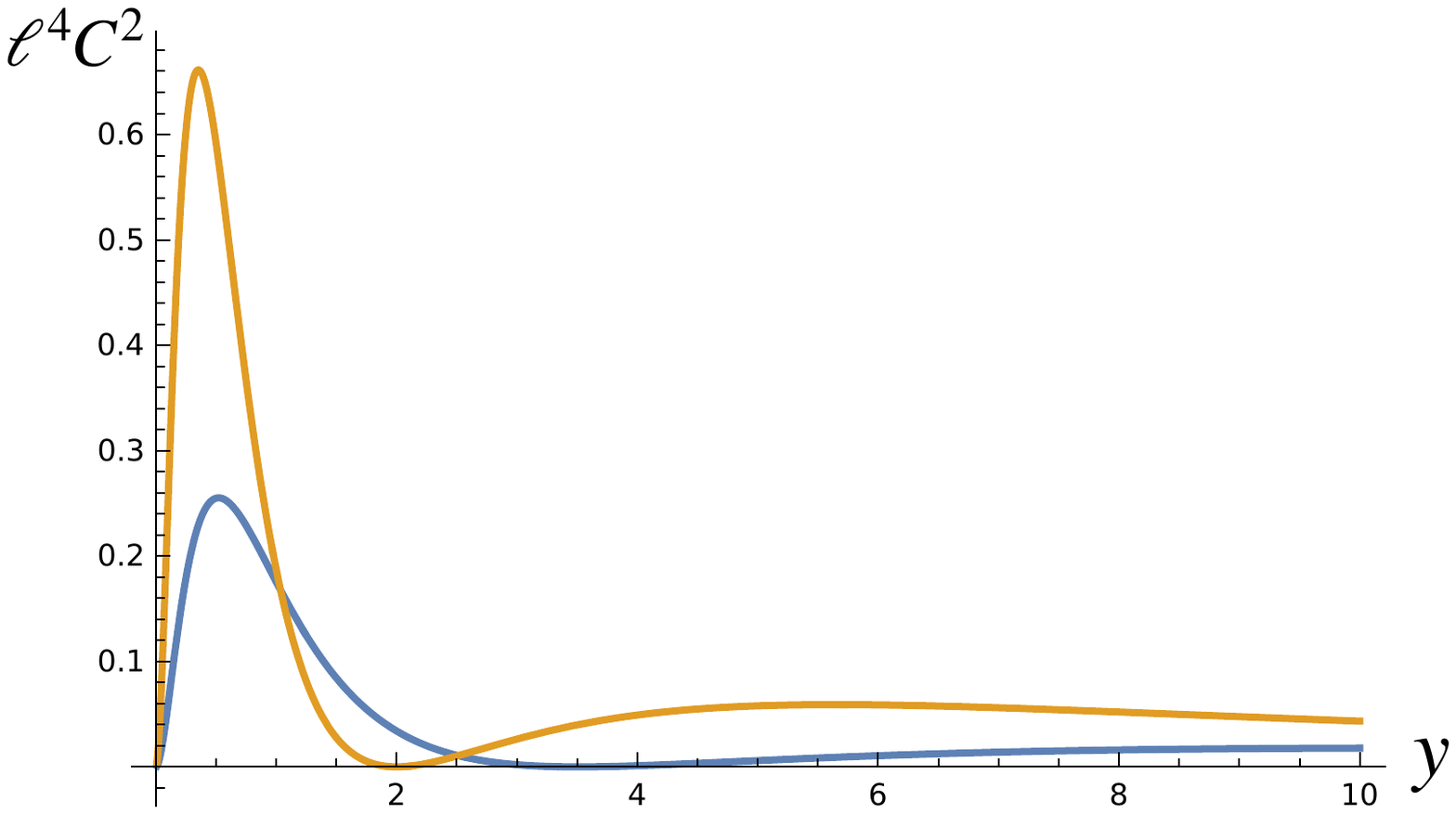}
\vspace{-1cm}
\caption{Dimensionless curvature invariant $\ell^{4}C^{2}$ as a function of $y$ in dimension $D=4$ with $c=0$ and $b=0$ (orange), $b=b_m$ (blue).}\label{PLTC2}
\end{figure}

Since we have relations between them \eqref{BIRel}, we can write the other basic curvature invariants explicitly in terms of $y$, namely 
\begin{equation}
\begin{split}
    q &= \mathcal{N}\bigg[ \Phi_0(y,\lambda_-) - \Phi_0(y,\lambda_+)  - \frac{D-1}{2} \bigg( \Phi_1(y,\lambda_-) - \Phi_1(y,\lambda_+)   \bigg)\bigg] \, ,\\
    v &=\mathcal{N}\bigg[\Phi_0(y,\lambda_-) - \Phi_0(y,\lambda_+) - \frac{(D-1)(D+2)}{2}\bigg(\Phi_1(y,\lambda_-) - \Phi_1(y,\lambda_+)  \bigg) \\
    & \qquad + \frac{(D-1)^2}{2}\bigg(\Phi_2(y,\lambda_-) - \Phi_2(y,\lambda_+) \bigg)   \bigg] \, .
\end{split}
\end{equation}
Where 
\begin{equation}
    \Phi_1(y,\lambda) = -y\frac{d}{dy}\Phi_0(y,\lambda) \, , \quad \Phi_2(y,\lambda) = y^2 \frac{d^2}{dy^2}\Phi_0(y,\lambda) \, .
\end{equation}
These functions $\Phi_i(y,\lambda)$ are also bounded for $\lambda >0$ \cite{NIST2010,Bateman1953}, then these basic invariants are bounded as well as polynomials constructed from them. 
This is important since the curvature invariants can always be written as some combination or some polynomial of these basic scalar invariants, to illustrate this, we include Figures \ref{PLTR2}, \ref{PLTS2}, and \ref{PLTC2}. Since our solutions depend on the number of dimensions $D$ and the parameters $b$ and $c$, we make the choice of looking at the $D=4$ and $c=0$ case, where we plot different values of $b$ from the Hayward solution ($b=0$) to the maximum allowed value in our model $b=b_m$ by the constrains. We choose to show how the  curvature invariants $R^2$, $S_{\mu\nu}S^{\mu\nu}$ and $C_{\mu\nu\alpha\beta}C^{\mu\nu\alpha\beta}$ change with respect to the dimensionless coordinate $y$. We have defined the traceless part of the Ricci tensor as $S_{\mu\nu} = R_{\mu\nu} - g_{\mu\nu}R/D$.


\section{The Metric}

Now that we have the solution to our master equation \eqref{MEQP}, we can write the function $f$ which enters the metric as 
\begin{equation}\label{eqF}
    f = 1 - F_D \Psi(y) \, , \quad F_D = \bigg(\frac{\tilde{\mu}}{\ell^{D-3}} \bigg)^\frac{2}{D-1} \, , \quad \Psi(y) = \ell^2 y^{2/(D-1)}p(y) \, .
\end{equation}
Let us emphasize that $y$, $F_D$ and $\Psi(y)$ are all dimensionless quantities. That we can write the metric function $f$ in this form is a direct consequence of the scaling property of the solution we discussed in section \eqref{sec2Ddil}. We will be using this representation because $\Psi(y)$ depends only on the parameters $b$, $c$ and on the number of dimensions $D$, while $F_D$ encodes the dependence on the mass parameter of some particular solution. 
From equation \eqref{eqF} we can see that the metric function $f$ has a finite value $f=1$ at $y=0$, as $y$ increases the function decreases until it reaches a minimum value defined by $d\Psi/dy = 0$, after which it monotonically increases towards 1 again. The metric displays different features depending on the value of the parameter $F_D$, for a sufficiently large value it has two horizons and is regular at the center. Their position is defined by the equation
\begin{equation}
    \Psi(y_H) = \frac{1}{F_D} \, .
\end{equation}
If the parameter $F_D$ is sufficiently small, no horizon is present. There also exists a critical value $F_D^*$ for which the inner and outer event horizon coincide. The critical value $y_H^*$ at which this happens can be found by solving both 
\begin{equation}
    \Psi(y) = \frac{1}{F_D}\, \quad \mathrm{and} \quad \frac{d\Psi}{dy} = 0 \, .
\end{equation}
The second equation determines the critical value $y=y_*$. Explicitly 
\begin{equation}\label{crY}
    \frac{2}{D-1}\bigg( \Phi_0(y,\lambda_-) - \Phi_0(y,\lambda_+)\bigg) + y\frac{d}{dy}\bigg( \Phi_0(y,\lambda_-)- \Phi_0(y,\lambda_+)\bigg) = 0\, .
\end{equation}
The following property can be obtained using the integral representation of $\Phi_0(y,\lambda)$
\begin{equation}
    y\frac{d}{dy}\Phi_0(y,\lambda) = \frac{1}{1+y} - \lambda\Phi_0(y,\lambda) \, ,
\end{equation}
and it is useful since we can use it in \eqref{crY} and obtain 
\begin{equation}
    (\lambda_+ - \frac{2}{D-1})\Phi_0(y_*,\lambda_+) = (\lambda_- - \frac{2}{D-1})\Phi_0(y_*,\lambda_-) \, .
\end{equation}
This equation is simplified in the case where $c=0$ to 
\begin{equation}
    \frac{2}{D-1}\Phi_0(y_*,\lambda_-) = \frac{1}{1+y_*} \, .
\end{equation}
Then to find the critical value of the mass of the black hole we use the relation
\begin{equation}
    F_D^* = \frac{1}{\Psi(y_*)} \, .
\end{equation}
This equation takes on a simple form for $c=0$
\begin{equation}
    F_D^* = y_*^{-\frac{2}{D-1}}(1+y_*) \, .
\end{equation}
In order to illustrate these values, Table \eqref{tablecrit} shows some of the critical values for $y_*$ and $F_D^*$ in $D=4$ and $D=5$ dimensions in the case where $c=0$ for different values of $b$.
\begin{table}[h!]
\hfill
\begin{tabular}{ |l|l| }
 \hline
 \multicolumn{2}{| c |}{$(y_*,F_D^*)$ in $D=4$}\\
 \hline
 (2.000, 1.890) & $b = 0$\\
 (2.922, 1.919) & $b = b_{m}/2$ \\
 (4.116, 1.992) & $b = b_{m}$  \\
 \hline
\end{tabular}
\hfill
\begin{tabular}{ |l|l| }
 \hline
 \multicolumn{2}{| c |}{$(y_*,F_D^*)$ in $D=5$}\\
 \hline
(1.000, 2.000)& $b = 0$ \\
(1.378, 2.026)  & $b = b_{m}/2$ \\
(1.817, 2.089)  & $b = b_{m}$   \\
 \hline
\end{tabular}
\hspace{1cm}
\caption{ \label{tablecrit}The left table gives critical values of $y_*$ and $F_D^*$ in $D=4$ spacetime dimensions for $c=0$ and different values of the parameter $b$, where $b_m=1/2$. The right table gives similar quantities for $D=5$, where $b_m=1/3$.}
\end{table}

\begin{figure}[!hbt]
    \centering
\includegraphics[width=0.7\textwidth]{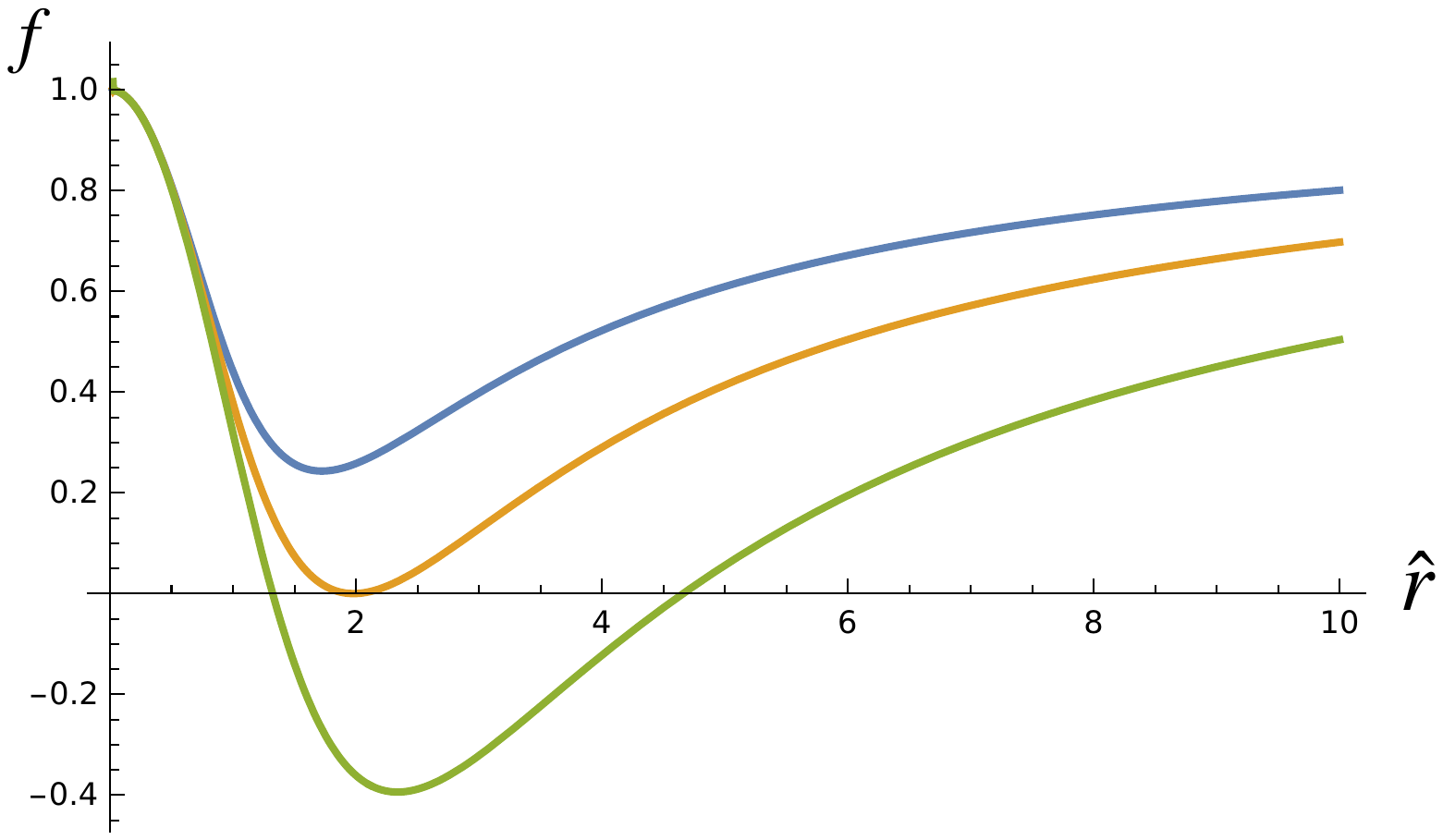}
\vspace{-1cm}
    \caption{Metric function $f$ in terms of dimensionless parameter $\hat{r}$ in $D=4$ dimensions for $b=b_m/2$, $c=0$ with $\hat{r}_\ins{g}=2$ (blue), $ \hat{r}_\ins{g}\approx 3.03$ (orange), and $ \hat{r}_\ins{g}=5 $ (green). }\label{PLTfD4}
\end{figure}

\begin{figure}[!hbt]
    \centering
\includegraphics[width=0.7\textwidth]{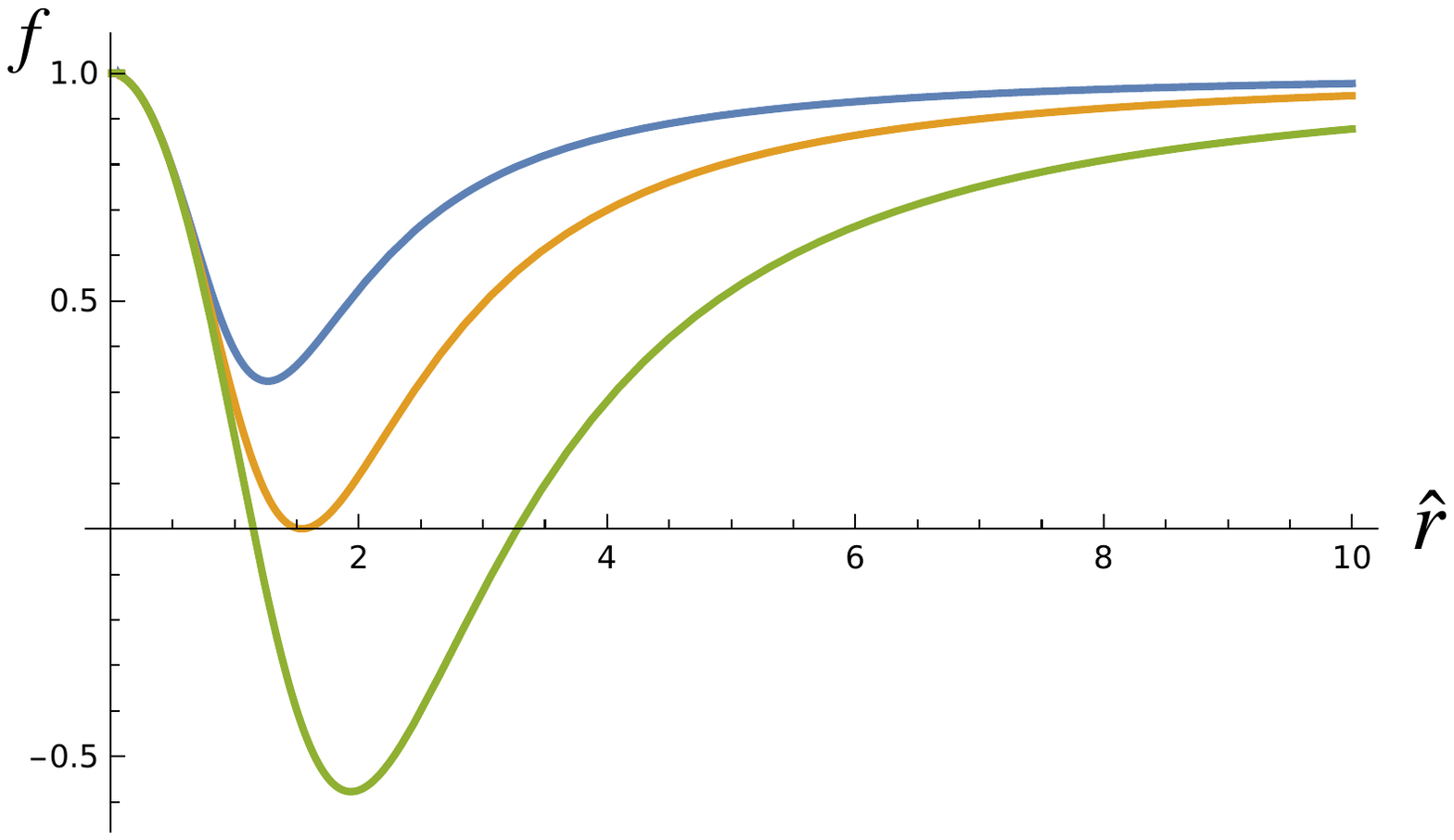}
\vspace{-1cm}
  \caption{Metric function $f$ in terms of dimensionless parameter $\hat{r}$ in $D=5$ dimensions for $b=b_m/2$, $c=0$ with $\hat{r}_\ins{g}=1.5$ (blue), $ \hat{r}_\ins{g}\approx 2.22$ (orange), and $ \hat{r}_\ins{g}=3.5 $ (green). }\label{PLTfD5}
\end{figure}

At this point, it is worth remembering that our metric ansatz is written in $(t,r)$ coordinates, therefore it will be instructive to write the metric function $f$ as a function of the $r$-coordinate. One may do so with the aid of the relations
\begin{equation}
\begin{split}
   \tilde{\mu} &= Z\mu = Zr_g^{D-3} \, , \\
   Z &= 1 - \frac{1}{2}(D-3)b + \frac{1}{2}(D-3)(D-2)c \, ,\\
   \hat{r} &= r/\ell \, , \quad \hat{r}_g = r_g/\ell \, , \quad
   F_D = Z^{\frac{2}{D-1}}\hat{r}_g^{\frac{2(D-3)}{D-1}} \, , \\
   y &= \frac{1}{Z}\frac{\hat{r}^{D-1}}{\hat{r}_g^{D-3}} \, .
\end{split}
\end{equation}
Figures \ref{PLTfD4} and \ref{PLTfD5} illustrate the shape of $f(\hat{r})$ in spacetimes with $D=4$ and $D=5$, respectively.

\subsection{Thermodynamical Aspects}
In order to study the thermodynamical aspects of the model, we first define a few standard equations. The Hawking temperature $T_H$ of a black hole is given in terms of the surface gravity $\kappa$  of the event horizon as 
\begin{equation}
    T_H = \hbar \frac{\kappa}{2\pi} \, .
\end{equation}
Keeping in mind that the speed of light is $c=1$ and the Boltzmann constant $k_B=1$ in our system of units. The surface gravity for the metric \eqref{sssastz} in $(t,r)$-coordinates and in the radial gauge where $N=1$ is given by
\begin{equation}
    \kappa = \frac{1}{2}\bigg|\frac{df}{dr}\bigg|_H = r_H|q|_H \, .
\end{equation}
Remember that $r_H$ is the radius of the event horizon, defined by the equation $f=0$. For our model, it is different form the parameter $r_g$ which is proportional to the asymptotic mass of the black hole unlike in the Einstein theory where they coincide. 
The change in energy $dE = dM$ and the change in entropy $dS$ are related according to the second law of thermodynamics 
\begin{equation}
    dM = TdS \, . 
\end{equation}
In terms of our mass parameter $\mu$ this explicitly reads 
\begin{equation}\label{eqS}
    dS = \frac{2\pi B}{\hbar \kappa}d\mu \, .
\end{equation}
We are able to find a relation between $\mu$ and $r_H$ thanks to the condition $f(r_H) = 0$. Making the parameters $\kappa = \kappa(r_H)$ and $\mu = \mu(r_H)$ functions of $r_H$ in \eqref{eqS} and taking the integral we get
\begin{equation}\label{intS}
    S = S_* + \frac{2\pi B}{\hbar}\int_{r_*}^{r_H} d\eta \frac{1}{\kappa(\eta)}\frac{d\mu(\eta)}{d\eta} \, .
\end{equation}
Where $r_*$ is the gravitational radius of the minimal mass black hole which is defined by the condition that the horizon is degenerate $r_+ = r_-$, which is consistent with the requirement that $\kappa(r_*) = 0$. 
The integral \eqref{intS} is rather difficult to solve analytically for our full solution \eqref{solP}, if needed, one may calculate it numerically. Instead, we will consider the simplest version of our model where $a=1$ and $b=c=0$, the Hayward metric. This way we will be able to analytically study the expression for $S$ which hopefully captures some of the aspects of the full model. In this case, the metric function takes the form
\begin{equation}
    f = 1 - \frac{\mu r^2}{r^{D-1} + \ell^2\mu} \, .
\end{equation}
Then we can write 
\begin{equation}
    \mu(r_H) =\frac{r^{D-1}_H}{r^2_H - \ell^2} \, ,
\end{equation}
and 
\begin{equation}
    \kappa(r_H) = \frac{(D-3)r_H^2 - (D-1)\ell^2}{2r^3_H} \, .
\end{equation}
The minimal black hole has a horizon radius 
\begin{equation}
    r_* = \sqrt{\frac{D-1}{D-3}}\ell \, ,
\end{equation}
with a minimal mass parameter $\mu_*$ given by
\begin{equation}
    \mu_* = \frac{(D-1)^{(D-2)/2}}{2(D-3)^{(D-3)/2}}\ell^{D-3} \, .
\end{equation}
Therefore we can write the change in entropy as 
\begin{equation}
    dS = \frac{4\pi B}{\hbar}\frac{r_H^{D+1}}{(r_H^2-\ell^2)^2}dr_H \, .
\end{equation}
We can introduce the Planck length $\ell_{\mathrm{Pl}}$ into our equations via the relation $\ell_{\mathrm{Pl}}^{D-2} = \hbar \varkappa/(8\pi)$. In $D=4$ we get
\begin{equation}
    S = S_* + \frac{\pi}{2\ell^2_{\mathrm{Pl}}(r_H^2 - \ell^2)} \bigg[ 2r_H^4 - 7\ell^2r_H^2 + 3\ell^4 + 4\ell^2(r_H^2 - \ell^2) \bigg(\ln(\frac{r_H^2}{\ell^2} - 1) - \ln 2\bigg) \bigg] \, .
\end{equation}
In the limit when we have a small $\ell$ such that $\ell \ll r_H $ we have 
\begin{equation}\label{SHAY}
    S = S_* + \frac{\pi r_H^2}{\ell_{\mathrm{Pl}}^2} + \frac{\pi \ell^2}{\ell^2_{\mathrm{Pl}}} \bigg(4 \ln \left( \frac{r_H}{\ell}\right) - 2 \ln 2 - \frac{5}{2}\bigg) + O(\ell^4) \, .
\end{equation}
In this same limit, the relation between $r_H$ and $r_g$ in $D$ dimensions is 
\begin{equation}
    r_H = r_g - \frac{\ell^2}{(D-3)r_g} + O(\ell^4) \, .
\end{equation}
In order to recover the Bekenstein-Hawking entropy term one should look at the leading term $\pi r_H^2 = \mathcal{A}/(4 \ell_{\mathrm{Pl}}^2)$ in \eqref{SHAY}, where $\mathcal{A}$ is the area of the horizon.

\section{Conclusions}

In this Chapter, we have constructed a regular black hole model that incorporates all the basic curvature invariants of a static spherically symmetric spacetime. We have shown that it generalizes both the Schwarzschild metric and quasitopological gravity models, which are recovered in the appropriate limits.

To achieve this, we generalized the reduced quasitopological action by introducing dependence on all the basic curvature invariants. Our model includes a length parameter, $\ell$, which determines the regime where our solution deviates from those of General Relativity. For the function $h(p)$, we demonstrated the existence of a covariant action that, after spherical reduction, yields the desired theory. Whether such a covariant formulation is possible for the more general case $h = h(p, q, u, v)$ remains an open and intriguing question that warrants further investigation.

One of the most notable features of models with a general function $h(p, q, u, v)$ is their scaling property. Specifically, there exists a new dimensionless coordinate $y$ such that the equations for the basic curvature invariants depend only on this variable. Consequently, if any curvature invariant is bounded as a function of $y$, this boundedness holds for any value of the mass parameter in the solution in accordance with Markov's limiting curvature conjecture.

In the final part of this Chapter, we considered a broad class of models with $h(p, q, u, v)$ for which the master equation is a second-order linear ordinary differential equation. We demonstrated that these models admit solutions describing non-singular black holes. We also analyzed the properties and thermodynamics of these regular black hole solutions.

	\chapter{Summary and Discussion}
\label{Concl}

\section{Conclusions}

In this thesis, we have investigated different approaches to constructing singularity-free black hole models, focusing on two classes of extensions to General Relativity: infinite derivative non-local gravity and higher-curvature quasitopological theories. 

Chapter~\ref{Ch2} began by generalizing Maxwell's electrodynamics in higher dimensions through the inclusion of nonlocal, infinite-derivative operators. This approach led to electromagnetic fields sourced by point-like charges that are regular at the origin, providing a concrete example of how infinite derivative models smooth ultraviolet divergences. In the gravitational context, a similar strategy was adopted in the linearized regime. By incorporating entire functions of the d'Alembert operator in the action, we obtained linearized solutions for static sources that exhibited regular curvature invariants in both four and higher dimensions.

In Chapter~\ref{Ch3}, we extended the nonlocal approach to linearized gravity. Considering a class of ghost-free, infinite-derivative theories of gravity, we computed the gravitational potential generated by static sources in both four and higher dimensions. The nonlocal modification replaces the usual Newtonian potential with a regular profile whose curvature invariants remain finite at the origin. The analysis further showed that within the class of theories known as $\mathrm{GF}_N$, different choices of the form factor like $\mathrm{GF_1}$ and $\mathrm{GF}_2$ yield qualitatively different regularization profiles, emphasizing the impact of the form factor structure.

Chapter \ref{Ch4} focused on regular black hole solutions arising in infinite derivative gravity, a class of nonlocal theories characterized by the presence of entire function form factors. We analyzed the static, spherically symmetric solutions sourced by localized distributions in both real and complexified spacetimes. This included a detailed construction of nonlocal modifications to the Schwarzschild and Kerr metrics, which we interpreted as perturbations due to nonlocality of the GR solutions. By employing nonlocal Green functions with exponential form factors, we showed that the resulting potentials are regular at the origin and asymptotically match the classical solutions at large distances. Furthermore, we examined the shift in the location of the event horizon induced by nonlocality and discussed the implications for causal structure and curvature regularization. These results demonstrate that nonlocal infinite derivative theories can serve as a consistent framework for obtaining singularity-free black hole geometries. While a complete description of a rotating black hole in the context of the full infinite derivative ghost-free gravity models is still lacking, our hope is that this model captures some of the features of what a full solution would look like.

In Chapter \ref{Ch5}, we explored an alternative route to regular black hole models based on higher curvature modifications of gravity. Specifically, we considered two-dimensional dilaton gravity theories obtained via spherical reduction from higher-dimensional actions. These models included an infinite series of higher-order curvature invariants, inspired by the quasitopological gravity framework. Using this approach, we constructed families of static, spherically symmetric solutions with regular cores, bounded curvature invariants, and well-defined thermodynamic properties. A notable feature of this construction is the ability to systematically tune parameters to interpolate between standard black holes, regular black holes, and black hole mimickers. The analysis also confirmed the asymptotic flatness of the solutions and their consistency with the limiting curvature conjecture.

The models presented in this work illustrate the feasibility of constructing physically motivated, singularity-free black hole spacetimes within modified theories of gravity. Both the nonlocal and higher-curvature approaches have proven effective in regulating divergences at the black hole core while preserving consistency with known large-distance behavior. While several challenges remain—such as understanding the full dynamics of gravitational collapse, incorporating rotation in general settings, and embedding these models in a complete quantum theory of gravity—the results of this thesis provide a concrete step forward in addressing the singularity problem in gravitational physics.



\bigskip 
\clearpage

\singlespacing                
 

\printbibliography[heading=bibintoc]
\appendix
 
\truedoublespacing            


	\chapter{Spherical Reduction of Einstein Gravity}
\label{app:SphReduction}

\section{Spherical Reduction}

In the absence of a cosmological constant, the Einstein-Hilbert action may be written in any number of spacetime dimensions $D$ as
\begin{equation}\label{EE_appA}
    S[g] = \frac{1}{2\mathcal{\varkappa}} \int d^Dx\sqrt{-g}R \, ,
\end{equation} 
where $\mathcal{\varkappa} = 8\pi G^{(D)} c^{-3}$ (keep in mind $c=1$ throughout this thesis) is the $D$ dimensional gravitational coupling and $G^{(D)}$ is the $D$ dimensional Newton constant, which has units of $[G^{(D)}] = [G^{(4)}][\mathrm{length}]^{D-4}$. By varying this action, one obtains the higher-dimensional Einstein field equations
\begin{equation}\label{Einst:eq_appA}
    R_{\mu \nu} - \frac{1}{2}g_{\mu\nu}R = 8\pi G^{(D)}T_{\mu\nu} \, .
\end{equation}
The right-hand side of the equation can be obtained by varying the corresponding matter action
\begin{equation}
    T_{\mu\nu} = \frac{2}{\sqrt{-g}}\frac{\delta S_{\mathrm{matter}}}{\delta g_{^{\mu\nu}}} \, .
\end{equation}
For now, we shall focus on vacuum solutions $T_{\mu\nu} = 0$ and consider a spherically symmetric $D$ dimensional ansatz of the form 
\begin{equation}\label{sphmetric}
    ds^2 = \tilde{g}_{ij}dx^idx^j + \rho^2d\Omega^2 \, , \quad i,j=0,1 \, .
\end{equation}
Where the two-dimensional metric $\tilde{g}_{ij}$ and the function $\rho$ depend on the $x^i$ coordinates. Furthermore 
\begin{equation}
    d\Omega^2 = \hat{g}_{ab}dx^adx^b \, ,\quad a,b = 2,...,(D-1) \, ,
\end{equation}
is the metric of a unit $(D-2)$-dimensional sphere whose area is given by the expression 
\begin{equation}
    \Omega_{(D-2)} = \frac{2\pi^{(D-1)/2}}{\Gamma((D-1)/2)} \, .
\end{equation}
The study of spherically symmetric solutions can be performed by using this ansatz in the gravity field equations. This in turn yields $G_{i}^j = 0$ and $G^b_a = \delta^b_a V = 0$, where
\begin{equation}\label{eqG2}
    G_i^j = -(D-2)\left[\frac{\rho_{:i}^{:j}}{\rho} + \delta_i^j \left(-\frac{\rho_{:k}^{:k}}{\rho} + \frac{D-3}{2\rho^2}(1-\rho^{:k}\rho_{:k}) \right) \right] = 0 \, ,
\end{equation}
\begin{equation}\label{eqV}
    V = -\frac{1}{2}\mathcal{R} + (D-3)\frac{\rho^{:k}_{:k}}{\rho} - \frac{(D-3)(D-4)}{2\rho^2}(1-\rho^{:k}\rho_{:k}) = 0 \, .
\end{equation}
Here $\mathcal{R}$ is the scalar curvature of the two-dimensional metric $\tilde{g}_{ij}$ and the colon symbol denotes a covariant derivative with respect to this metric. These equations are, in fact, related via the equation
\begin{equation}
    G_{i \, :k}^k +(D-2)G_i^k \frac{\rho_{:k}}{\rho} = (D-2)\frac{\rho_{:i}}{\rho}V \, .
\end{equation}
Therefore, equation \eqref{eqV} is a consequence of \eqref{eqG2}. This is due to the $D$-dimensional conservation law $G_{\alpha \, ; \beta}^{\beta} = 0$ which is, in turn, a consequence of the Bianchi identity. To find solutions that are spherically symmetric, one would have to find solutions for equation \eqref{eqG2}. Alternatively, one may substitute the metric ansatz \eqref{sphmetric} directly into the Einstein-Hilbert action, which, after integrating over the angular variables, takes the form
\begin{equation}
    S_{\mathrm{S}} = \frac{\Omega_{D-2}}{2\mathcal{\kappa}}\int d^2x \sqrt{-\tilde{g}} \left[ \mathcal{R}\rho^{D-2} + (D-2)(D-3)\rho^{D-4}(1+ \rho^{:i}\rho_{:i})\right] \, .
\end{equation}
This is a reduced form of the action that depends on the two-dimensional metric $\tilde{g}_{ij}$ and the scalar function $\rho$, both of which depend on the $x^i$ coordinates. In this way, the original $D$-dimensional problem is reduced to a two-dimensional one. It is very interesting to note that is possible to obtain the same full vacuum solutions as one would obtain by solving the field equations by performing a variation $\delta \tilde{g}_{ij}(x^k)$ and $\delta \rho(x^k)$ over the reduced action; of course, this is not possible in general since the possible variations $\delta g_{\mu\nu}(x^{\alpha})$ of the full action is much greater than the allowed variations of the reduced action. Luckily, it can be shown that it is possible to obtain the full gravity equations via this so-called spherical reduction for spherical symmetry. In what follows, the subscript $\mathrm{S}$ will indicate quantities obtained by means of this reduction.

\section{Static Spherical Reduction}

The spherically reduced vacuum Einstein equations imply that  
\begin{equation}
    \xi^{\mu} = \delta^{\mu}_i e^{ij} \rho_{:j} \, ,
\end{equation}  
is a Killing vector of the metric \eqref{sphmetric}, where $ e^{ij} $ is the two-dimensional Levi-Civita tensor. In the general case, the metric can be non-static; however, if one searches for vacuum spherically symmetric solutions, an additional reduction is possible.  
Assuming the spacetime is also static, it admits an $ \mathrm{SO(D-1) \times R^1} $ symmetry group. An explicit form of this metric is  
\begin{equation}\label{SSSmetric}
    ds^2 = -N(r)^2 f(r) dt^2 + f(r)^{-1} dr^2 + \rho(r)^2 d\Omega_{D-2} \, .
\end{equation}  

With the three arbitrary functions $N(r)$, $f(r)$ and $\rho(r)$. This is an interesting ansatz for several reasons, one of which is that there exist four independent curvature invariants we will denote by $p$, $q$, $u$, and $v$, such that the non-zero $D$-dimensional Riemann curvature tensor components take the form 
\begin{equation}
\begin{split}
    & R_{jl}^{\ \ ik} = v\delta^{ik}_{jl} \, , \quad  R_{ta}^{\ \ tb} = u\delta^{b}_{a} \, , \\
    &  R_{ra}^{\ \ rb} = q\delta^{b}_{a} \, , \quad   R_{ab}^{\ \ cd} = p\delta^{cd}_{ab} \, .
\end{split}
\end{equation}
In terms of our metric, they take the form 
\begin{equation}\label{BCI}
\begin{split}
    & p = \frac{1- (\rho ')^2f}{\rho^2}\, , \quad v = -\frac{1}{2}\left(f''+ 3f'\frac{N'}{N} + 2f\frac{N''}{N} \right) \, , \\
    & q = -\left( \frac{f'\rho'}{2\rho} + f\frac{\rho''}{\rho}\right) \, , \quad u = -\left( \frac{f'\rho'}{2\rho} + f\frac{\rho'N'}{\rho N}\right) \, .
\end{split}
\end{equation}
Here, the prime symbol denotes a derivative with respect to the variable $r$. All of the other scalar invariants constructed from the curvature tensors can be written in terms of these four functions; therefore, we will call them basic curvature invariants. A couple of examples of this are 
\begin{equation}\label{SSSRicci}
    R = 2v +(D-2)[2q + 2u + (D-3)p] \, ,
\end{equation}
\begin{equation}
    C_{\alpha \beta \mu \nu}C^{\alpha \beta \mu \nu} = \frac{4(D-3)}{D-1}(p-q-u+v)^2 \, .
\end{equation}
Where $C_{\alpha \beta \mu \nu}$ is the $D$-dimensional Weyl tensor. One can substitute \eqref{SSSRicci} into the Einstein-Hilbert action, and after integration of the angular part and some integration by parts, obtain that 
\begin{align}
    S_{\mathrm{SS}} &= B \int dtdrN\rho^{D-2}[(D-3)p + 2q] \\
    &=B \int dtdr \frac{N}{\rho'}[\rho^{D-3}(1-(\rho')^2f)]' \\
    &= B\int dtdr\frac{N}{\rho'}[\rho^{D-1}p]' \, ,
\end{align}
where we have the constant $B$ defined by 
\begin{equation}\label{eq_B}
    B = \frac{(D-2)\Omega_{D-2}}{2\mathcal{\varkappa}} \, .
\end{equation}
We may, once again, integrate by parts to obtain 
\begin{equation}\label{SSAction}
    S_{\mathrm{SS}} = -B \int dtdr\rho^{D-1}\left(\frac{N}{\rho'}\right)' p \, .
\end{equation}
We can now search for arbitrary functions $N$, $f$ and $\rho$ which minimize this action. Variation of this action over $f(r)$ yields the following equation
\begin{equation}\label{eqNSS}
    \left(\frac{N}{\rho'}\right)' = 0 \,  \quad \to \quad N = C\rho' \, ,
\end{equation}
where $C$ is some arbitrary constant that may be set to 1 so time is properly normalized at infinity. The second equation may be obtained by variation over $N(r)$
\begin{equation}\label{eqFSS}
    [\rho^{D-3}(1- (\rho ')^2f)]' = 0 \quad \to \quad f = \frac{1}{(\rho')^2}\left(1 - \frac{\mu}{\rho^{D-3}}\right) \, .
\end{equation}
This is the same function as the one in the Schwarzschild-Tangherlini metric, which describes a $D$-dimensional static black hole. The constant $\mu$ comes as a constant of integration, and it is related to the radius of the black hole via $r_{\mathrm{g}} = \mu^{1/(D-3)}$, it is also related to the mass of the black hole $m$ 
\begin{equation}
    \mu = \frac{2\mathcal{\varkappa}m}{(D-2)\Omega_{(D-2)}} = \frac{m}{B} \, .
\end{equation}
This reduces to the well known expression $\mu = r_{\mathrm{g}} = 2G^{(4)}m$ in four dimensions. 
There is one possible variation left in \eqref{SSAction}, the one over $\rho(r)$. Instead of doing that, let us briefly analyze the form of the action
\begin{equation}
    S_{\mathrm{SS}} = - B \int dtdr H(N,\rho') \frac{\partial}{\partial r} F[f,\rho,\rho'] \, .
\end{equation}
Notice that the previous two equations are $ H' = 0$ and $ F'=0$, and that the third equation (taking the variation with respect to $\rho$) would correspond to  
\begin{equation}
    -\left(\frac{\partial H}{\partial \rho'}F' \right)' + \left( \frac{\partial F}{\partial \rho'} H'\right)' - \frac{\partial F}{\partial \rho}H' = 0 \, .
\end{equation}
We can see that this is just a linear combination of the first two equations; therefore, it does not represent an independent constraint on the solutions. Since it is sufficient to solve the first two equations, we may choose 
\begin{equation}
    \rho = r \, ,
\end{equation}
without losing generality. This choice is particularly nice since we obtain 
\begin{equation}
    N = 1 \quad \mathrm{and} \quad f = 1 - \frac{\mu}{r^{D-3}} \, .
\end{equation}
This corresponds to the Schwarzschild-Tangherlini metric.

\section{Radial Gauge Fixing}

The previous metric ansatz \eqref{SSSmetric} exhibits gauge freedom, as it remains invariant under transformations of the form $ r \to \tilde{r}(r) $. Notably, we can impose the gauge condition $ \rho = r $ from the outset without compromising our ability to obtain the full vacuum solutions through a spherical reduction of the action. In these radial coordinates, the metric assumes the form:
\begin{equation}
    ds^2 = N^2fdt^2 + f^{-1}dr^2 + r^2d\Omega_{d-2} \, ,
\end{equation}
 which now only contains two arbitrary functions $N(r)$ and $f(r)$. After spherical reduction, the Einstein-Hilbert action takes the form 
 \begin{align}
     S_{\mathrm{SSR}} &= B \int dt dr N'r^{D-1}p \\
     &= b\int dtdrN'[r^{D-3}(1-f)] \, .
 \end{align}
 Whose extrema are met when 
 \begin{equation}
     N' = 0 \quad \mathrm{and} \quad [r^{D-3}(1-f)]'=0 \, .
 \end{equation}
 Where the equation on the left corresponds to a variation over $f(r)$ while the equation on the right corresponds to a variation over $N(r)$. These equations are completely equivalent to the equations \eqref{eqNSS} and \eqref{eqFSS} evaluated in the radial gauge $\rho = r$, just as promised. This procedure is called radial static spherical reduction, and quantities obtained via this procedure are denoted by the subscript SSR.

\section{Scalar Curvature Invariants}

The explicit form of the Ricci scalar $R$ is given in the main body of the thesis in terms of the basic curvature invariants \eqref{BCI} in the gauge $\rho = r$. Some higher-order curvature invariants are
\begin{align}
R_{\mu}^{\nu} R_{\nu}^{\mu}
=\big[v+(D-2)q\big]^2+\big[v+(D-2)u\big]^2+(D-2)\big[(D-3)p+q+u\big]^2\,,
\end{align}
\begin{equation}
\begin{split}
R_{\mu}^{\nu} R_{\nu}^{\sigma}R_{\sigma}^{\mu}
=&\big[v+(D-2)q\big]^3+\big[v+(D-2)u\big]^3\\
&+(D-2)\big[(D-3)p+q+u\big]^3\,,
\end{split}
\end{equation}
\begin{equation}
C_{\alpha\beta\mu\nu} C^{\alpha\beta\mu\nu}=\frac{4(D-3)}{D-1}(p-q-u+v)^2\, .
\end{equation}

It is also possible to reverse these relations and express invariants $p$,$q$, $u$, and $v$ in terms of $R$, $C_{\alpha\beta\mu\nu} C^{\alpha\beta\mu\nu}$, $R_{\mu}^{\nu} R_{\nu}^{\mu}$, and $R_{\mu}^{\nu} R_{\nu}^{\sigma}R_{\sigma}^{\mu}$. However, this representation requires solving a cubic algebraic equation and is quite involved.

	\chapter{The Kerr metric}
\label{app:KerrMetric}

In 1963, Roy P. Kerr derived the most general axially symmetric and asymptotically flat vacuum solution to Einstein's equations \cite{PhysRevLett.11.237}. This describes a rotating mass and was made possible with the aid of the Goldberg-Sachs theorem \cite{Gold-Sachs}. This solution became known as the Kerr metric, which describes a rotating black hole. Various coordinate sets highlight different aspects of the metric. Here, we present it in one of the most commonly used forms, the Boyer-Lindquist coordinates:
\begin{equation}
\begin{aligned}\label{KerrBL} 
  ds^2 = & - \left(1 -\frac{2mr}{r^2+a^2\cos^2\theta} \right)dt^2 - \frac{4mra\sin^2\theta}{r^2 + a^2\cos^2\theta}dtd\phi  
+ \left( \frac{r^2 + a^2\cos^2\theta}{r^2 - 2mr +a^2} \right)dr^2 \\ &+ (r^2 + a^2\cos^2\theta)d\theta^2 
+ \left( r^2 + a^2 + \frac{2mra^2\sin^2\theta}{r^2+a^2\cos^2\theta}\right)\sin^2\theta d\phi^2 \, .
\end{aligned} 
\end{equation}
This form of the metric is especially beneficial because it has only one off-diagonal term. This allows us to determine that $m$ is the mass and $J = ma$ is the angular momentum, and $a \leq m$. As they align with the same parameters for the approximate metric of a rotating isolated body in the asymptotic limit. Furthermore, these coordinates effectively illustrate the concepts of the "event horizon" and "ergosphere" \cite{visser2008kerrspacetimebriefintroduction}. In these coordinates, the Krestchmann invariant takes the form
\begin{equation}
K = \frac{48m^2(r^2-a^2\cos^2\theta)((r^2+a^2\cos^2\theta)^2 - 16r^2a^2\cos^2\theta)}{(r^2 + a^2\cos^2\theta)^6}.
\end{equation}
Notice that it matches \eqref{KreschSchw} as $a\to0$. However, it diverges at $r=0$ and $\theta = \pi/2$. Remarkably, for a rotating black hole, this divergence occurs on a ring with a radius of $a$.

\section{Event Horizon}
We now study some of the important surfaces from this geometry. Let us start by taking the $g_{rr}$ component of the metric and noting that it is singular at 
\begin{equation}
  r_{\pm} = m \pm \sqrt{m^2 - a^2} \, .
\end{equation}
This is known as a coordinate singularity, but it is not a physically singular region of space-time. Moreover, as $a \to 0$, then $r_{+} \to 2m$, which is the location of the event horizon for the Schwarzschild black hole. This suggests that $r_{\pm}$ are the locations of two event horizons for the Kerr metric. To confirm this, one can verify the existence of a 4-vector $L^{\mu} = (L^t,0,L^{\theta},L^{\phi})$ that lies exactly on the 3-surfaces $r= r_{\pm}$, and satisfies
\begin{equation}
g_{\mu\nu}L^{\mu}L^{\nu} = 0 \, .
\end{equation}
Therefore, $L^{\mu}$ is a null vector corresponding to photon orbits along the $r= r_{\pm}$ surfaces. These photons never fall in or escape to infinity. In this context, these surfaces define the boundary from which null curves cannot escape to infinity, which we refer to as an event horizon. To explain further, we can define:
\begin{equation}
 L_{\pm}^{\mu} = (1,0,0,\Omega_{\pm}) \, , \quad \Omega_{\pm} = \frac{a}{r_{\pm}^2 + a^2} \, .
\end{equation}
Which satisfies 
\begin{equation}
g_{\mu\nu}(r_{\pm})L^{\mu}_{\pm}L^{\nu}_{\pm} = 0 \, .
\end{equation}
Interestingly, $\Omega_{\pm}$ is often referred to as the "angular velocity" of the event horizons, suggesting that the horizons rotate in a manner similar to a solid body.

\section{Ergosphere}
There is another surface of physical importance that can be described using the Boyer-Lindquist coordinates. Suppose you suddenly gain superpowers and decide to visit a rotating black hole. If you want to behold this majestic entity at a safe distance, ideally, you'd want to remain stationary. In other words, your 4-velocity should be:
\begin{equation}
u^{\mu} = (1,0,0,0) \, .
\end{equation}
You might be curious about what happens if you approach the horizon of a black hole, especially after traveling such a long distance. In such a case, you would follow a path known as a timelike geodesic, which implies that
\begin{equation}
g _{\mu \nu}u^{\mu}u^{\nu} <0 \, .
\end{equation}
Now, as you approach, you'll want to stop and stand still to enjoy the view. This would mean that $g_{tt}<0$ should be satisfied. However,
\begin{equation}
g_{tt} = - \left(1- \frac{2mr}{r^2 + a^2\cos^2\theta} \right) \, ,
\end{equation}
and there is a region where this quantity becomes positive, namely
\begin{equation}
r_E^{-} < r <r_E^{+} \, , \quad r_E^{\pm} = m \pm \sqrt{m^2 - a^2\cos^2\theta} \, .
\end{equation}
This means standing still would be impossible, as it would require moving along a space-like geodesic, which is not feasible. In other words, you'd have to co-rotate with the black hole but could still escape this region. The area between the outer event horizon $r_+$ and $r_E^+$ is known as the ergosphere, and the $r_E^{\pm}$ are known as the ergosurfaces. It's important to note that as the rotation ceases ($a\to0$), the ergosurface aligns with the event horizon. Therefore, Schwarzschild black holes do not have an ergosphere.

\section{Killing Vectors}

The last aspect to highlight is the presence of two Killing vectors $\xi^{\mu}_t$ and $\xi^{\mu}_{\phi}$. Killing vectors are vector fields that work as infinitesimal generators of isometries; they are defined via the Killing equation $\xi_{(\mu ; \nu)} = 0$. It's apparent from equation \eqref{KerrBL} that it's independent of both time and the azimuthal angle. This suggests the existence of two Killing vectors.
\begin{equation}
\xi_{t}^{\mu} = (1,0,0,0)\, , \quad \xi_{\phi}^{\mu} = (0,0,0,1) \, .
\end{equation}
Any linear combination of these two vectors also results in a Killing vector. The crucial characteristic of the time translation Killing vector $\xi_{t}^{\mu}$ is that it becomes a unit timelike vector at infinity. On the other hand, the $\xi_{\phi}^{\mu}$ Killing vector is distinguished by its integral lines being closed \cite{BHFrolZel}.

\section{Coordinate Systems}

\begin{figure}[!hbt]
    \centering
      \includegraphics[width=0.7\textwidth]{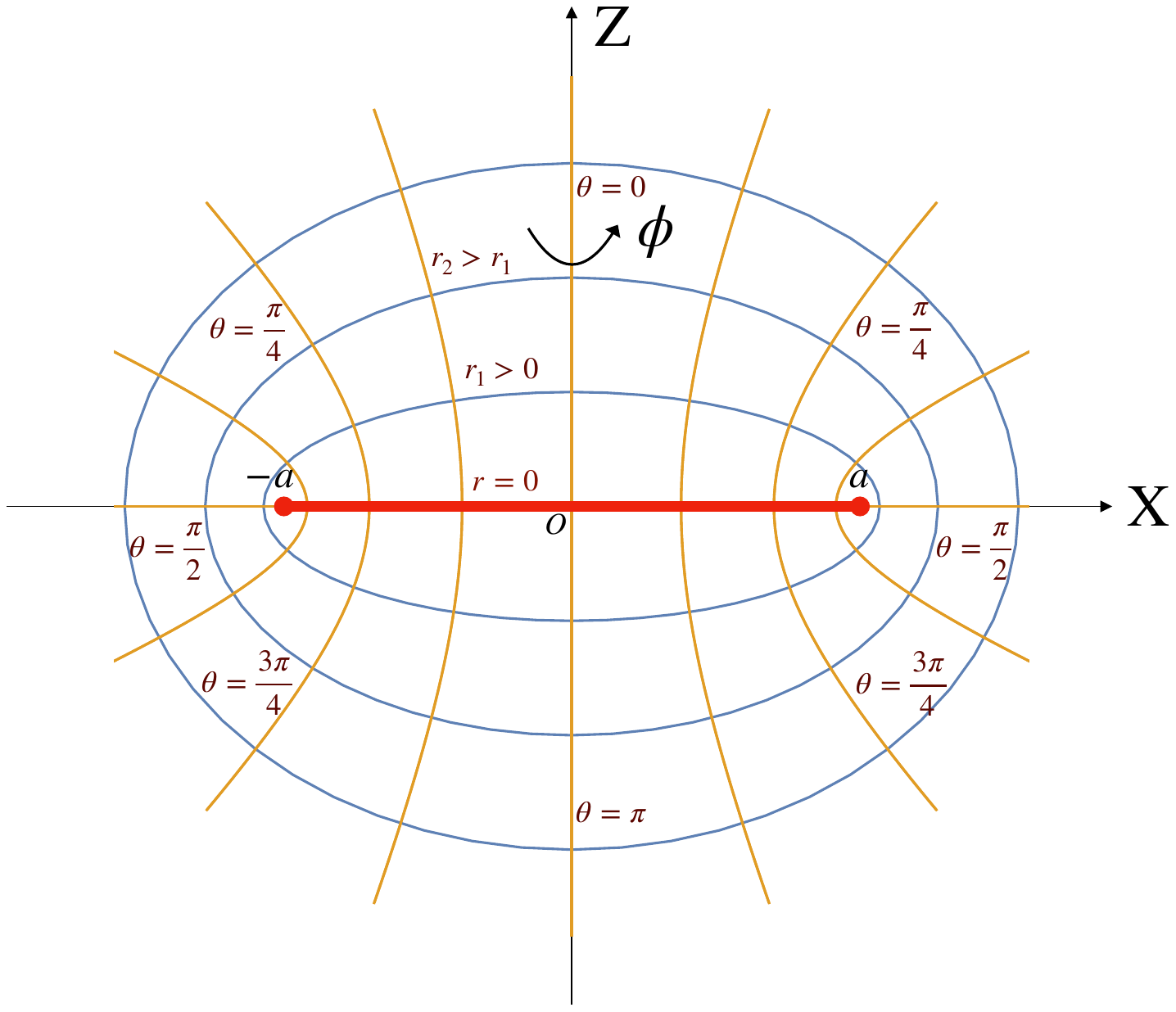}
    \caption{\label{SpheFig} Coordinate lines of the oblate spheroidal coordinates $(r,\theta)$ in the plane $Y=0$ ($\phi=0$)}
\end{figure}

\begin{figure}[!hbt]
    \centering
      \includegraphics[width=0.7\textwidth]{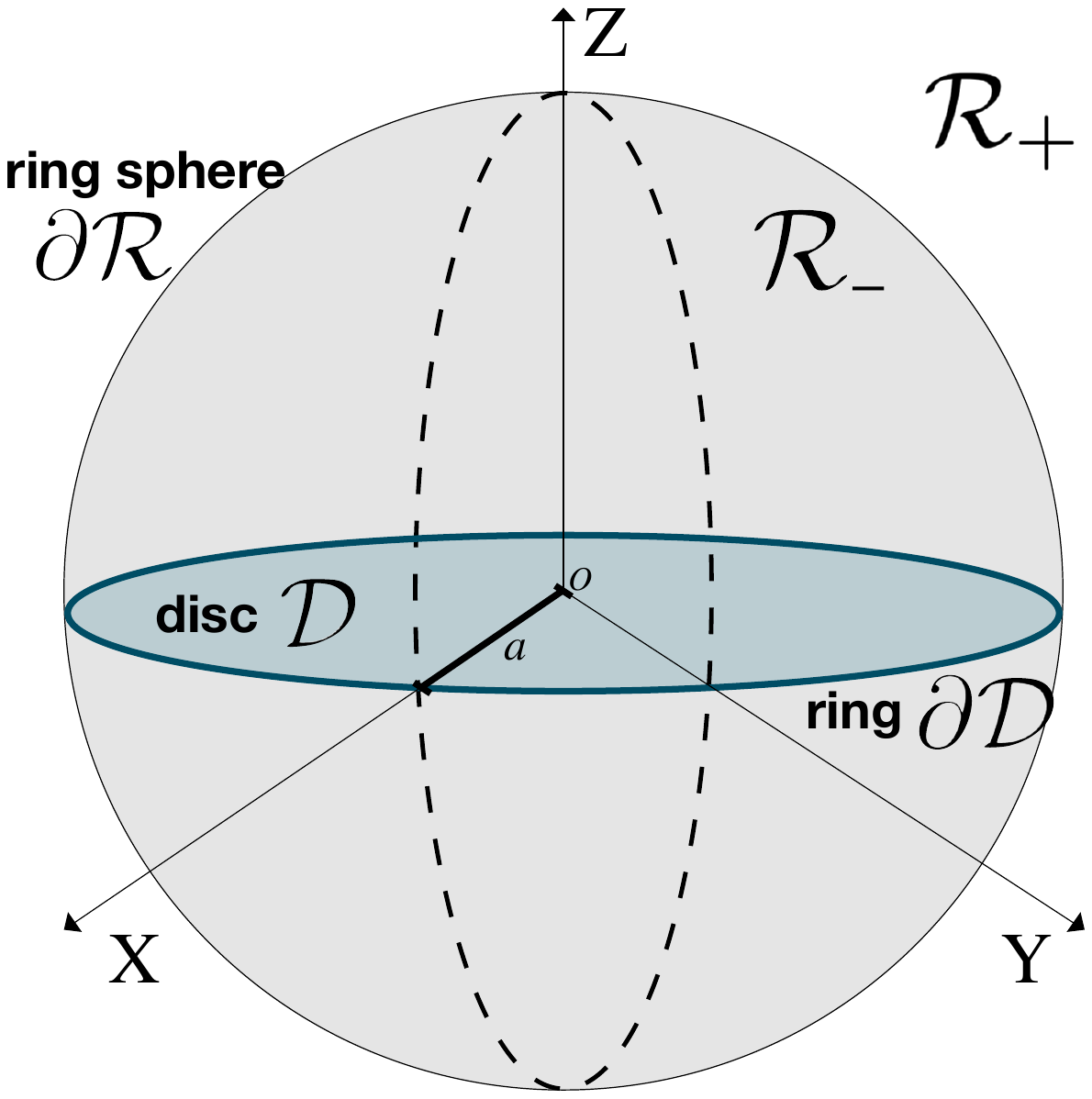}
    \caption{\label{C2} The ring $\pa\mathcal{D}$, the disc $\mathcal{D}$ and the ring sphere $\pa\mathcal{R}$.}
\end{figure}

We now find it convenient to talk about different coordinate systems which are used in chapter \eqref{Ch3}. The above form of the metric is written in oblate spheroidal coordinates, which are related to Cartesian coordinates via
\begin{equation}
  \begin{split}
    X &= \sqrt{r^2 + a^2}\sin\theta \cos\phi \, , \\
    Y &= \sqrt{r^2 + a^2}\sin \theta \sin \phi  \, ,\\
    Z &= r \cos\theta \, .
  \end{split}
\end{equation}
In the flat spacetime limit $m,a \to 0$, the Kerr metric is 
\begin{equation}
    ds_0^2 = -dt^2 + \frac{r^2 + a^2\cos^2\theta}{r^2 + a^2}dr^2 + (r^2 + a^2\cos^2\theta)d\theta^2 + (r^2 + a^2)\cos^2\theta d\phi^2 \, .
\end{equation}
The coordinates take values $r \geq 0$, $\theta \in [0,\pi]$ and $\phi \in [0,2\pi]$. The coordinate lines of this coordinate system are illustrated in Figure \eqref{SpheFig} for the $(r,\theta)$ plane, where $Y = 0$ and $\phi = 0$. For instance, the $r = \text{const}$ surfaces for all $r>0$ are oblate spheroids
There are a few important surfaces that are worth highlighting. For $r=0$ and $\theta \in [0,\pi]$, $\phi \in [0,2\pi]$ there is a disc $\mathcal{D}$ of radius $a$ in the $Z=0$ plane. The coordinate $\theta$ is discontinuous on this disc; it covers its upper half for $(0,\pi/2)$ and the lower half for $(\pi/2,\pi)$. Naturally, the boundary $\partial \mathcal{D}$ is a ring of radius $a$. The axis of symmetry at $X=Y=0$ is described by $\theta = 0$ and $\theta = \pi$. If $\theta = 0$, then $Z=r$ is positive, while for $\theta = \pi$ one has that $Z=-r$ is negative. Finally, there is a sphere $\partial \mathcal{R}$ of radius $a$ which is defined by $r = |a\cos\theta|$. This is an important surface as it divides space into two regions $\mathcal{R}_-$ and $\mathcal{R}_+$ for $r<|a\cos\theta|$ and $r>|a\cos\theta|$. These surfaces are illustrated in Figure \eqref{C2}
Another set of coordinates, which are used as an intermediate step, are the cylindrical coordinates $(\rho,z,\phi)$ related to Cartesian coordinates by
\begin{equation}
    \rho = \sqrt{X^2 + Y^2} \, , \quad z=Z \, .
\end{equation}
The Minkowski metric takes the form 
\begin{equation}
    ds_0^2 = -dt^2 + d\rho^2 + \rho^2d\phi^2 + dz^2 \, ,
\end{equation}
in these coordinates, which are also related to the oblate spheroidal coordinates via 
\begin{equation}
    \rho = \sqrt{r^2 + a^2}\sin\theta \, , \quad z = r\cos\theta \, .
\end{equation}
The ring $\partial\mathcal{D}$ is defined by $\rho = a$ and $z=0$ in this coordinate system.

\section{Canonical Coordinates and Hidden Symmetries}

To delve deeper into the symmetries of the Kerr metric, it's helpful to introduce a set of coordinates referred to as the "canonical" coordinates. Specifically, this involves performing the following transformation:
\begin{equation}
y = a\cos\theta \, , \quad \psi = \phi/a \, , \quad \tau = t-a\phi \, .
\end{equation}
The metric becomes 
\begin{equation}\label{KerrCan}
ds^2 = \frac{1}{\Sigma}(-\Delta_r(d\tau + y^2d\psi)^2 + \Delta_y(d\tau - r^2d\psi)^2) + \Sigma\left(\frac{dr^2}{\Delta_r} +\frac{dy^2}{\Delta_y}\right)
\end{equation}
where 
\begin{equation}
\Sigma = r^2 + y^2 \, , \Delta_r = r^2-2mr + a^2 \, , \Delta_y = a^2 - y^2 \, .
\end{equation}
This form of the metric is useful as it highlights the so-called \textit{hidden symmetries} \cite{LivreKerr}. These are defined as conserved quantities of the geodesic equations that are second order or higher in momentum. The generators of these symmetries are tensors, known as Killing tensors and defined by the equation $K_{(\mu_1 .. \mu_n ; \alpha)} = 0$. In the case of the metric \eqref{KerrCan}, the said tensor takes this form.
\begin{equation}
K^{\mu}_{\ \ \nu} = \begin{pmatrix}
0 & 0 & 0 & r^2y^2 \\
0 & y^2 & 0 & 0 \\
0 & 0 & -r^2 & 0 \\
1 & 0 & 0 & -(r^2 - y^2)
\end{pmatrix} \, .
\end{equation}
This tensor, together with the Killing vectors $\xi_{\tau}^{\mu} = (1,0,0,0)$ and $\xi_{\psi}^{\mu} = (0,0,0,1)$ are responsible for the complete integrability of the geodesic equations of motion, as they guarantee that we have enough first integrals.

	\chapter{Kerr-Schild Double Copy Formalism}
\label{app:KerrSchild}

\section{Kerr-Schild Metrics}
When the Kerr metric was first derived, it was presented in the form 
\begin{equation}
ds^2 = ds_0^2 + \frac{2mr}{r^2 + a^2\cos^2\theta}(d\hat{t} + a\sin^2\theta d\hat{\phi} + dr)^2 \, ,
\end{equation}
Where $ds_0^2$ is the Minkowski metric in these coordinates, and it is given by 
\begin{equation}
\begin{aligned}
ds_0^2 &= -d\hat{t}^2 + dr^2 + (r^2 + a^2\cos^2\theta)d\theta^2 \\
&\quad + (r^2 + a^2)\sin^2\theta d\hat{\phi}^2 + 2a\sin^2\theta d\hat{\phi}d\hat{t} \, .
\end{aligned}
\end{equation}
This is a coordinate system in which the metric has three non-diagonal terms, which may make it seem more complicated than the Boyer-Lindquist coordinates, but notice something interesting about this form: it is split into one part that is independent of mass and one which isn’t. In other words, the metric can be interpreted as flat spacetime plus a perturbation. Explicitly 
\begin{equation}\label{KerrSchildappB}
ds^2 = ds_0^2 + \Phi(\ell_{\mu}dx^{\mu})^2
\end{equation}
Where $ds_0^2$ is the flat spacetime, $\Phi$ is the perturbation and $\ell_{\mu}$ is a special type of null vector. That the first time the metric was written down was in this form is in fact is not a coincidence, when exploring the possible solutions to Einstein’s vacuum field equations for a rotating mass the guiding principle was to follow the (back then recently discovered) Goldberg-Sachs theorem \cite{Gold-Sachs}.
\subsection{Goldberg-Sachs Theorem}

Before delving into the theorem, let's define some key terms. We start with the conformal (also known as Weyl) tensor:
\begin{equation}
\begin{aligned}
    C_{\mu\nu\alpha\beta} &= R_{\mu\nu\alpha\beta} +\frac{1}{6}R\left(g_{\mu\alpha}g_{\nu\beta} - g_{\mu\beta}g_{\nu\alpha} \right) \\
    &\quad -\frac{1}{2}\left(g_{\mu\alpha} R_{\nu \beta} + g_{\nu\beta}R_{\mu\alpha} - g_{\mu\beta}R_{\nu\alpha} - g_{\nu\alpha}R_{\mu\beta} \right)
\end{aligned}
\end{equation}

It has been demonstrated that this tensor possesses four special vectors, also known as the principal null vectors (PNVs), $\kappa^{\mu}\neq0$, which satisfy the following algebraic properties:
\begin{equation}
\begin{aligned}
    \kappa_{[\mu}C_{\nu]nm[\alpha}\kappa_{\beta]}\kappa^{n}\kappa^m &= 0 \, , \\
    \kappa_{\mu}\kappa^{\mu} &= 0 \, .
\end{aligned}
\end{equation}
Here, the square brackets denote antisymmetrization. In essence, these vectors can be regarded as "eigenvectors" of the conformal tensor and correspond to four congruences. A spacetime is called \textit{algebraically special} if at least two of these vectors coincide everywhere. Alternatively, such vectors can be demanded to satisfy:
\begin{equation}
\begin{aligned}
    \kappa^{\mu}\kappa^{\alpha}C_{\mu\nu\alpha[\beta}\kappa_{\sigma]} &= 0 \, .
\end{aligned}
\end{equation}
For completeness, let's consider a set of null geodesics defined by $x^{\mu} = x^{\mu}(\lambda)$
where $\lambda$ is the affine parameter along the geodesics. Then,
\begin{equation}
\kappa^{\mu} = \frac{\partial x^{\mu}}{\partial \lambda} \, , \quad \kappa^{\mu}_{;\nu}\kappa^{\nu} = 0 \, .
\end{equation}
We will additionally demand that a congruenece of such null vectors is shear-free, where by shear we mean the tendency of a cross sectional area enclosing the null congruence to become distorted, it is defined by the tensor  \cite{Frolov1979,sommers1976}
\begin{equation}
\sigma_{\mu \nu} = \kappa_{(\mu ; \nu)} - \frac{1}{2}\kappa^{\alpha}_{\ ;\alpha}h_{\mu \nu} \, ,
\end{equation} 
where $h_{\mu\nu} = g_{\mu\nu} + \kappa_{\mu}N_{\nu} + \kappa_{\nu}N_{\mu}$ is a transversal metric that acts a spatial projector operator and $N^{\mu}$ is an auxiliary null vector which satisfies $\kappa_{\mu}N^{\mu}$ = -1. The shear free condition is $\sigma_{\mu\nu}\sigma^{\mu\nu} = \sigma^2 = 0$ or
\begin{equation}
\kappa_{(\mu;\nu)}\kappa^{(\mu;\nu)} - \frac{1}{2} \left(\kappa^{\mu}_{\ ; \mu}\right)^2  = 0  \, .
\end{equation}

\begin{theorem}
A vacuum metric $R_{\mu\nu} = 0$ is algebraically special if and only if it contains a shear-free null geodesic congruence. 
\end{theorem}

This theorem \cite{Gold-Sachs} holds significant importance as it establishes a direct relationship between vacuum solutions to Einstein’s field equations and a special type of null eigenvectors of the conformal tensor. Notably, the only physically relevant metrics discovered before the Kerr metric, namely, Schwarzschild and plane-fronted waves, are both algebraically special. This realization enabled systematic exploration of new solutions with the PNVs as a starting point, precisely what occurred with Kerr.

Both the Schwarzschild and Kerr metrics can be expressed in the form of equation \eqref{KerrSchildappB} where the null vectors $\ell^{\mu}$ are, in fact, PNVs of each spacetime \cite{KerrSchild}. Exploring spacetimes in a form that explicitly utilizes their algebraic properties reveals intriguing insights. 

\section{Double Copy Mechanism}
The Kerr-Schild metrics have garnered recent attention due to the connection they make between stationary vacuum Einstein solutions and vacuum Maxwell equations \cite{BHdoublecopy,doubleLW,ball2023,Chawla_2023,LUNA2015272,dcgluon}. This relationship can be outlined as follows: starting with the metric \eqref{KerrSchildappB}, the Einstein vacuum field equations yield
\begin{equation}\label{DC1}
\partial_{\nu}\left(\partial^{\alpha}(\Phi\ell^{\nu}\ell_{\beta}) + \partial_{\beta}(\Phi\ell^{\nu}\ell^{\alpha}) - \partial^{\nu}(\Phi\ell^{\alpha}\ell_{\beta})\right) = 0 \, .
\end{equation}
With the background metric $\eta_{\mu\nu}$ used to lower and raise indices. Given our assumption of spacetime stationarity, we can adopt a gauge where $\partial_0(\Phi\ell^{\nu}\ell^{\alpha}) = 0$ and fix $\ell_0 = 1$. This simplifies the equation \eqref{DC1} to the vacuum Maxwell equations when $\beta = 0$:
\begin{equation}
\partial_{\nu}\left[\partial^{\alpha}(\Phi\ell^{\nu}) - \partial^{\nu}(\Phi\ell^{\alpha}) \right] = 0 \, .
\end{equation}
Defining $A^{\mu} \equiv \Phi\ell^{\mu}$ renders the above equation equivalent to
\begin{equation}
\partial_{\nu}F^{\alpha\nu} = 0 \, ,
\end{equation}
where $F^{\alpha\nu} = \partial^{\alpha}A^{\nu} - \partial^{\nu}A^{\alpha}$ represents the electromagnetic strength tensor in flat spacetime. The Kerr-Schild metric not only employs an explicit use of the principal null vectors of the conformal tensor but also naturally selects the electromagnetic gauge field $A^{\mu}$. Some in the literature refer to this field as the single copy of the metric $g_{\mu\nu}$.
This procedure extends to non-flat backgrounds \cite{Bahjat-Abbas:2017htu}, though here we only consider a Minkowski background. Before presenting examples to illustrate this mechanism, let's make a few annotations. For a gauge field to serve as a single copy, it must be geodesic and null:
\begin{equation}
A_{\mu}A^{\mu} = 0 \, , \quad A^{\mu}\partial_{\mu}A^{\nu} = gA^{\nu} \, ,
\end{equation}
where $g$ is a scalar function. Combining these requirements, we realize that
\begin{equation}
-F^{\nu}_{\ \mu}A^{\mu} = gA^{\nu} \, .
\end{equation}
Hence, the condition for the field $A^{\mu}$ to be geodesic and null is equivalent to demanding that it be an eigenvector of its own electromagnetic strength field. This latter requirement is more succinct and thus more practical. The process of using two copies of the gauge field $A^{ \mu}$ to construct a metric is known as the double copy mechanism in the literature.
This mechanism provides a means of relating solutions from electrodynamics to solutions in general relativity and vice versa. One can start by expressing a spacetime in explicit Kerr-Schild form and then compute the gauge field $A^{\mu}$. Alternatively, and of primary interest, one can derive solutions for the field of charge in spacetime and construct the corresponding metric.
\section{Examples}
Navigating the bidirectional nature of the established relations might seem daunting, so let's illustrate the process through a few examples, focusing on deriving Einstein solutions from Maxwell solutions. Suppose $A^{\mu}$ represents a solution of Maxwell’s equations, such as a Lienard-Wiechert field, which also serves as an eigenvector of its strength tensor. In this scenario, the corresponding Kerr-Schild double copy would be
\begin{equation}
g_{\mu\nu} = \eta_{\mu\nu} + \frac{1}{\Phi}A_{\mu}A_{\nu} \, .
\end{equation}
As the function $\Phi$ remains arbitrary, additional conditions are necessary. One viable condition is to demand that the corresponding stress-energy tensor $T^{\mu\nu}$ be traceless, a sensible criterion for an electromagnetic stress-energy tensor \cite{Landau:1975pou}. This condition yields a differential equation that can be solved to determine $\Phi$. Notably, for the most physically relevant scenarios, this differential equation simplifies to the Poisson equation:
\begin{equation}\label{Poiss}
\Delta\Phi = 0 \, .
\end{equation}
This result serves as the starting point for constructing Kerr-Schild metrics via the double copy mechanism. To find the null vectors $\ell^{\mu}$, we examine the structure of the gauge field $A^{\mu}$ from which we derive our metric, ensuring they satisfy all algebraic properties defined by the Goldberg-Sachs theorem.
\subsection{Schwarzschild Double Copy}
For our first example, let's consider a point-like source at the origin in $(t,r,\theta,\phi)$ coordinates. Solving \eqref{Poiss}, we find $\Phi = \frac{2 \kappa m}{r}$ (with $\kappa$ as the coupling constant), and the corresponding Lienard-Wiechert field for a point-like source in these coordinates is
\begin{equation}
A^{\mu} = \frac{2\kappa q}{r}(1,1,0,0) \to \ell^{\mu}=(1,1,0,0) \, .
\end{equation}
Constructing a metric with these components,
\begin{equation}
ds^2 = ds_0^2 + \frac{2\kappa m}{r}(dt+dr)^2 \, ,
\end{equation}
we find that the Schwarzschild black hole is the double copy of a real stationary charge. By eliminating the cross term, we can rewrite this metric in the usual form using the transformation
\begin{equation}
dt = dt_S + \frac{\Phi}{1-\Phi}dr \, .
\end{equation}
The subscript “$S$” indicates the "Schwarzschild" time coordinate. Notably, the coordinates $(t,r,\theta,\phi)$ cover both the exterior and interior of the black hole, remaining regular at the event horizon.

\subsection{Kerr Double Copy}
For the second example, consider a point-like source located in the complex Minkowski plane $x^{\mu}+iy^{\mu}$. Let's transform our background metric to spheroidal coordinates and take $r \to r+ia\cos\theta$ (for a similar procedure see \cite{adamo2016kerrnewmanmetricreview}). We find
\begin{equation}
\Phi = \frac{2\kappa m}{r+ia\cos\theta} = 2\kappa m\frac{r-ia\cos\theta}{r^2 + a^2\cos^2\theta} \, .
\end{equation}
We're interested only in the real part of this function, equivalent to placing our source at the position $-i\vec{a}$. Now, let's relate the complex Lienard-Wiechert field of a charge in complex Minkowski space with the function $\Phi$. Such a gauge field can be expressed as
\begin{equation}
A^{\mu} =\frac{2\kappa qr}{r^2+a^2\cos^2\theta}(1,-1,0,-\frac{a}{r^2+a^2}) \, ,
\end{equation}
yielding $\ell^{\mu} = (1,-1,0,-\frac{a}{r^2+a^2})$. The Kerr-Schild double copy is then
\begin{equation}
ds^2 = ds_0^2 + \frac{2\kappa m r}{r^2+a^2\cos^2\theta}(dt-dr-\frac{a}{r^2+a^2}d\phi)^2 \, ,
\end{equation}
which represents the Kerr metric.



\end{document}